\documentclass[]{aa}           
\usepackage[usenames,dvipsnames]{xcolor}                     
\usepackage{graphicx}       
\usepackage[varg]{txfonts}        
\usepackage{lscape}
\usepackage{rotating}
\usepackage{mathrsfs}
\usepackage{ulem}
\usepackage{natbib}
\bibpunct{(}{)}{;}{a}{}{,} % to follow the A&A style
\begin{document}        
        
\def\ang{{\rm\AA}}        
        
\def\n{\noindent}        
        
\def\ph{\phantom}

\newcommand{\fuse}{\rm{FUSE}}
\newcommand{\stis}{\rm{STIS}}
\newcommand{\cosp}{\rm{COS}}
\newcommand{\hst}{\emph{HST}}
\newcommand{\eso}{\emph{ESO}}
\newcommand{\uves}{\rm{UVES}}
\newcommand{\aat}{\emph{AAT}}
\newcommand{\flames}{\emph{FLAMES}}
\newcommand{\xshooter}{{\sc XSHOOTER}}
\newcommand{\ergs}{ergs\,cm$^{-2}$\,s$^{-1}$\,\AA$^{-1}$}
\newcommand{\halpha}{\mbox{H$\alpha$}}
\newcommand{\hbeta}{\mbox{H$\beta$}}
\newcommand{\hgamma}{\mbox{H$\gamma$}}
\newcommand{\hdelta}{\mbox{H$\delta$}}
\newcommand{\bralpha}{\mbox{Br$\alpha$}}
\newcommand{\brgamma}{\mbox{Br$\gamma$}}
\newcommand{\lyalpha}{\mbox{Ly$\alpha$}}
\newcommand{\lybeta}{\mbox{Ly$\beta$}}
\newcommand{\lygamma}{\mbox{Ly$\gamma$}}
\newcommand{\lydelta}{\mbox{Ly$\delta$}}
\newcommand{\htwo}{\mbox{H$_{2}$}}
\newcommand{\lb}{$\lambda$}
\newcommand{\mdot}{$\dot{\rm M}$}
\newcommand{\gcmcube}{g\,cm$^{-3}$}
\newcommand{\gcmcarre}{g\,cm$^{-2}$}
\newcommand{\vsini}{$v\sin\,i$}
\newcommand{\vrad}{$v_{\rm rad}$}
\newcommand{\vmc}{$v_{\rm MC}$}
\newcommand{\flux}{ergs\,cm$^{-2}$\,s$^{-1}$}
\newcommand{\teff}{T$_{\rm eff}$}
\newcommand{\logteff}{$\log T_{\rm eff}$}
\newcommand{\logg}{$\log g$}
\newcommand{\loggc}{$\log g_{\rm c}$}
\newcommand{\met}{[Fe/H]}
\newcommand{\vturb}{$v_{\rm turb}$}
\newcommand{\vmax}{$v_{\rm max}$}
\newcommand{\vmin}{$v_{\rm min}$}
\newcommand{\vcl}{$v_{\rm cl}$}
\newcommand{\vinf}{$\rm v_{\infty}$}
\newcommand{\finf}{$f_{\infty}$}
\newcommand{\vt}{$v_{turb}$}
\newcommand{\vrot}{$v_{rot}$}
\newcommand{\vmac}{$v_{\rm mac}$}
\newcommand{\radius}{$R_{*}/R_{\odot}$}
\newcommand{\rstar}{$R_{*}$}
\newcommand{\mstar}{M$_{*}$}
\newcommand{\zsun}{${\rm Z}_{\odot}$}
\newcommand{\lsun}{${\rm L}_{\odot}$}
\newcommand{\zzsun}{${\rm Z/Z}_{\odot}$}
\newcommand{\rsun}{R$_{\odot}$}
\newcommand{\msun}{M$_{\odot}$}
\newcommand{\tlusty}{{\sc Tlusty}}
\newcommand{\synspec}{{\sc Synspec}}
\newcommand{\cmfgen}{{\sc CMFGEN}}
\newcommand{\fastwind}{{\sc FASTWIND}}
\newcommand{\wmbasic}{{\sc WM-BASIC}}
\newcommand{\ullyses}{{\sc ULLYSES}}
\newcommand{\xshootu}{{\sc XShootU}}
\newcommand{\Av}{$A_{\rm v}$}
\newcommand{\ebv}{$\rm E(B-V)$}
\newcommand{\Rv}{$R_{\rm v}$}
\newcommand{\Mv}{$M_{\rm v}$}
\newcommand{\msp}{M$_{\rm spec}$}
\newcommand{\mev}{M$_{\rm evol}$}
\newcommand{\mshrd}{M$_{\rm sHRD}$}
\newcommand{\mhrd}{M$_{\rm HRD}$}

\newcommand{\grid}{{\sc Ostar2002}}
\newcommand{\kms}{km\,s$^{-1}$}
\newcommand{\cun}{cm$^{-1}$\ }
\newcommand{\cdeux}{cm$^{-2}$\ }
\newcommand{\ctrois}{cm$^{-3}$\ }
\newcommand{\ci}{\mbox{C~{\sc i}}}
\newcommand{\cii}{\mbox{C~{\sc ii}}}
\newcommand{\ciii}{\mbox{C~{\sc iii}}}
\newcommand{\civ}{\mbox{C~{\sc iv}}}
\newcommand{\cv}{\mbox{C~{\sc v}}}
\newcommand{\hi}{\mbox{H~{\sc i}}}
\newcommand{\hyd}{\mbox{H}}
\newcommand{\he}{\mbox{He}}
\newcommand{\hei}{\mbox{He~{\sc i}}}
\newcommand{\heii}{\mbox{He~{\sc ii}}}
\newcommand{\oii}{\mbox{O~{\sc ii}}}
\newcommand{\oiii}{\mbox{O~{\sc iii}}}
\newcommand{\oiv}{\mbox{O~{\sc iv}}}
\newcommand{\ov}{\mbox{O~{\sc v}}}
\newcommand{\ovi}{\mbox{O~{\sc vi}}}
\newcommand{\ovii}{\mbox{O~{\sc vii}}}
\newcommand{\nii}{\mbox{N~{\sc ii}}}
\newcommand{\niii}{\mbox{N~{\sc iii}}}
\newcommand{\niv}{\mbox{N~{\sc iv}}}
\newcommand{\nv}{\mbox{N~{\sc v}}}
\newcommand{\fe}{\mbox{Fe~}}
\newcommand{\neon}{\mbox{Ne~}}
\newcommand{\p}{\mbox{P~}}
\newcommand{\piv}{\mbox{P~{\sc iv}}}
\newcommand{\pv}{\mbox{P~{\sc v}}}
\newcommand{\pvi}{\mbox{P~{\sc vi}}}
\newcommand{\siii}{\mbox{Si~{\sc ii}}}
\newcommand{\siiii}{\mbox{Si~{\sc iii}}}
\newcommand{\siiv}{\mbox{Si~{\sc iv}}}
\newcommand{\siv}{\mbox{S~{\sc iv}}}
\newcommand{\sv}{\mbox{S~{\sc v}}}
\newcommand{\svi}{\mbox{S~{\sc vi}}}
\newcommand{\feiii}{\mbox{Fe~{\sc iii}}}
\newcommand{\feiv}{\mbox{Fe~{\sc iv}}}
\newcommand{\fev}{\mbox{Fe~{\sc v}}}
\newcommand{\fevi}{\mbox{Fe~{\sc vi}}}
\newcommand{\msunyr}{M$_{\odot}$\,yr$^{-1}$}
\newcommand{\pcyg}{P~Cygni}
\newcommand{\xit}{$\xi_{t}$}
\newcommand{\epv}{$\times\,10^{5}$}
\newcommand{\ev}{$\times\,10^{-5}$}
\newcommand{\evi}{$\times\,10^{-6}$}
\newcommand{\evii}{$\times\,10^{-7}$}
\newcommand{\eviii}{$\times\,10^{-8}$}
\newcommand{\eix}{$\times\,10^{-9}$}
\newcommand{\ex}{$\times\,10^{-10}$}
\newcommand{\logL}{$\log L/L_{\odot}$}
\newcommand{\lognHI}{$\log N$(H~{\sc i})}
\newcommand{\lognHII}{$\log N({\rm H}_{2})$}
\newcommand{\logLX}{$\log L_{\rm X}/L_{\rm bol}$}

\title{Massive stars in the Small Magellanic Cloud}
\subtitle{Evolution, rotation, and surface abundances
\thanks{This research is based on observations made with the NASA/ESA Hubble Space Telescope obtained from the Space Telescope Science Institute, which is operated by the Association of Universities for Research in Astronomy, Inc., under NASA contract NAS 5-26555. These observations are associated with programmes GO 7437, GO 9434, and GO 11625}
\thanks{Based on data products from observations made with ESO Telescopes at the La Silla Paranal Observatory under programmes ID 67.D-0238, 70.D-0164, 074.D-0109, 079.D-0073, and 079.D-0562}
%\thanks{Appendix B is only available in electronic format.}
}

\author{J.-C. Bouret\inst{1},  
               F. Martins \inst{2}, D. J. Hillier\inst{3}, , W. L .F. Marcolino\inst{4}, H. J. Rocha-Pinto\inst{4}, C. Georgy\inst{5},  T. Lanz\inst{6},   
                 I. Hubeny\inst{7}}        
 
 \offprints{J.-C. Bouret}

 \institute{Aix Marseille Univ, CNRS, CNES, LAM, Marseille, France \\
 \email{Jean-Claude.Bouret@lam.fr}
 \and
      LUPM, Universit\'e de Montpellier, CNRS, Place Eug\`ene Bataillon, F-34095 Montpellier Cedex 05, France
      \and
             Department of Physics and Astronomy \& Pittsburgh Particle physics, Astrophysics, and Cosmology Center (PITT PACC), University of Pittsburgh,  Pittsburgh, PA 15260, USA  
        \and
        Universidade Federal do Rio de Janeiro, Observat\'orio do Valongo. Ladeira Pedro Ant\^onio, 43, CEP 20080-090,
           Rio de Janeiro, Brazil
        \and
       Department of Astronomy, University of Geneva, Maillettes 51, CH-1290, Versoix, Switzerland
                        \and
        Observatoire de la C\^ote d'Azur, Universit\'e C\^ote d'Azur, 06304 Nice, France
                                    \and
  Steward Observatory, University of Arizona, AZ, USA
    }
%       
        
%\date{ }        
  \date{Received ; Accepted }        

\abstract{The evolution of massive stars depends on several physical processes and parameters. Metallicity and rotation are among the most important, but their quantitative effects are not well understood.}
{To complement our earlier study on main-sequence stars, we study the evolutionary and physical properties of evolved O stars in the Small Magellanic Cloud (SMC). We focus in particular on their surface abundances to further investigate the efficiency of rotational mixing as a function of age, rotation, and global metallicity.}
  % methods heading (mandatory)
   {We analysed the UV and optical spectra of 13 SMC O-type giants and supergiants using the stellar atmosphere code \cmfgen\ to derive photospheric and wind properties. We compared the inferred properties to theoretical predictions from evolution models. For a more comprehensive analysis, we interpret the results together with those we previously obtained for O-type dwarfs. }
  % results heading (mandatory)
 {Most dwarfs of our sample lie in the early phases of the main sequence. For a given initial mass, giants are farther along the evolutionary tracks, which confirms that they are indeed more evolved than dwarfs. Supergiants have higher initial masses and are located past the terminal-age main-sequence in each diagram. We find no clear trend of a mass discrepancy, regardless of the diagram that was used to estimate the evolutionary mass. Surface CNO abundances are consistent with nucleosynthesis from the \mbox{CNO} cycle. Comparisons to theoretical predictions reveal that the initial mixture is important when the observed trends in the N/C versus N/O diagram are to be reproduced. A trend for stronger chemical evolution for more evolved objects is observed. Above about 30~\msun, more massive stars are on average more chemically enriched at a given evolutionary phase. Below 30~\msun, the trend vanishes. This is qualitatively consistent with evolutionary models. A principal component analysis of the abundance ratios for the whole (dwarfs and evolved stars) sample supports the theoretical prediction that massive stars at low metallicity are more chemically processed than their Galactic counterparts. Finally, models including rotation generally reproduce the surface abundances and rotation rates when different initial rotational velocities are considered. Nevertheless, for some objects, a stronger braking and/or more efficient mixing is required.
 }
{}


\authorrunning{Bouret et al.}        
\titlerunning{Massive stars at low metallicity}        
        
\maketitle        
     
%%%%%%%%%%%%%%%%%%%%%%%%%%%%%%%%%%%%%%%%%%%%%%%%%%%%%%%%%%%%%%%%%%%%%%%%        
        
\section{Introduction}        
\label{intro_sect} 
Although the prime properties of a (single) massive star are set by its initial mass and metallicity, its ultimate destiny and actual path in the Hertzsprung-Russell diagram (HRD) critically depend on mass loss and rotation \citep{puls08, langer12}. 
Furthermore, rotation, the amount of mixing, the metal content, and the mass-loss rate form a feedback loop where more rotation leads to more mixing and changes in the mass-loss rate \citep[e.g.][]{gagnier19}, which then affect the rotation rate, and so on. However, the relative roles that these factors play in every phase of single massive-star evolution remain unclear, and evolutionary sequences linking 
various types of massive stars are poorly defined. 

The effects of rotation include an increase in mass-loss rate, a change in the evolutionary tracks in the HRD, a lowering of the effective
gravity, an extension of the main-sequence (MS) phase, and the mixing of CNO-cycle processed material up to the stellar surface \citep{maeder00, heger00}.
A direct consequence of including rotation in stellar evolution calculations is that for a given metallicity, each point in the HRD cannot be uniquely associated with a unique zero-age main-sequence (ZAMS) mass.

In stellar evolution codes different settings and prescriptions for rotation-related quantities, such as the transport of chemicals and angular momentum (purely diffusive or advecto-diffusive), instabilities (e.g.Tayler-Spruit dynamo for the internal magnetic fields), shear instabilities, and meridional circulation affect the global structure of a massive star, alter the stellar evolution and the path of a star in the HRD. 
\citep{brott11a, ekstrom12, langer12}.

In the past decade, several studies have investigated the link between surface abundance patterns and rotation, yielding controversial results.
The ${\sl VLT}$-FLAMES\footnote{Very Large Telescope-Fibre Large Array Multi-Element Spectrograph} survey \citep{evans05, evans06} and ${\sl VLT}$-FLAMES Tarantula Survey \citep[VFTS,][]{evans11} found that the majority of the massive OB stars they studied in the Large Magellanic Cloud (LMC) follow the predictions of models well, including rotation. However, the remaining 30-40\% of their sample stars present surface nitrogen abundances that cannot be understood in the current framework of rotational mixing for single stars,
for instance populations of slowly rotating nitrogen-enriched stars, or 
rapidly rotating OB-type stars that show little or no nitrogen enhancement
\citep[e.g.][]{hunter07, hunter08, hunter09, rivero12, grin17}. 

At higher metallicity, subsequent studies of surface abundances of OB stars in the Galaxy lead to somewhat different results. \cite{hunter09} did not identify clear signs of chemical enrichment at the surface of their B stars, in contrast to \citet{morel06,morel08}. This highlighted the need for samples that include stars that are relatively evolved off the ZAMS to test the effects of rotation. \cite{martins15} analysed a large sample of 74 (single) Galactic O-type stars of all spectral types and luminosity classes  (20-50 \msun\ mass range) that were observed in the context of the MiMeS survey of massive stars \citep{wade16}. In this sample, the surface abundances could be explained by models with rotation in a large majority of stars \citep[see also][]{martins17}.

To understand the effect of rotational mixing on the chemical evolution of massive stars, studies of stars in the SMC are particularly important because the metallicity in the SMC is only 0.2 \zsun\ and half that of the LMC. Rotational mixing is indeed expected to be more efficient at lower metallicities \citep[e.g.][]{meynet05, georgy13}, because of $\it i)\rm$ steeper internal gradient of the angular velocity and $\it ii)\rm$ reduced loss of angular momentum caused by lower mass-loss rates.
 Furthermore, the interplay between stellar rotation and the stellar wind, which blurs the initial rotational velocity properties of the stars within only a few million years in the case of Galactic massive stars, is expected to be reduced with decreasing metal content. The initial conditions of rotational velocity remain better preserved during the MS life of O-type stars at low metallicity. 

Studies of massive star properties at the lower metallicity of the SMC are still relatively rare. Although the fraction of slowly rotating SMC stars with strong nitrogen surface enrichment is similar to the LMC \citep{hunter09}, only  upper limits on the nitrogen abundance of rapidly rotating B-type stars (with little or no nitrogen enhancement) could be obtained. \citet{dufton20} also reported a similar number of nitrogen-rich slowly rotating stars in NGC~346 and in the LMC.

\cite{ bouret13} analysed ultraviolet (UV) and optical spectra of 23 O dwarfs in the SMC and showed that a majority of stars, regardless of their mass, have
abundance ratios (N/C versus N/O) that match the predictions of stellar evolution models. The fraction of stars close to the ZAMS 
that show unexplained differences to the standard evolution is 9\%, which is far smaller than the 40\% outliers of OB-type core-hydrogen burning stars found in the LMC \citep{hunter09, grin17}.

Following this initial work, we now focus on more evolved objects. We adopt the same observational strategy and 
modelling method for optimal consistency between both studies. 

The paper is organised as follows. In Sect. \ref{sect_obs} we present the observational data sets used in this work. In Sect. \ref{sect_analysis} we introduce our modelling strategy. We discuss the evolutionary status of the sample stars in Sect. \ref{sect_evol}, their masses in Sect. \ref{sect_mass}, and their rotation in Sect. \ref{sect_vsini}. The properties of their surface abundances are discussed
in Sect. \ref{sect_abund0}. We then give a summary in Sect. \ref{sect_summary}.

\section{Observations}
\label{sect_obs}
Following \cite{bouret13}, the present work primarily relies on the analysis of UV spectra. Here we use data obtained for our Hubble Space Telescope ({\sl HST\/}) \cosp\ programme
%\LEt{you opted for British English spelling conventions. Please change to "To understand the effect of rotational mixing on the chemical evolution of massive stars, studies of stars in the SMC are particularly important because the metallicity in the SMC is only 0.2 \zsun\ and half that of the LMC. Rotational mixing is indeed expected to be more efficient at lower metallicities \citep[e.g.][]{meynet05, georgy13}, because of $\it i)\rm$ steeper internal gradient of the angular velocity and $\it ii)\rm$ reduced loss of angular momentum caused by lower mass-loss rates.me" throughout} 
GO 11625 (I. Hubeny, PI), completed with spectra obtained
with {\sl HST\/}/\stis\ and available at the Mikulski Archive for Space Telescopes (MAST\footnote{http://archive.stsci.edu}). These spectra were supplemented with optical spectra whenever available (see Table \ref{tabone}). 

For the purpose of the analysis, we would preferably only work with single stars. The main reason is that predictions of the surface chemical patterns 
resulting from binary evolution are still very uncertain \citep[see][for a review]{langer12}.
On the other hand, we lacked information, a priori, about the
multiplicity of the sample stars.

For programme GO 11625, we verified for each selected target that there was no other bright nearby star that might be in the \cosp\ aperture, which is 2.5 arcsec in diameter and subtends $\sim$ 0.8pc at the distance of SMC.
For stars from \stis\ programme GO 7437, \cite{heap06} concluded that the targets in the acquisition images did not appear to be multiple. However, they acknowledged that despite the \stis\ spatial resolution of 0.1'', which is much better than that of \cosp, systems whose separations are smaller than 6300 AU cannot be resolved. 
However, a detailed analysis of the spectra and of the spectral energy distribution (SED) can reveal a companion star if the companion star is not too faint.

\begin{table*}
\centering \caption{Observations summary for the sample stars.}
\begin{tabular}{lccccccccc}
\hline\hline
Star ID\tablefootmark{a}                                & Spec.                 &  \multicolumn{3}{c}{UV}                                                 &  \multicolumn{3}{c}{Optical}  \\
\cline{3-9}\\
                                                                &Type                   &  Telescope      & Date                  & Spect. Coverage       &\vline &               Telescope               &  Date                   & Spect. Coverage \\
                                                                &                               & Instrument      &                               &               (\AA)        &\vline     &               Instrument      &                               &       (\AA)                   \\
\hline
\object{AV 75}                                          &  O5.5 I(f)            &  \stis          & 1999/05/24            & 1150 -- 1700          &\vline &          ESO/UVES               &   2001/09/09          & 3731 -- 5000  \\ 
                                                                &                               &                       &                                   &                            &\vline         &                                       &                                 & 6438 -- 10250 \\
\object{AV 15}                                          &  O6.5 I(f)            &  \stis          & 1999/05/26            & 1150 -- 1700          &\vline         &       ESO/UVES                &   2001/09/29            & 3731 -- 5000   \\     
                                                                &                               &                       &                               &                               &\vline         &                                       &                               & 6438 -- 10250 \\
\object{AV 232}\tablefootmark{b}                &  O7 Iaf                       &  \stis          & 2003/07/05            & 1150 -- 1718          &\vline         &        ESO/UVES                &   2001/09/28          & 3731 -- 5000     \\ 
                                                                &                               &                         &                               &                               &\vline         & ESO/UVES                      &   2007/07/23          & 4720 -- 6835  \\
\object{AV 83}                                          & O7 Iaf                        & \stis           & 1999/05/26            & 1150 -- 1700          &\vline         &  ESO/CASPEC             &   1997/11/19          &  3900 -- 5170 \\
                                                                &                               &                       &                               &                               &\vline         &                                       &   1997/11/20          & 4586 -- 8293 \\
\object{AV 327}                                         &  O9.5 II-Ibw          &  \stis                  & 1999/11/06            & 1150 -- 1700          &\vline         & ESO/CASPEC            &   1997/11/19          & 3910 -- 5170   \\
                                                                &                               &                       &                               &                               &\vline &                                         &   1997/11/20          & 4586 -- 8293 \\
\object{AV 77}                                          &  O7 III                       &  \cosp          &  2010/04/22           & 1132 -- 1798          &\vline         &               ...                     &         ...                     &     ...     \\ 
\object{AV 95}                                          &  O7 III((f))          &  \stis          & 1999/05/16            & 1150 -- 1700          &\vline & ESO/UVES                        &   2004/11/28          & 3731 -- 5000  \\
                                                                &                               &                       &                               &                               &\vline         &                                       &                               & 4583 -- 6686 \\
\object{AV 69}                                          & OC7.5 III((f))                &  \stis                  & 1999/05/17            & 1150 -- 1700          &\vline         &  AAT/UCLES                    &   1997/10/15          & 3847 -- 5008  \\
                                                                &                               &                       &                               &                               &\vline         &                                       &  1997/10/17           & 4586 -- 8293 \\
\object{AV 47}                                          &  O8 III((f))          &  \stis          & 1999/05/20            & 1150 -- 1700          &\vline         &  ESO/UVES                       &   2002/12/08          & 3281 -- 6686  \\
\object{AV 307}                                         &  O9 III                       &  \cosp                  & 2010/08/02            & 1132 -- 1798          &\vline         &               ...                     &       ...                     &    ...       \\ 
\object{AV 439}                                         &  O9.5 III                     &  \cosp                  & 2010/08/02            & 1132 -- 1798          &\vline         &               ...                     &      ...                      & ...     \\ 
\object{AV 170}                                         &  O9.7 III                 &   \stis            & 1998/11/12            & 1150 -- 1700          &\vline         & AAT/UCLES                     &   1997/11/15          &  3847 -- 5008  \\
                                                                &                               &                       &                               &                               &\vline         &                                       &   1997/10/17          &  4586 -- 8293 \\
\object{AV 43}                                          &  B0.5 III                     &  \cosp                 & 2010/04/10            & 1132 -- 1798          &\vline         &  2dF                          &   1998/09/ 25         & 3907 --  4900            \\ 
\hline
\end{tabular}
\tablefoot{
\tablefoottext{a}{The primary identification is the AV number \citep{azzopardi75, azzopardi82}}.\\
\tablefoottext{b}{Cross-ID for this star is Sk 80 \citep{sanduleak68}, MPG 789 \citep{massey89}, or NGC346-001 \citep{evans06}}
 }
\label{tabone}
\end{table*}

\subsection{UV and optical data}
\label{sect_uv_opt_spec}
The sample for our {\sl HST\/}/\cosp\ programme GO 11625 was extracted from the \cite{massey02} UBVR CCD survey of the Magellanic Clouds \citep[see][]{bouret13}.
We observed each target in a sequence of four science exposures, using gratings G130M and G160M with central wavelengths 
(\lb1291, \lb1327) and (\lb1577, \lb1623) for the adopted settings, respectively. With this sequence, the full UV spectrum from  1132 to 1798~\AA\ is obtained at a resolving power of R $\approx$ 20,000.
Co-addition of overlapping segments improves the final signal-to-noise ratio (S/N), which typically is $\approx$ 20 - 30 per resolution element for a single {\sl HST\/} orbit.

For {\sl HST\/}/\stis\ programme GO 7437, stars were observed through the 0.2$\arcsec$ -- 0.2$\arcsec$ aperture using the far-UV MAMA detector in the E140M mode.
The spectral interval ranges from 1150 \AA\ to 1700 \AA and is covered in a single exposure. The effective spectral resolving power of R $\approx$ 46,000, while
the typical S/N per binned data point ranges from 55 to 110 at 1300 \AA\ (the wavelength with the highest sensitivity), and from 25 to 50 at 1600 \AA.
We refer to \cite{walborn00} and \cite{heap06} for a more detailed description of the observations and reduction.

The {\sl HST\/}/\stis\ spectrum of \object{AV 232} was retrieved from the MAST archives. This spectrum was obtained for programme GO 9434 (PI: J. Lauroesch),
designed for the study of the highly ionised hot component of the interstellar medium \citep[see][]{jenkins06}. The same detector, gratings, and aperture settings 
as above were used.
The exposure times were set such that an S/N above 30 per resolution element was achieved at 1240, 1393, and 1550 \AA. 

In the early days of \stis\ (1999), order 86, which contains \niv\ \lb1718, did not fall on the detector, and stars such as those observed for programme GO 7437 lack this important spectral region.
Fortunately, the specific location of different echelle orders on the UV MAMA detector has changed significantly over time, and for spectra after the fourth service mission on {\sl HST\/}, order 86 was recorded on the detector. The spectrum of \object{AV 232} therefore does show \niv\ \lb1718.
More recently, the standard \stis\ reduction pipeline stopped extracting order 86 because the regions used to estimate the inter-order background fall off the top of the detector.

Whenever possible, we supplemented the UV spectra with optical spectra that are available from various sources. 
For all stars from {\sl HST\/}/\stis\ programme 7437, optical coverage exist, either obtained with  {\sl ESO}/CASPEC or with  {\sl AAT}/UCLES. They were extensively presented 
in  \cite{walbornSMC}, and  \cite{heap06}, to which we refer for more details. 
For four stars of this programme (\object{AV 15},  \object{AV 47},  \object{AV 75},  \object{and AV 95}), {\sl ESO}/UVES \citep{Dekker2000} spectra are also available from the {\sl ESO} Phase 3 data products archive, and we chose to use them as they achieve better S/N for a higher spectral resolving power (typically about 45,000).

\object{AV 232} was observed with {\sl ESO}/UVES for our programme 079.D.0073 (PI: E. Depagne). The spectrum covers the spectral range $\sim$ 4720 \AA\ -- 6830 \AA. 
The reduction process was performed using the UVES context within MIDAS \citep[see][]{Ballester2000}, and it included flat-fielding,  bias, and sky-subtraction, and a relative wavelength calibration. Cosmic-ray removal was performed with an optimal extraction method for each spectrum. The S/N is $\approx$ 130.
Another {\sl ESO}/UVES spectrum of \object{AV 232} obtained in programme 067.D-0238 (PI: P.~Crowther) was used to extend the optical coverage down
to 3900 \AA\ (the full spectral coverage of this spectrum is $\approx$ 3730 \AA\ -- 5000 \AA). The S/N of this spectrum is $\approx$ 125. More details about the instrumental settings and reduction process can be found in \cite{crowther02}.
This star was also part of the ESO {\sl VLT}/FLAMES survey of massive stars in the SMC \citep[][to which we refer for a more detailed description of the properties of the FLAMES data and their reduction]{evans06}. 
This optical spectrum was modelled and presented in \cite{mokiem06, mokiem07}. 

The optical spectrum for \object{AV 43} has a lower spectral resolution and was obtained with the {\sl AAT}/2dF by \cite{evans04}.  
These authors provided details about the data and their reduction. Three stars with  {\sl HST\/}/\cosp\ spectra have no optical spectra available (see Table \ref{tabone}), hence they are more prone to uncertainties in the analysis.

The optical spectra obtained with {\sl ESO}/UVES were normalised interactively using a cubic-spline fit to the pre-defined continuum regions. We examined the regions around important He and \mbox{CNO} lines
(see Sect \ref{sect_param}) and the wings of the Balmer lines with particular care. Details about the normalisation of the spectra obtained with {\sl ESO}/CASPEC or {\sl AAT}/UCLES can be obtained in \cite{walborn00} and in \cite{evans04} for the {\sl AAT}/2dF spectrum.

The non-contemporaneity of the observations in the UV and in the optical may affect the results of the spectroscopic analysis because the significance of a parameter set derived from data
obtained several years apart may be questioned (assuming a single set of parameters can indeed be obtained at all). 
Intrinsic variability for instance could jeopardise our ability to derive such a single set of parameters representing the stellar properties.  
Variability like this is unknown for our targets. An analogy with Galactic cases indicates that the changes in the
stellar and wind properties are not enough to prevent the determination of a single parameter set, where the changes are bracketed within the uncertainties on each physical parameter 
\citep[e.g.][]{bouret12}. 
Intrinsic variability aside, binarity could also be a cause for confusion for systems that were observed years apart in different spectral bands, as in our case. 
We have no indication for such binarity in
our target stars, except for \object{AV 77,} for which the UV spectrum clearly shows the presence of two components (we have no optical spectrum to further confirm this). 
Confirmation that other stars of our sample are in fact binary systems would require more observations either in the UV or optical or in both, in particular, to search for radial velocity or spectral variations. 
The \ullyses\ programme on {\sl HST\/} (the Hubble UV Legacy Library of Young Stars as Essential Standards) or/and its ground-based optical to near-infrared (NIR) range counterpart on {\sl ESO}/XSHOOTER, which are currently performed, should provide the necessary additional data that we lack at this point.

\section{Spectroscopic analysis}
\label{sect_analysis}
\subsection{Model atmosphere}
\label{sect_model}
The spectroscopic analysis was performed using model atmospheres and synthetic spectra calculated with the code \cmfgen\ \citep{hillier98, hillier03}. 
\cmfgen\ computes non-local thermodynamic equilibrium (NLTE) line-blanketed model atmospheres, solving the coupled radiative transfer and statistical equilibrium equations 
in the comoving frame of the fluid in a spherically symmetric outflow.
To facilitate the inclusion of extensive line blanketing, a formalism of super-levels is adopted, allowing the incorporation of many energy levels from ions of many different 
species in the model atmosphere calculations. More specifically, in this work we included ions of H, He, C, N, O, Ne, Mg, Si, P, S, Ar, Ca, Fe, and Ni (see Table \ref{tabions}).

\begin{table*}
\centering \caption{Number of levels and super-levels for each atomic species included in models.}
\begin{tabular}{lccclcc}
\hline\hline
Ion             &       Full levels             & Super-levels  &  & Ion                &       Full levels          & Super-levels\\
\hline
H I             &       30                      & 30  & \vline & P V            &       62                      & 16      \\
He I            &       69                      & 69  & \vline & S III          &       78                      & 39 \\
He II   &       30                      & 30   &\vline & S IV           &         108                     & 40 \\
C II            &       322                     & 92  &\vline & S V             &       144                     & 37 \\
C III           &       243                     & 99  & \vline & S VI           &         58                      & 28 \\
C IV            &       64                      & 64 &\vline & Ar III           &       346                     & 32 \\
N II            &       442                     & 358  &\vline & Ar IV          &       382                     & 50 \\
N III           &       287                     & 57  &\vline & Ar V            &         376                     & 64 \\
N IV            &       70                      & 44   &\vline & Ar VI          &       81                      & 21 \\ 
N V             &       49                      & 41  &\vline & Ca III  &         232                     & 44 \\
O II            &       274                     & 155  & \vline & Ca IV &         378                     & 43 \\
O III           &       104                     & 36  & \vline &  Ca V          &       613                     & 73 \\
O IV            &       64                      & 30  & \vline & Ca VI  &         288                     & 48    \\
O V             &       56                      & 32   &\vline & Fe II          &       100                     & 100\\
O VI            &       31                      & 25  & \vline & Fe III         &       607                         & 65 \\
Ne II           &       48                      & 14   &\vline & Fe IV  &       1000                    & 100 \\
Ne III  &       80                      & 32   &\vline & Fe V   &       1000                    & 139  \\
Ne IV   &       112                     & 39      &\vline & Fe VI       &       1000                    & 59  \\
Ne V            &       166                     & 37   &\vline & Fe VII       &       41                     & 252 \\
Mg II   &       44                      & 36   &\vline & Ni III         &       150                     & 24 \\
Si II           &       56                      & 56   &\vline & Ni IV          &       200                     & 36 \\
Si III          &       50                      & 50  & \vline & Ni V           &         183                     & 46 \\
Si IV   &       66                      & 66   &\vline & Ni VI          &       182                  & 40 \\
P IV            &       178                     & 36  & \vline & Ni VII              &          37                     & 308 \\

\hline
\end{tabular}
\tablefoot{The ions listed here are used (or not) according to the effective temperature of the models.}
\label{tabions}
\end{table*}

The radiative acceleration was calculated from the solution of the level populations and was used to compute iteratively the hydrodynamical density structure of the inner atmosphere. The inferred velocity law
in this quasi-hydrostatic region was then smoothly connected to a $\beta$ velocity law in the wind. 
The wind mass-loss rate, density, and velocity are related by the continuity equation. Wind clumping is implemented in \cmfgen\ assuming the optically thin clumps formalism \citep[see][for more details]{hillier99}. 

After the atmosphere model converged, a formal solution of the radiative transfer equation was computed in the observer's frame \citep{busche05}, thus providing the
synthetic spectrum for a comparison to observations. At this step, we used a radially dependent turbulent velocity \citep[see][]{hillier98}.

Finally, we accounted for the effects of shock-generated X-ray emission in the model atmospheres. Because no measured X-ray fluxes are  available for our targets, we adopted the 
standard Galactic luminosity ratios log$\rm L_{X}/L_{bol} \approx -7$ \citep{sana06}, with the bolometric luminosities determined in this study.
This scaling ratio might not  be valid at low metallicity, depending on the physical nature of the shock 
generating the X-rays (radiative or adiabatic) and the wind density \citep{owocki13}. 
Depending on the actual relation between the metallicity and the mass-loss rate,  log$\rm L_{X}/L_{bol}$ could change accordingly. 

\subsection{Stellar parameters}
\label{sect_param}

The photospheric and wind parameters were determined using the same procedure and diagnostics as in
\cite{bouret13, bouret15}, to which we refer for more details. Below, we simply
outline some details specific to the present work. % \LEt{please reformat this so that it becomes proper text, not a bulleted list}.

%\begin{itemize}
-- {\bf Stellar luminosity}:
We used the flux-calibrated \cosp\  and \stis\ spectra together with optical and NIR photometry \citep[][and references therein]{bonanos10} to constrain the luminosity, and interstellar extinction as a side-product
(the values we derived for $E(B-V)$ are listed in Table \ref{tabtwo}).
In general, it is possible to constrain the intrinsic stellar luminosity by comparing  the theoretical SED predicted by a model for a set of fundamental parameters (mostly \teff, \logg, \rstar, \mdot, Z) 
to  the observed spectro-photometry when extinction amounts and laws, as well as a distance, are taken into account. We adopted a distance modulus to the SMC of 18.91$\pm$0.02 \citep{harries03}. 

\begin{table*}
\centering \caption{Interstellar extinction derived for the stellar sample.}
\begin{tabular}{llccccc}
\hline\hline
Star ID                   & Spec.                               &$V$\tablefootmark{a}           &   $B-V$\tablefootmark{a}                        &   $E(B-V)$            & Refs. \\
                                &Type                           &               &                       &  Galactic + SMC &              \\
\hline
\object{AV 75}                                          &  O5.5 I(f)            &   12.76 &  $-0.15$      &   0.14                        & 1, 2    \\ 
\object{AV 15}                                          &  O6.5 I(f)            &   13.18 &  $-0.19$              &   0.10                        & 1, 2                \\        
\object{AV 232}                                         &  O7 Iaf                       &   12.31 &  $-0.19$              &   0.05                        & 2, 4          \\ 
\object{AV 83}                                          & O7 Iaf                        &   13.37 &  $-0.12$              &   0.13                        &   1 \\
\object{AV 327}                                         &  O9.5 II-Ibw          &   13.09 &  $-0.16$              &   0.06                        & 1, 2                 \\ 
\object{AV 77}                                          &  O7 III                       &   13.95 &  $-0.16$              &   0.15                        & 1, 2                \\ 
\object{AV 95}                                          &  O7 III((f))          &   13.83 &  $-0.19$              &   0.04                        & 1                                     \\ 
\object{AV 69}                                          & OC7.5 III((f))                &   13.33 &  $-0.18$              &   0.11                        &   1, 2 \\
\object{AV 47}                                          &  O8 III((f))          &   13.50  &  $-0.18$             &   0.05                        & 1, 2                \\ 
\object{AV 307}                                         &  O9 III                       &   14.02 &  $-0.14$              &   0.10                        & 1               \\ 
\object{AV 439}                                         &  O9.5 III                     &   14.68 &  $-0.19$              &   0.05                        & 1, 2                 \\ 
\object{AV 170}                                         &  O9.7 III                      &   14.11       &  $-0.29$      &   0.06                        & 1, 2    \\ 
\object{AV 43}                                          &  B0.5 III                      &   14.08       &  $-0.13$              &   0.08                        & 1, 3              \\ 
\hline
\end{tabular}
\tablefoot{
\tablefoottext{a}{Photometry extracted from SIMBAD database, operated at CDS, Strasbourg, France. The sources for the photometry are
1. \cite{bonanos10}, 2. \cite{cutri03}, 3. \cite{massey02}, and 4. \cite{evans06}. The colour excess, $E(B-V)$, is divided into two contributions from the Galactic foreground and the SMC field (see above).}
 }
\label{tabtwo}
\end{table*}

\begin{figure*}[htbp]
   \centering
   \includegraphics[width=18.2cm]{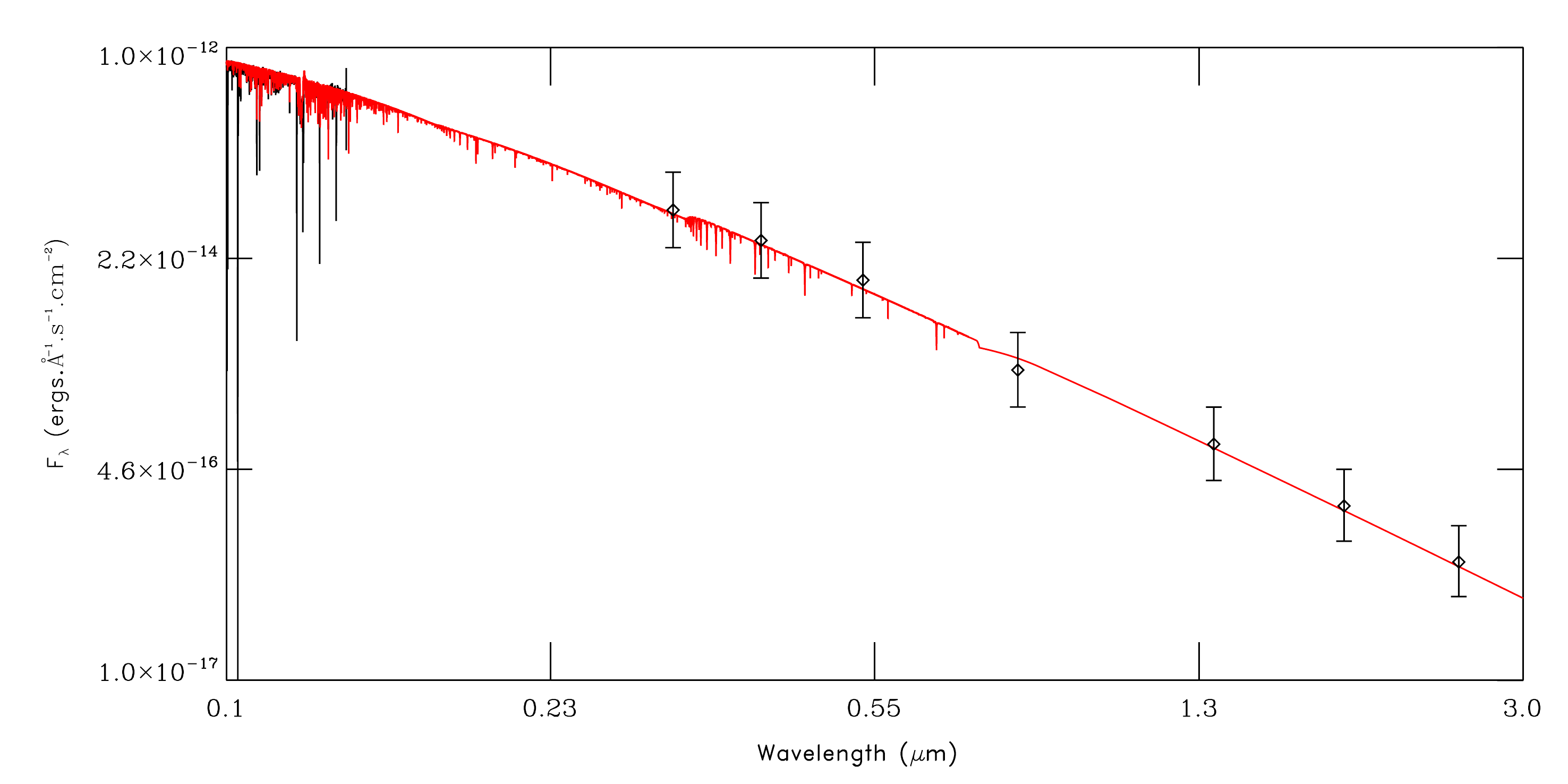}
      \caption{Observed vs. theoretical SEDs (black vs. red) for the O8 giant \object{AV 47}.}
         \label{Fig_photom}
   \end{figure*}

After accounting for a correction for reddening caused by the Galactic foreground \citep{cardelli89}, we find that the synthetic SEDs match the observed very well, from the near-IR range (JHK) down to optical (UBVI).  This is expected because overall, the SMC extinction laws are quite similar to the Galactic laws in this wavelength range.
However, in the UV range, an additional contribution, specific to SMC conditions, was required to account for the flux distribution, which we did using a correction for reddening caused by the SMC interstellar medium 
\citep{gordon03}. While this correction is valid for the SMC bar, some of our targets are located in the outskirts (\object{AV 439} and \object{AV 307}) of this galaxy, 
which might explain why a residual mismatch between the modelled and observed fluxes shortward of $\approx$ 1250 \AA\ is stronger in these objects.   
Figure \ref{Fig_photom} shows a typical fit to the reconstructed SED using UV flux-calibrated spectrum and photometric data, with a theoretical SED computed with \cmfgen.

For every star of the sample, we adopted a ratio of total-to-selective extinction $R_{\rm V}=3.1$ \citep[see also][]{massey09}, consistent with our work on SMC dwarfs \citep[see][for a discussion on adopting a different
 $R_{\rm V}$]{bouret13}. We did not find obvious indications for deviation from this standard $R_{\rm V}$ with our procedure, even though the extinction is not expected to be accurately described by a single-parameter functional form in the UV \citep{fm07}. 

-- {\bf Effective temperature and surface gravities}:
When optical data were available, we used the ionisation balance of helium lines (equivalent width and line profiles) to derive \teff. Two lines,  \hei\ \lb5876 and \heii\ \lb4686,\ 
were discarded because they often suffer from non-negligible wind contribution in evolved stars.
%Further,
Additionally, we used spectral diagnostics in the UV range, such as ionisation ratios of iron ions (\feiii\ to \fevi), to estimate the effective temperature.
For several stars, and in contrast to \cite{bouret13}, we were unable to use the ratios  \civ\ \lb1169 to \ciii\ \lb1176 because these lines showed strong signs of wind contamination (the two are sometimes
blended).  
 
 For stars with both UV and optical spectra, the large number of diagnostics together with the good S/N of the data means that the typical uncertainty on \teff\ can be as good as 1000~K. 
 However, for stars with no optical spectra, the uncertainty on \teff\ can be as large as 1500~K.

During the \teff\ determination process, the relative strength of optical helium lines was used to constrain the helium abundance. For \object{AV 77}, \object{AV 307}, and \object{AV 439} (UV spectra only),  
 the helium abundance was set to adjust the strength of \heii\ 1640 \AA.  
 
In Table \ref{tab2}, \teff\ refers to UV-based \teff\ only for these three stars. For the other stars, the quoted \teff\ corresponds to that providing the best fit over the whole UV-optical spectral range. 

When optical spectra are available, the principal source of uncertainty in the determination of surface gravities is rooted in the use of \'echelle spectra whose rectification is uncertain, especially in the vicinity of 
(broad) Balmer lines. This translates into typical uncertainty  of 0.1 -- 0.15 dex (including \object{AV 43} with a 2dF spectrum, i.e. with lower spectral resolution). 
For stars with UV spectra only, we adopted typical values for the corresponding spectral types \citep{massey09} because photospheric features in the UV range show little sensitivity to changes
in surface gravities \citep[but see][for a discussion of using lines from different iron ions in the UV range]{heap06}. The uncertainty was also adopted to be 0.2 dex in this case. 
Note that UV wind lines do show some sensitivity to \logg\ but we cannot use it for quantitative purposes
until the wind driving is fully understood from first principles.
For consistency with our previous study \citep{bouret13}, we accounted for the effect of centrifugal forces caused by rotation (see below), and corrected the effective gravity derived from the 
spectroscopic analysis following the approach outlined in \cite{repolust04}. These corrected \logg\  values (Table \ref{tab2}) were used to derive the spectroscopic masses.

-- {\bf Surface abundance}: 
We used the numerous photospheric lines of iron ions (\feiii\ to \fevi) in the UV to constrain the iron content of the stars. 
The sensitivity of iron line strengths and line ratios to
temperature, gravity, iron abundance, and/or microturbulent velocity \citep[see also][]{heap06} has been discussed previously by \cite{bouret15}.
For the photospheric parameters listed in Table \ref{tab2}, the iron lines in the UV spectra of our sample stars are well reproduced with models having Fe/Fe$_{\odot}$ = 0.2, further confirming earlier metallicity determinations in the SMC \citep{venn99}.
For consistency with \cite{bouret13}, we used abundances for Mg and Si from \cite{hunter07}, while
the standard solar abundances from \cite{asplund09}, scaled down by a factor of 5 (or equivalently, by $-$0.7~dex), were adopted 
for all other elements included in the models (\mbox{Ne, P, S, Ar,Ca, Ni}), but excluding \mbox{CNO}. %\LEt{a single sentence does not constitute a paragraph. Please either add to this or merge}}.
 
\mbox{CNO} surface abundances and associated errors were determined following the same procedure as in
\cite{martins15}. When the fundamental parameters were constrained, we ran several models in which we changed only the 
\mbox{CNO} abundances. For a set of selected lines of given element, a $\chi^{2}$ analysis was performed 
in which these selected lines were combined. The computed $\chi^{2}$ values were then renormalised to a minimum value of
1.0, which defines the best-fit abundance. 1$\sigma$ uncertainties are defined by $\chi^{2}$ = 2.0.

%\begin{itemize}
-- {\it Carbon}:
\civ\ \lb1169 and \ciii\ \lb1176 serve as prime indicators for carbon abundances. For stars of the latest type in the sample, where \ciii\ is the dominant ionisation stage of carbon, \ciii\ \lb1247 was given more weight than \ciii\ \lb1176 because
the latter shows some sensitivity to the wind density \cite[see also][]{bouret15}. 
Powerful diagnostic lines in the optical for stars later than O 7.5 are \ciii\ 4070, and  \ciii\ 4153, \ciii\ 4157, and \ciii\ 4163 \citep{martins15, martins17}. However, they are at best very weak or even absent in our optical spectra, and could seldom be used. 
Overall, in the optical the several diagnostic lines that are available for carbon \citep{martins15} are weaker than the UV lines. 
While \ciii\ \lb\lb\lb4647, 4650, and 4651 are often detected as the strongest carbon lines in the optical, they are notoriously sensitive to the details of the atomic physics and atmosphere model \citep{martins12b}.  
Although the effect of Fe metal lines found by \cite{martins12b} should be lower than for galactic O stars because the Fe metal abundance is down a factor of five, we used them only as consistency checks. 

-- {\it Nitrogen}: 
In the UV, the photospheric lines \niii\ \lb\lb1183, 1185 and \niii\ \lb\lb1748, 1752 (for stars observed with \cosp) were used as primary diagnostics.
For stars with \cosp\ spectra, \niv\ \lb1718 is also detected but was mostly used as consistency checks because it is potentially affected by the stellar wind (and hence the mass-loss rate and clumping filling-factor). 
In the optical, \niv\ 4058 is rarely detected although it is quite frequently seen in Galactic evolved O-type stars with similar spectral types \citep{bouret12}. 
Other lines that are often used as main diagnostics of nitrogen in the optical include \niii\ 4511, \niii\ 4515, \niii\ 4518, and \niii\ 4524. These are often weak in several stars, such that we opted not to rely on them except for a few cases such as \object{AV 95, AV 232, and AV 327}.
We opted to use \niii\ \lb\lb4634, 4642 as the main diagnostic for the nitrogen abundance in the optical \cite[but see][for a thorough discussion of their formation processes, in particular, their sensitivity to the background metallicity, nitrogen abundance, and wind strengths]{rivero11, rivero12}.

-- {\it Oxygen}:
In the UV, the \oiv\ \lb\lb1338, 1343 lines generally serve as a diagnostic for the oxygen abundance, with \ov\lb1371 as a secondary indicator \citep[e.g.][]{bouret13}. However, 
they are affected by wind contribution for stars with significant mass-loss, such as several of the giants and supergiants of our sample. \oiii\ \lb\lb1150,1154 could be used instead, but these lines are on the shortest side of the
UV spectra where the flux level is more uncertain.
Only one line is strong enough to be detected (depending on the \teff) in the optical spectra (\oiii\ \lb5592). 
In all cases, the oxygen content was tuned to improve the fit quality to the lines listed above. However, because of the paucity of sensitive and reliable diagnostics, the values we derive
suffer from larger uncertainties than those for carbon and nitrogen.  

-- {\bf Rotation and macro-turbulence}:
For comparison to observations, the synthetic spectra were convolved with appropriate instrumental profiles, as well as rotational and macro-turbulent profiles
that contribute to line-broadening. A  pure isotropic Gaussian profile was used to mimic the effect of macroturbulence \cite[but see][for a warning about adopting a Gaussian formulation of macro-turbulence]{aerts09}. 

To derive the projected rotational velocity (\vsini) of the stars, we first used optical spectra, when available. We used the 
IACOB-BROAD \citep{simon14} analysis tool to derive the projected rotational velocity (\vsini) and the contribution of macro-turbulence to the total line-broadening (\vmac).
Depending
on the spectral type and spectrum quality of the target, we 
relied on diagnostic lines such as \hei\ \lb4471, \hei\ \lb4713, \hei\ \lb4922, \siii\ \lb4552, and \heii\ \lb4541.
In several instances, however, the optical spectra suffered from significant nebular contamination and/or an insufficient S/N such that unambiguous and accurate determination of \vsini\ and \vmac\ with this method was not guaranteed.  
We refer to \cite{ramirez13} for a thorough discussion of the different biases and problems with this method in the case of the VFTS.

We verified that the derived \vsini\ and \vmac\ yielded consistent UV spectral profiles. Many UV spectra show partial or fully resolved
components of the \ciii\ \lb1176 multiplet, indicating moderate (\vsini\ < 100 \kms) rotational velocities, with a typical precision of 20 \kms. Other useful constraints in the UV are provided by the \oiv\lb\lb1338-1343 and numerous iron lines.
 
For three stars with no optical spectra, we simply convolved the synthetic spectra  with rotational profiles (plus instrumental) and varied \vsini\ until we achieved a good fit to the observed photospheric profiles, adopting a standard limb-darkening law for the convolution. 
We refer to \cite{hillier12} for a discussion about the effect of limb darkening on some spectral lines of fast rotators. For these stars the line-broadening parameter therefore is \vsini\ alone, which means that the values we quote are upper limits on the true \vsini.

-- {\bf Wind parameters}:
For stars with saturated (UV) P~Cygni profiles, the wind terminal velocities, \vinf, were measured from the maximum velocity associated with the saturated black portion of the wind profiles. In cases of unsaturated P~Cygni profiles, \vinf\ was estimated from the blue-most edge of the absorption component of UV P~Cygni profiles. The typical uncertainty for this determination of \vinf\ is 100 \kms\ (depending on the maximum microturbulent velocity we adopt). For stars with underdeveloped P~Cygni profiles, we confirmed our values by scaling down the wind terminal velocities found in Galactic stars of the same spectral types.
More precisely, we used extracts from Table 1 in \cite{kudritzki00}, and applied a scaling \vinf\ $\propto {\rm Z}^{n}$ with $n =0.13$ \citep{leitherer92}.

Mass-loss rates were derived from the analysis of UV P~Cygni profiles of \civ\ and sometimes \siv\ resonance doublets, plus the \halpha\ line for stars with optical spectra. 
For these stars, a unique value of \mdot\ allows a good fit to both the UV lines and \halpha.
In all cases, less weight was given to the \nv\ \lb1238-1242 doublet as it is blended with the adjacent Ly$\alpha$ line, and is also notoriously sensitive to the interclump medium and to the X-ray flux, which is not constrained for our sample stars \citep[e.g.][]{zsargo08}.

The $\beta$ exponent of the wind velocity law was derived from the fit of the shape of the P~Cygni profile. 
Clumping parameters, $f$ and $v_{\rm cl}$, were derived in the UV domain, following \cite{bouret05}. 
We found no need to tune $v_{\rm cl}$ to improve the fit to observed lines, therefore we adopted a canonical $v_{\rm cl}$ = 30 \kms\ \citep{bouret05}.  
Some photospheric lines in the optical also show some sensitivity to the adopted filling factor (and scaled \mdot). For photospheric \hi\ and \he\ lines,
for instance, this is essentially caused by a weaker wind contribution (emission) in clumped models, thus producing deeper absorption than smooth-wind models.
The wind properties for our giant and supergiant SMC stars will be
extensively discussed in a forth-coming paper.
%\end{itemize} 

Table \ref{tab2} summarises the fundamental parameters we derived for the giants and supergiants of programmes {\sl HST\/} GO 7437 and 11625 introduced in Sect. \ref{sect_uv_opt_spec}. 
In the following sections, we merge these 13 stars with the sample of 23 dwarfs presented in \cite{bouret13} and discuss the properties of the whole (unevolved $+$ evolved stars) sample. 
Table \ref{tab_abund_full} presents the surface abundances and ratios measured in the present analysis together with those measured for the dwarf sample. Details about the other parameters concerning the unevolved sample can be found in \cite{bouret13}.

The SEDs and spectra of the evolved sample stars are satisfactorily fit by single-star models, with the exception of 
\object{AV 77}. For this star alone, the spectra  distinctively indicate binarity. Its UV spectrum clearly shows the presence of two 
components. Nevertheless, we cannot rule out that more members of the evolved sample are in fact binary systems, until
more observations are obtained, in particular to search for radial velocity variations. The sample of (23) dwarf stars likely contains up to six binaries \citep{bouret13}. Appendix B provides the best-model fits for each object in the present study. 

\begin{table*}[htbp]
\centering
\caption{Fundamental stellar parameters.}
\begin{tabular}{llccccccccccc}
\hline
\hline
Star                            & Sp. Type                  & \teff     & \logg & $\log \frac{L}{L_{\odot}}$      & log(\mdot)    & $\beta$       &$f$            & \vinf   & \vsini& \vmac         & \msp          \\
                    &                               & [kK]      & [cgs] &                                               &                       &                   &            & [\kms]&[\kms]         &[\kms]                 & [\msun]                       \\ \hline
\object{AV 75}          &  O5.5 I(f)            & 38.5  & 3.51  & 5.94                                  & -6.30           & 0.8           & 0.10  & 2050  & 120           & 70                                    & 51.1    $\pm$ 8.2       \\
\object{AV 15}          & O6.5 I(f)                     & 39.0  & 3.61  & 5.83                                    & -6.46         & 1.3           & 0.10    & 2050  & 120           & 85                                    & 47.2    $\pm$ 8.7       \\
\object{AV 232}         & O7 Iaf                        & 33.5  & 3.16  & 5.89                                    & -5.92         & 1.5           & 0.07    & 1350  & 75                    & 92                                     & 35.3  $\pm$ 8.2 \\
\object{AV 83}          & O7 Iaf                        & 32.8  & 3.26  & 5.54                                    & -6.14         & 2.0           & 0.10    & 940   & 80            & -                                             & 22.1    $\pm$ 8.3       \\
\object{AV 327}         & O9.5 II-Ibw           & 30.0  & 3.12  & 5.54                                  & -7.37           & 1.0           & 0.10  & 1500  & 95            & 83                                             & 22.8  $\pm$ 5.8 \\
\object{AV 77}          & O7 III                        & 37.5  & 3.74  & 5.40                                    & -7.88         & 1.0           & 0.10    & 1400  & 150           & -                                     & 28.0    $\pm$ 12.2  \\
\object{AV 95}          & O7 III((f))           & 38.0  & 3.70  & 5.46                                  & -7.40           & 0.9           & 0.10  & 1700  & 55            & 86                                    & 28.1    $\pm$ 5.6 \\
\object{AV 69}          & OC7.5 III((f))                & 33.9  & 3.50  & 5.61                                    & -6.51         & 1.0           & 0.10    & 1800  & 70            & -                                             & 39.7    $\pm$ 7.3       \\            
\object{AV 47}          & O8 III((f))           & 35.0  & 3.75  & 5.44                                  & -8.18           & 1.0           & 0.10  & 2000  & 60            & 44                                    & 42.2    $\pm$ 5.1                       \\
\object{AV 307}         & O9 III                        & 30.0  & 3.50  & 5.15                                    & -8.82         & 0.9           & 0.10    & 1300  & 60            & -                                     & 22.5    $\pm$ 13.2 \\
\object{AV 439}         & O9.5 III                      & 31.0  & 3.54  & 5.16                                    & -7.70         & 0.8           & 0.20    & 1000  & > 260         & -                             & 22.0  $\pm$ 9.3     \\
\object{AV 170}         & O9.7 III                      & 30.5  & 3.41  & 5.14                                    & -8.82         & 0.9           & 0.10    & 1200  & 70            & 52                            & 16.6  $\pm$ 8.1     \\
\object{AV 43}          & B0.5 III                      & 28.5  & 3.37  & 5.13                                    & -8.00         & 0.9           & 0.20    & 1200  & 200           & < 120                                 & 22.4    $\pm$ 11.0       \\     
  \hline
\end{tabular}
 \label{tab2}
 \begin{list}{}{}
\item For stars with UV + optical spectra, uncertainties on \teff\ are $\pm$ 1000~K,  $\pm $ 0.1 dex for \logg, $\pm $ 0.1 dex on \logL. As for the wind quantities, 
an uncertainty of $\pm$ 0.2 dex was estimated for the mass-loss rates (given in \msunyr), while \vinf\ was measured within $\pm$ 100 \kms. 
For stars without optical data, \logg\ is accurate within $\pm $ 0.15 -- 0.2, while uncertainties
on \teff\ are $\pm$ 1500~K. For both \vsini\ and \vmac, uncertainties are $\pm$ 10 -- 20 \kms.
\end{list}
 \end{table*}

  %%%%%%%%%%%% 
\section{Evolutionary properties}
\label{sect_evol}
\begin{figure}[tbp]
   \centering
   \includegraphics[width=9.2cm]{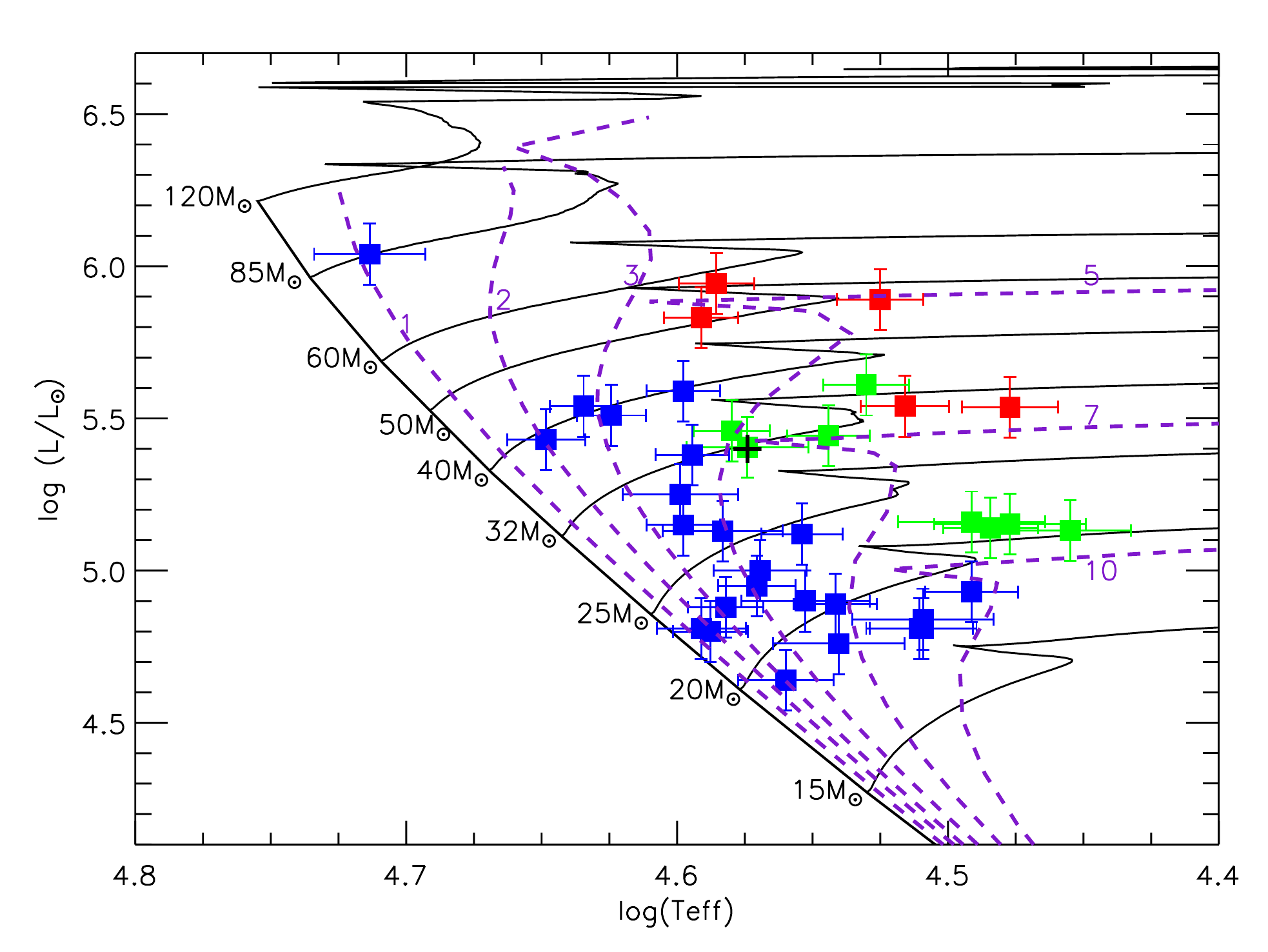}
   \includegraphics[width=9.2cm]{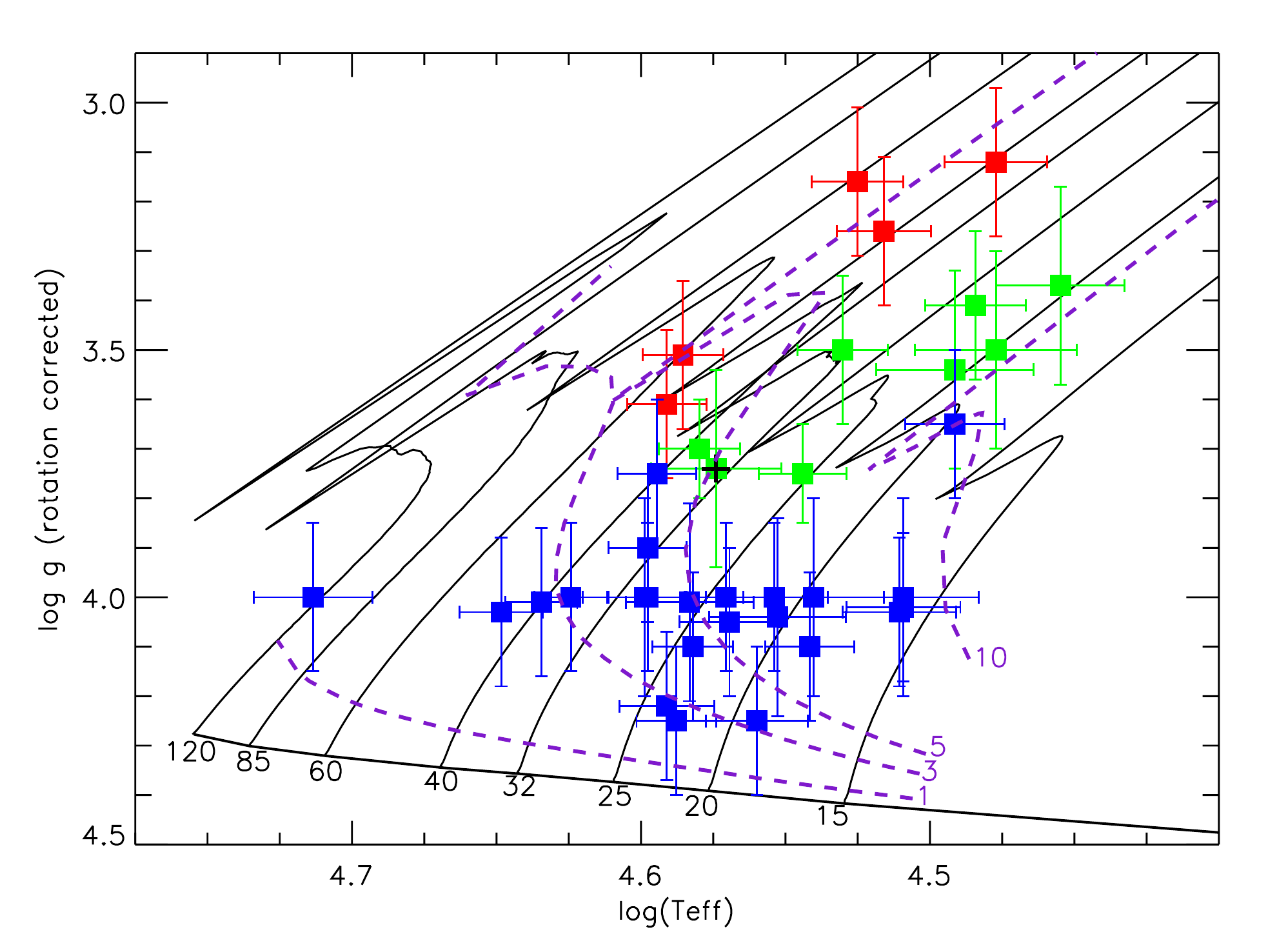}
      \caption{HRD (top) and KD (bottom) of the full sample stars. The ordinate in the KD is defined as the measured surface gravity corrected for centrifugal rotation (see text).
      Blue squares indicate luminosity class V stars (luminosity class III, green, and luminosity class I stars, red). The black plus indicates the location of \object{AV 77}, the only clear binary of our sample.
Evolutionary tracks (black lines) and isochrones (dashed purple lines) show models with an initial rotational velocity of roughly 300 \kms\ \citep{georgy13}. The tracks are annotated in solar masses and the isochrones in Myr.}
         \label{fig_tefflogg}
   \end{figure}

In Fig. \ref{fig_tefflogg} we show the location of the individual stars in the classical HRD (top) and in the Kiel diagram (KD, bottom). The latter depends only on 
\teff\ and \logg, which quantities are independent of distance and reddening, and can be directly determined from spectroscopy.  
To build the KD, we corrected the measured surface gravities for the centrifugal acceleration \citep[see][for a derivation of a first-order analytical form for this correction]{repolust04}.  

To build the HRD, we used the flux-calibrated UV spectra together with UVBIJHK photometry to derive \logL\ for each star. This quantity is well constrained  because of the very moderate extinction towards the stars in our SMC sample and because of the accurate distance modulus (see Sect. \ref{sect_param}). 
In this sense, the physical properties and evolutionary status of our sample stars, derived from location in the HRD, are more secure than for Galactic (field) stars, where distances are both uncertain and varied. 
Therefore we expect that the HRD on one hand and a distance-independent plot such as the KD 
\citep[or rather equivalently, a spectroscopic HR diagram as introduced by][]{langer14} on the other hand lead to the same conclusions concerning 
the evolutionary properties of the sample stars. 

We compared the locations of the stars in the HRD and KD to evolutionary tracks and isochrones for models with rotation predicted by the Geneva code for metallicity typical of SMC \citep{georgy13}. 
These models start on the ZAMS with an equatorial surface velocity of 
$v_{init}$ = 0.4$v_{crit}$, ($v_{crit}$ being the critical surface velocity), which for the mass range
we considered is close to $\approx$ 300 \kms. 

This initial rotation rate corresponds to velocities of between 110 and 220 \kms\ at the middle of the 
MS \citep{ekstrom12}. It is therefore a good match to the average \vsini\ $\approx$ 110 \kms\ 
of the full sample, although it must be kept in mind that the initial rotational velocity for evolutionary models is 
without projection effects. This may be of importance for stars such as the fastest two rotators of our sample (\vsini\ $\geq$
300 \kms), which  are also close to the ZAMS, and it can therefore be expected that 
the typical velocity assumed in this grid of models is probably lower than the actual initial rotational velocities for these dwarfs.

The error bars on the position of stars in the HRD and KD reflect the uncertainties on the effective temperature, surface gravities, and luminosities from the analysis above.
The  bulk of the sample stars has masses ranging from about 17 to 55 \msun\ in both diagrams. The isochrones indicate that the star ages range from 1 Myr up to 10 Myr. We refer to \cite{bouret13} for a full discussion of the evolutionary status and ages of the dwarfs, in particular, concerning the stars in NGC 346.

As a whole, the sample stars cover the regions of the HRD and KD where O-type stars are expected. Although 
the two observing programmes were designed to cover the full range in spectral type and luminosity,  there is a significant gap in the distribution, in particular, between the hottest two stars of the sample. 
The different luminosity classes are relatively well segregated in both diagrams. The observed overlap between stars of different luminosity classes may be the consequence of a misclassification. We recall that we collected 
the spectral types and luminosity classes from \cite{massey02} and \cite{walborn00}, which did not systematically use the same criteria to assign a spectral type and luminosity class, nor did they use the same spectroscopic and photometric data. 
These uncertainties are possibly related to the sensitivity to abundance and metallicity effects of the luminosity classification criteria for O-type stars \citep[see][for a careful presentation of these effects]{massey09, walborn14}. 

Another possible cause for the observed overlap between stars of different luminosity classes is binarity. If binarity is important, luminosities may be overestimated for some stars, leading to a greater luminosity spread. However, only one star in the evolved sample  (\object{AV 77}) shows unambiguous signs of binarity.

In the HRD, all dwarfs are located in the first part of the H-burning phase, well before the reversal of \teff\ on the MS (the terminal-age main-sequence, TAMS). The ages indicated by the isochrones are shifted towards higher values than those in \cite{bouret13}, where we relied on the tracks and isochrones of \cite{brott11a} with an average $v_{\rm rot}$ = 180 \kms.
This is due to different input parameters such as the overshooting.
Higher initial rotation (about 300 \kms\ in our models) indeed produces a longer MS phase, with a
shift of the TAMS to lower temperatures \citep[e.g.][]{maeder01}.
For the three dwarfs that passed the 7 Myr isochrone, a timescale argument excludes that they are at a later evolutionary stage (looping back to the blue for a lower initial mass track). 
This is further supported by the moderate surface nitrogen enrichments we measure (cf. Sect \ref{sect_abund0}). Models with rotation \citep{brott11a, ekstrom12, georgy13} predict that such enrichment should indeed occur during the MS phase, and is not expected to be delayed to post–MS phases. 
%\LEt{if you intend to use this abbreviation, please introduce it at first occurrence and then use it consistently throughout. Otherwise, please continue to spell out (either is fine). This applies to all abbreviations throughout, please check and change as required. I do not indicate this in every case to avoid cluttering the ms } 

The supergiants are clearly separated into two mass bins, one above 45 \msun\ and one around 30 \msun\ (three and two stars, respectively).
All but one (the hottest) are beyond the turn-off, in the part where the tracks evolve to the red at constant luminosities. 
A closer look shows, however, that this apparently least evolved supergiant, in the second half of the MS next to the \mstar\ = 50 \msun, could also be located on the \mstar\ = 44 \msun\ and have passed the TAMS. This reading of the HRD is supported by the location
of the star in a \logg\ -- log(N/C) diagram, showing that the chemical abundances at the surface of the star underwent significant processing compatible with a post–MS phase (see Sect. \ref{sect_age}).
The two $\sim$30 \msun\ stars are located farther along the evolutionary track than the more massive supergiants, indicating that they are more evolved, as further indicated by the nearby 7 Myr isochrone.

The KD also shows a dichotomy for the supergiants, although with different mass distributions. The KD indicates that the
five supergiants have masses in the range 40 -- 60 \msun. The two supergiants with the highest \logg\ are less evolved than their
fellows with lower \logg, which are all well beyond the TAMS.
Supergiants are the only luminosity class for which the KD indicates systematically higher masses than the HRD, by more than 8 \msun\ on average (see next section). 
Finding some of the most massive stars of our sample among the supergiants could be expected because the most evolved O stars are usually biased toward higher masses (the advanced evolutionary stages of lower mass stars occur at cooler temperatures).
%We expect to find some of the most massive stars of our sample among the supergiants because the most evolved O stars are usually biased toward higher masses (the advanced evolutionary stages of lower mass stars occur at cooler temperatures).
Overall, our sample contains six to eight stars that are more massive than 40 \msun\ (dwarfs  and supergiants), regardless of whether the HRD or the KD is used to determine the stellar mass.  
This is noteworthy because this mass limit seems to be connected to the properties of stellar evolution for a metallicity typical of the SMC \cite[see e.g.][]{ramachandran19, dufton20}. 

A direct reading of the HRD indicates that the giants of our sample gather along two evolutionary tracks, with initial masses about 20 \msun\ (for the stars O9 IIII and later) and 32 \msun\ (the O7-O8 III stars), respectively. 
Those with M $\approx$ 32  \msun\ lie half-way between dwarfs and supergiants, that is, they lie in the second part of the MS and after the TAMS. The less massive giants (M $\approx$ 20 \msun), however, are well past the TAMS, and as such should be as evolved as supergiants. Analysis of their surface abundance ratios confirms this advanced evolutionary status (Sect. \ref{sect_age}).

In the KD diagram, the dichotomy between the O7-8 and the O9-B0.5 is still present, although the dispersion is higher. 
The first group of stars lies in the mass range bracketed by the 25 \msun\ and 40 \msun\ tracks, in the second part of the MS,  reminiscent of the findings by \cite{martins17} for Galactic O7-8 giant stars. The later type giants have masses between 20 \msun\ and 25 \msun, and appear to be clearly more evolved (past the TAMS for the 20 \msun\ track). 

Most dwarfs (slightly less than 90\%) have higher \logg\ values than giants, while the lowest \logg\ are found for supergiants. In the KD, supergiants are therefore farther along the evolutionary tracks than the giants and the dwarfs for a given initial mass, which confirms that they are more evolved (see e.g. the 40 \msun\ track in the bottom plot of Fig. \ref{fig_tefflogg}).
However, there is a grey zone in which some dwarfs have \logg\ similar to giants (\logg\ = 3.65 and \logg\ = 3.75). One of these two dwarfs is probably a binary \citep{bouret13,dufton19}, while the other is ascribed a luminosity class IV by \cite{walborn00}.
Furthermore, half of the stars in the giants sample have surface gravities 3.25 $\leq$ \logg\ $\leq$ 3.50, which is expected to be typical of supergiants in most calibrations \citep[although these are for Galactic stars, e.g.][]{martins05}.  
The giants located past the TAMS are thus evolved objects with lower masses and luminosities than supergiants. That they have weaker winds explains their luminosity class classification \cite{martins21} %(see Martins \& Palacios, in press).

The three stars with the highest surface gravities lie between the 2 and 4 Myr isochrones in the KD and are therefore among the youngest in our sample. This status is qualitatively supported by their location in the HRD, although an age determination based on their location in the HRD is inaccurate because the lower part of the HRD shows notoriously poor sensitivity to age as a result of the small separation between isochrones, and the position of these stars overlaps with different isochrones within their uncertainties.
Two of these objects are fast rotators, with \vsini\ $\geq$ 300 \kms, and the third is classified as a O7 Vz star \citep[see][]{bouret13}. It is misleading to link their higher surface gravity to a younger evolutionary stage than the other O-type dwarfs in our sample, however, because three other, hotter dwarfs share the same age range while displaying lower \logg\ \citep[see][for a thorough discussion of the O Vz phenomenon, and related classification criteria]{sabinsanjulian14}.

% 
%%%%%%%%%%%%%%%%%%%%%%%%%%%%%%%%%%%%%%%%%%%%%%%%%%%%%%%%%%%
\section{Stellar masses}
\label{sect_mass}
\begin{figure}[tbp]
   \centering
   \includegraphics[width=9.2cm]{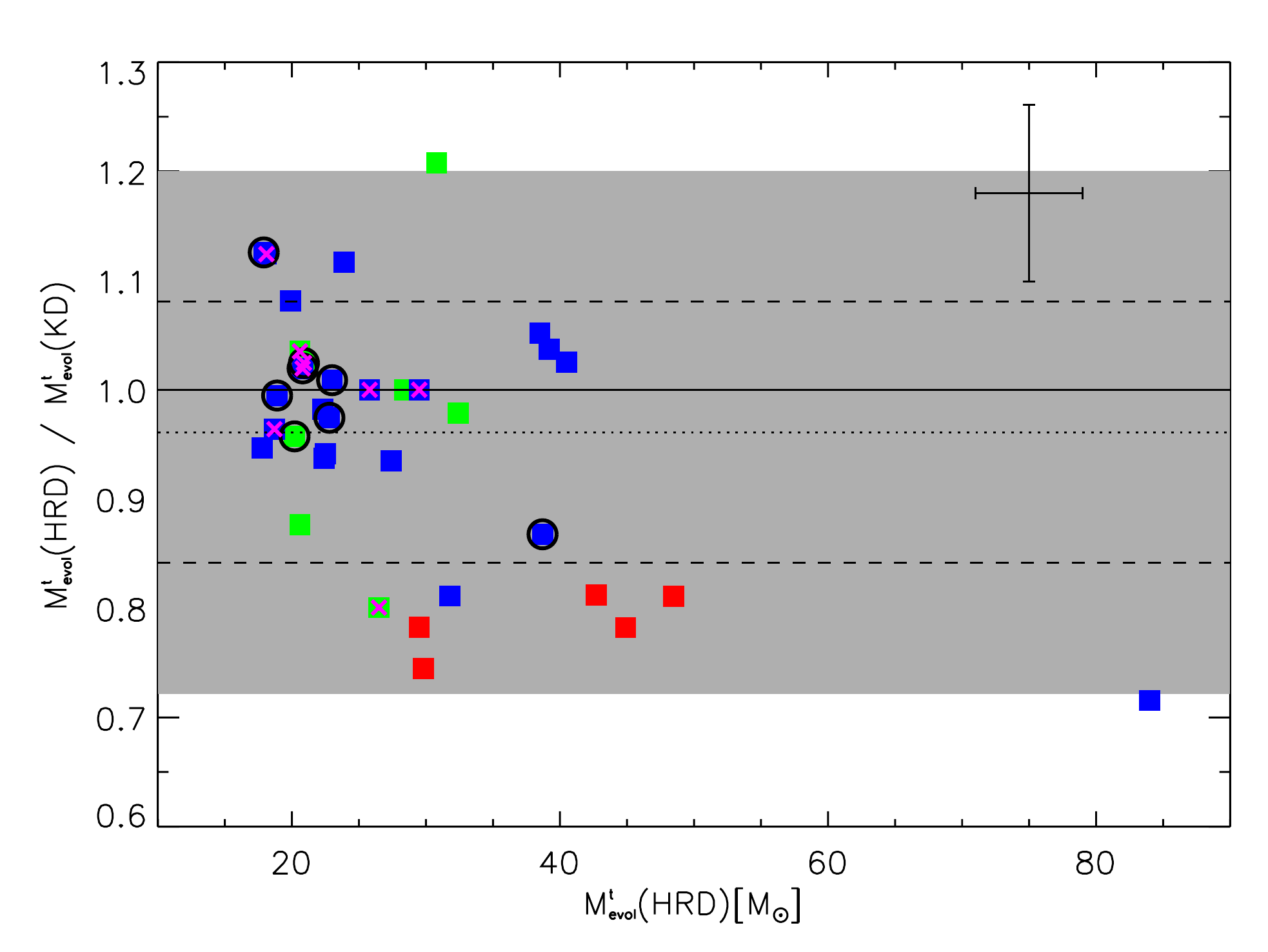}
       \caption{Ratio of evolutionary masses as derived from the HRD, and KD, respectively. The symbols and colour-coding are the same as in Fig. \ref{fig_tefflogg}.
       The full line indicates the one-to-one relation, the dotted line indicates the mean value of the mass ratio. The dashed lines and the shaded area correspond to a difference of $\pm$ 1$\sigma$, and 2$\sigma$   from the mean value, respectively. Black circle indicate stars with \vsini\ $\ge$ 200 \kms, while pink crosses indicate stars whose analysis relied on UV spectra only (no optical data available). A typical error bar for the mass ratio is indicated.}
         \label{fig_masse1}
   \end{figure}

A comparison of the evolutionary masses derived from the classical HRD and the KD is presented in Fig.~\ref{fig_masse1} by the ratio M$^{t}_{\rm HRD}$ to M$^{t}_{\rm KD}$ as a function of M$^{t}_{\rm HRD}$. 

A majority of stars present a ratio M$^{t}_{\rm HRD}$ / M$^{t}_{\rm KD}$ within 1$\sigma$ of the mass ratio centered at unity, three stars show a ratio M$^{t}_{\rm HRD}$ / M$^{t}_{\rm KD}$ = 1. Twelve objects show discrepancies higher than 1$\sigma$ (i.e. 25\% of the sample), which include the five supergiants, three dwarfs, and two giants. 
The outlier beyond the 2$\sigma$ limit in Fig.~\ref{fig_masse1} is \object{MPG 355}. 
This star is found to be significantly less massive in the HRD (85 \msun) than in the KD (120 \msun), although the masses may marginally agree within the error bars.
The optical spectrum of \object{MPG 355} has also been modelled by \cite{massey09}. Interestingly, their HRD indicates a mass of 75 \msun, while a KD would still yield a mass of 120 \msun, although the star is marginally farther along its evolutionary sequence in their case.

Figure \ref{fig_masse1} also shows that the majority of stars with \vsini\ $\leq$ 200 \kms, as well as those for which no optical analysis could be performed to derive the surface gravity, only show very moderate differences of M$^{t}_{\rm HRD}$ over M$^{t}_{\rm KD}$, regardless of the evolutionary status, as indicated by the luminosity class. 
Further confirmation requires a larger sample, however, because we are hampered here by small number statistics. 

The only systematic property we observe is that the five supergiants have lower M$^{t}_{\rm HRD}$  than M$^{t}_{\rm KD}$, and that the difference is higher than 1$\sigma$. Uncertainties in either the surface gravities or the distance plus reddening correction are not expected to be larger for these five stars, for which we have a full spectral coverage with very good S/N in most cases. 
Overall, a small majority of stars tend to exhibit M$^{t}_{\rm HRD}$ lower than M$^{t}_{\rm KD}$ (20 stars out of 36 for the full sample).  
Although very marginal, this is reminiscent of the results by \cite{sabin17} for the LMC, but this was obtained with different evolutionary tracks \citep[i.e.][]{brott11a}.

In contrast to the results of \cite{markova18}, we do not observe a systematic trend for higher ratios M$^{t}_{\rm HRD}$ / M$^{t}_{\rm KD}$ as the HRD mass increases. 
We recall that although \cite{markova18} used the HRD and the spectroscopic HRD \citep[sHRD,][]{langer14} to derive evolutionary masses for a sample of Galactic stars, neither KD nor sHRD require knowledge of stellar distances, and masses are almost independent of stellar radii.

\begin{figure*}[tbp]
   \centering
   \includegraphics[width=9.cm]{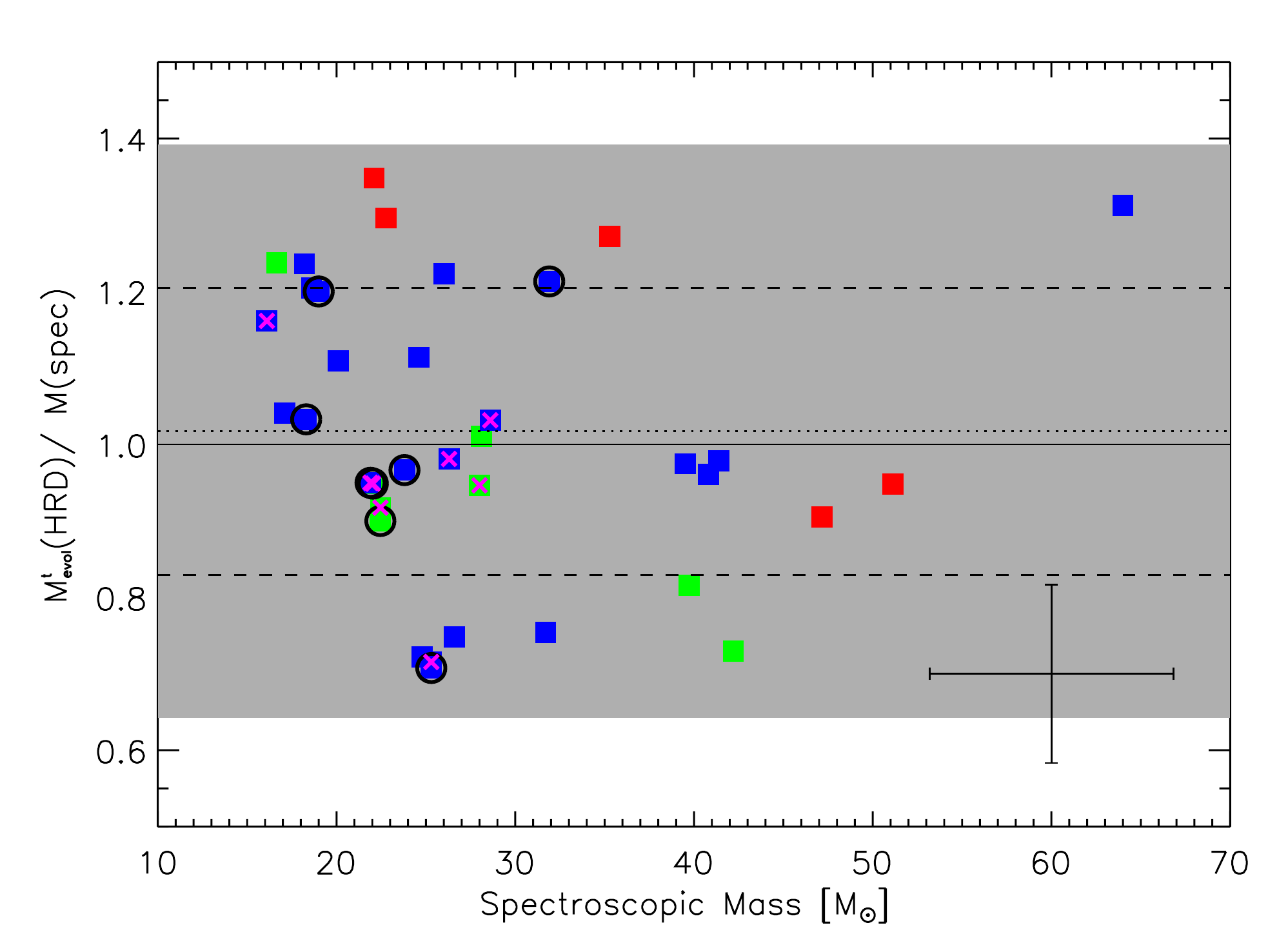}
      \includegraphics[width=9.cm]{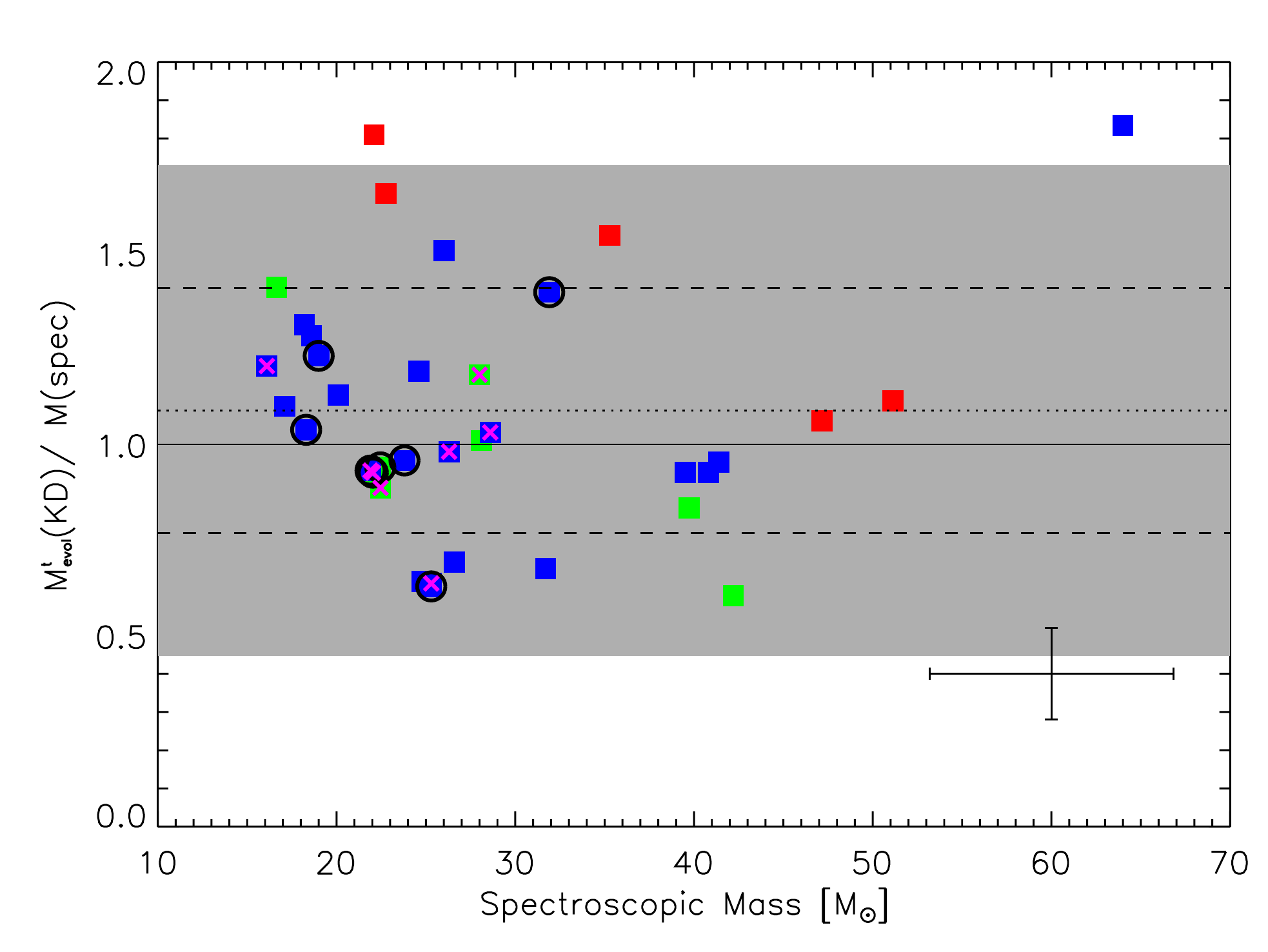}
  \caption{Ratio of evolutionary masses (as derived from the HRD and KD, left and right, respectively) to spectroscopic mass. The symbols and colour-coding are the same as in Fig. \ref{fig_masse1}. The dashed lines and the shaded area correspond to a difference of $\pm$ 1$\sigma$ and 2$\sigma$ from the one-to-one value, respectively. Typical error bars for the mass ratios are indicated.}
         \label{fig_masse2}
   \end{figure*}

We now focus on a comparison of spectroscopic masses to evolutionary masses derived from the HRD and KDs (Fig. \ref{fig_masse2}). 
As usual, the spectroscopic masses were obtained by combining the measured \logg, corrected for centrifugal acceleration \citep{repolust04}, with the stellar radius 
(from \teff\ from the spectroscopic analysis, and the luminosity from the modelling the SED). 

No obvious trend emerges from one plot or the other; the mass ratios are evenly distributed about the 1:1 line. It is noteworthy that fast rotators do not present any specific peculiarities in the mass discrepancy, which concurs with the conclusion of \cite{markova18} for Galactic stars. 

There might be a weak trend for a stronger mass discrepancy with lower spectroscopic masses (typically below 30 \msun) for the evolutionary masses from the HRD and KD. 
This is reminiscent, although not as clear, of the results of \cite{markova18}, who found a trend for a stronger mass discrepancy with lower spectroscopic masses for M$_{\rm evol}$(HRD) and for masses derived from the sHRD \citep{langer14} in their sample of Galactic stars. 

The spectroscopic versus evolutionary mass discrepancy is shown in the left plot in Fig. \ref{fig_masse2}. 
All stars fall within 2$\sigma$ of the 1:1 M$_{\rm evol}$(HRD)/M$_{\rm spec}$ mass ratio, although for 15 stars (i.e. $\approx$ 41\% of the sample) M$_{\rm evol}$(HRD) differs from M$_{\rm spec}$ by more than 1$\sigma$.  
For M$_{\rm evol}$(KD)/M$_{\rm spec}$, the right plot in Fig. \ref{fig_masse2} shows that for 10 stars ($\approx$ 28\% of the sample), this ratio differs by more than 1$\sigma$, and in one supergiant (and \object{MPG 355}), this ratio differs by more than 2$\sigma$.
In both cases, our values are therefore fairly close to the normal distributions, according to which 68\% of the objects should lie within $\pm 1 \sigma$.

The stellar distance is known to a high accuracy for SMC stars, therefore the error in \logg\ is the dominant source of error in the spectroscopic mass, hence in the ratio of the evolutionary mass to the spectroscopic mass. These uncertainties are partially rooted in the difficulties associated with rectifying echelle spectra, especially in the vicinity of the broad H lines. 
On the other hand, it is striking that the stars for which no optical data are available, hence whose \logg\ should be most affected by uncertainties, do not show a conspicuous trend of a larger mass discrepancy. 
In addition, in the specific cases of the M$_{\rm evol}$(KD) versus M$_{\rm spec}$ mass discrepancy, the intrinsically weak dependence of the KD on the distance suggests that the observed mass discrepancy is not a consequence of poorly constrained distances (or alternatively but somewhat counter-intuitively, that the error on the distance and reddening correction that goes into the calculation of the luminosity used in the classical HRD is smaller than the weak distance dependence of KD on the distance). 

A possible explanation for the larger M$^{t}_{\rm HRD}$ (or M$_{\rm evol}$(KD))  compared to M$_{spec}$ is that the spectroscopic masses are underestimated because our \cmfgen\ models do not account for turbulent pressure when the photospheric stratification is calculated, the neglect of which leads to lower \logg\ than the actual value \citep[see e.g.][]{massey13, markova18}. This effect is expected to be stronger for higher mass stars and/or for evolved stars with larger atmospheric scale height. The evidence for this trend is lacking in our sample.

%%%%%%%%%%%%%%%%%%%%%%%%%%%%%%%%%%%%%%%%%
\section{Projected rotational velocities}
\label{sect_vsini}

\begin{figure}[tbp]
   \includegraphics[width=9.cm]{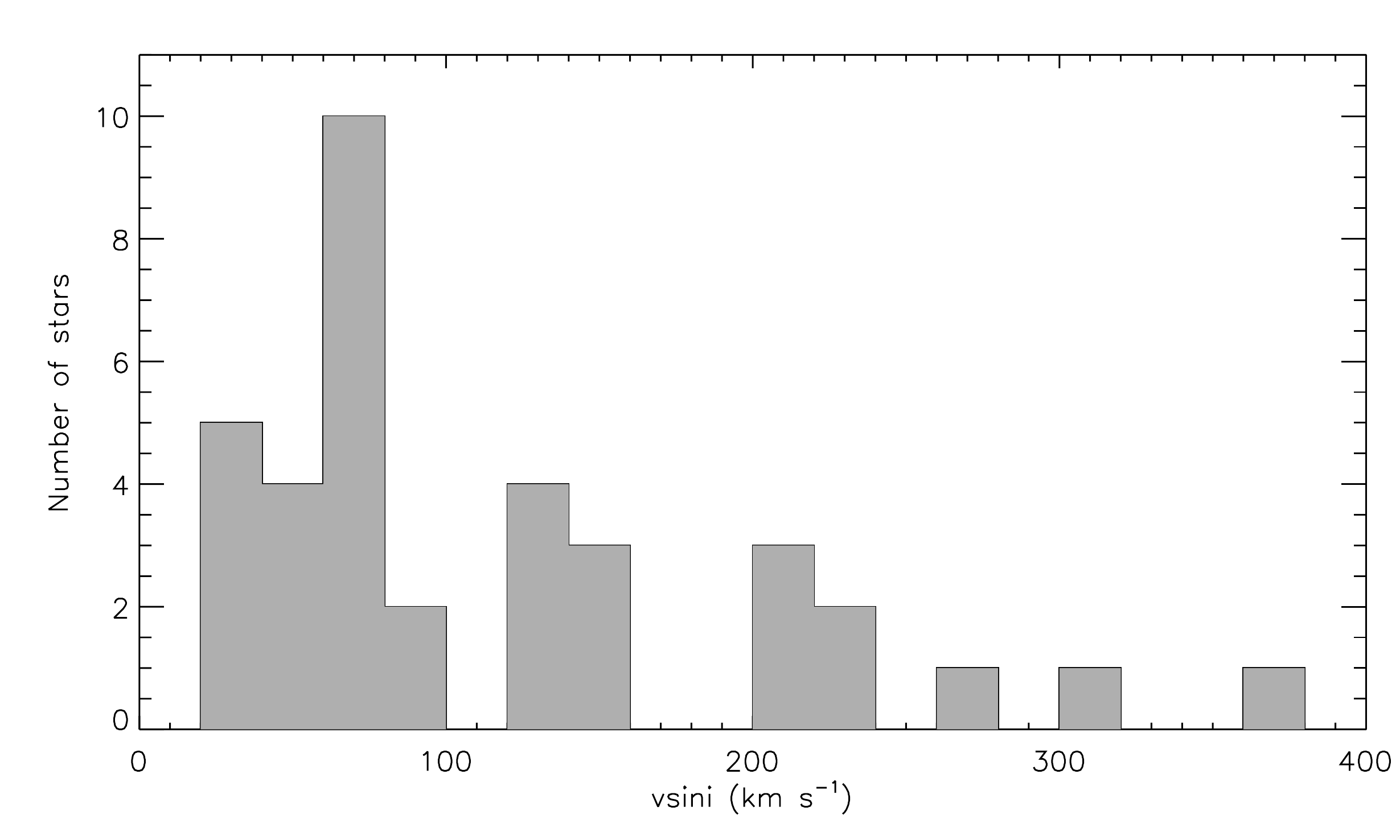}
    \includegraphics[width=9.cm]{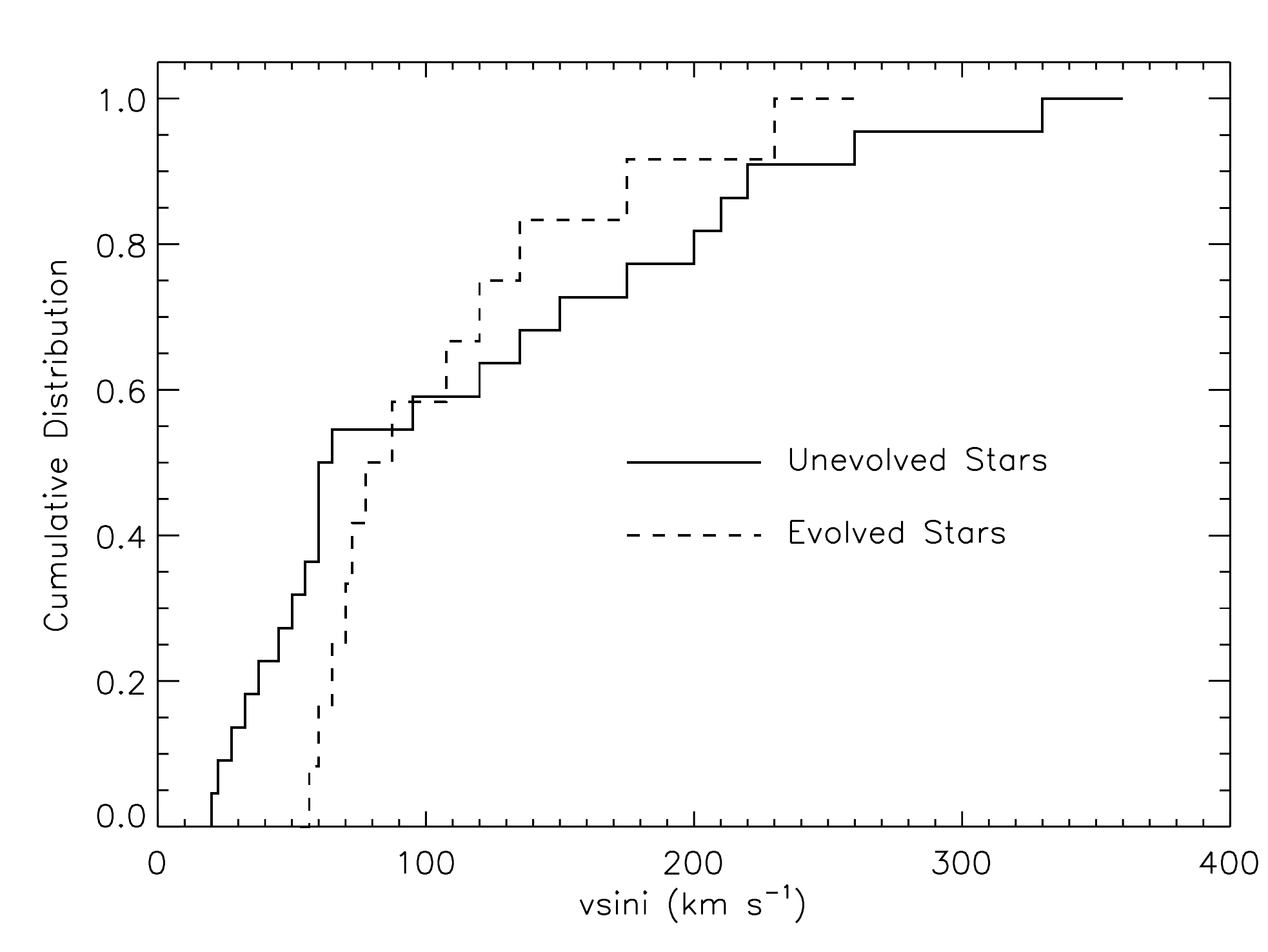}
 \caption{Top: Histogram of the project rotation velocities of our sample stars. The adopted bin is 20 \kms.
 Bottom: Cumulative distribution functions of \vsini\ values for the sample of unevolved SMC stars 
 (full line), and evolved stars (dashed line).}
         \label{fig_vsin}
   \end{figure}

The top panel of Fig. \ref{fig_vsin} shows the distribution of \vsini\ in our sample. A wide range of values is covered, from 20 to more than 350 \kms, with a clear bias toward low to moderate values (e.g. 25 stars have \vsini\ $\leq$ 120 \kms). The general shape of this histogram is likely a consequence of the way in which the sample is constructed.
Although rotation velocity was not a selection criterion of the \cosp\ sample, it was meant
to ensure that we would have a broad distribution of rotation velocities, especially compared to the 
the \stis\ sample alone, which was biased toward stars with narrow-line spectra, and did not
include stars rotating faster than 120 \kms\ \citep{heap06}. The actual distribution of true rotational velocities is possibly different owing to the projection factor.
Based on considerations of their mass discrepancies (spectroscopic versus evolutionary from the HRD), \cite{heap06} argued that the stars in the \stis\ sample were preferentially viewed pole-on \citep[see also][]{hillier03}. 
This is important when measured surface abundances as a function of rotation are compared to  
 theoretical predictions (see Sect. \ref{sect_rotation}). Using projected values rather than true rotational velocities may for instance introduce a bias when polar and equatorial regions of a rapidly rotating stars are or are not homogeneously enriched. 

To further examine the rotation rates, we split the full sample into two groups, 15 stars with \logg\ $\geq$ 3.95 (i.e. stars on the MS), and 21 stars for which 3.1 < \logg\ < 3.95 (i.e. post–MS stars \citep[see e.g.][]{hunter09, dufton18}).  The bottom panel in Fig. \ref{fig_vsin} shows the cumulative distribution functions (CDFs) of the unevolved and evolved sub-samples. This plot shows that there is a larger fraction of unevolved stars for projected rotation velocities below 100 \kms. The group of
unevolved stars also has a fraction of $\sim$ 26\% of stars with \vsini\ $\geq$ 200 \kms, while the fraction is 
$\sim$ 15 \% for the evolved stars. This is reminiscent of the findings by \cite{mokiem06}, who
found that their sample of SMC stars contains relatively more rapidly rotating unevolved stars than evolved stars. Furthermore, 
the difference in our case is significantly higher than what was found by \cite{penny09} in their analysis of SMC stars observed with \fuse, and taken at face value, this seems to support the predictions of stellar evolution that stars
slow down as they evolve on the MS \citep{brott11a,ekstrom12}. 
However, a Kolmogorov-Smirnof (KS) test in the combined sub-samples does not allow us to firmly conclude
that the distribution of \vsini\ for the unevolved stars is different from that of the evolved stars (significance level of 0.05 for a p-value = 0.016).
This is caused for a large part by the large spread of \vsini\ in SMC dwarfs. 

We are interested in the comparison of the surface abundances as a function of metallicity, therefore we
further compared the distribution of projected rotation velocities of our sample stars to that of the sample of Galactic stars analysed by \cite{martins15}, who used the same tools and method to measure \vsini\ and \mbox{CNO} abundances. 
Using the same criteria as above to define unevolved and evolved stars, we built two
sub-samples of 19 unevolved and 45 evolved stars. 

\begin{figure*}[tbp]
   \includegraphics[width=9.cm]{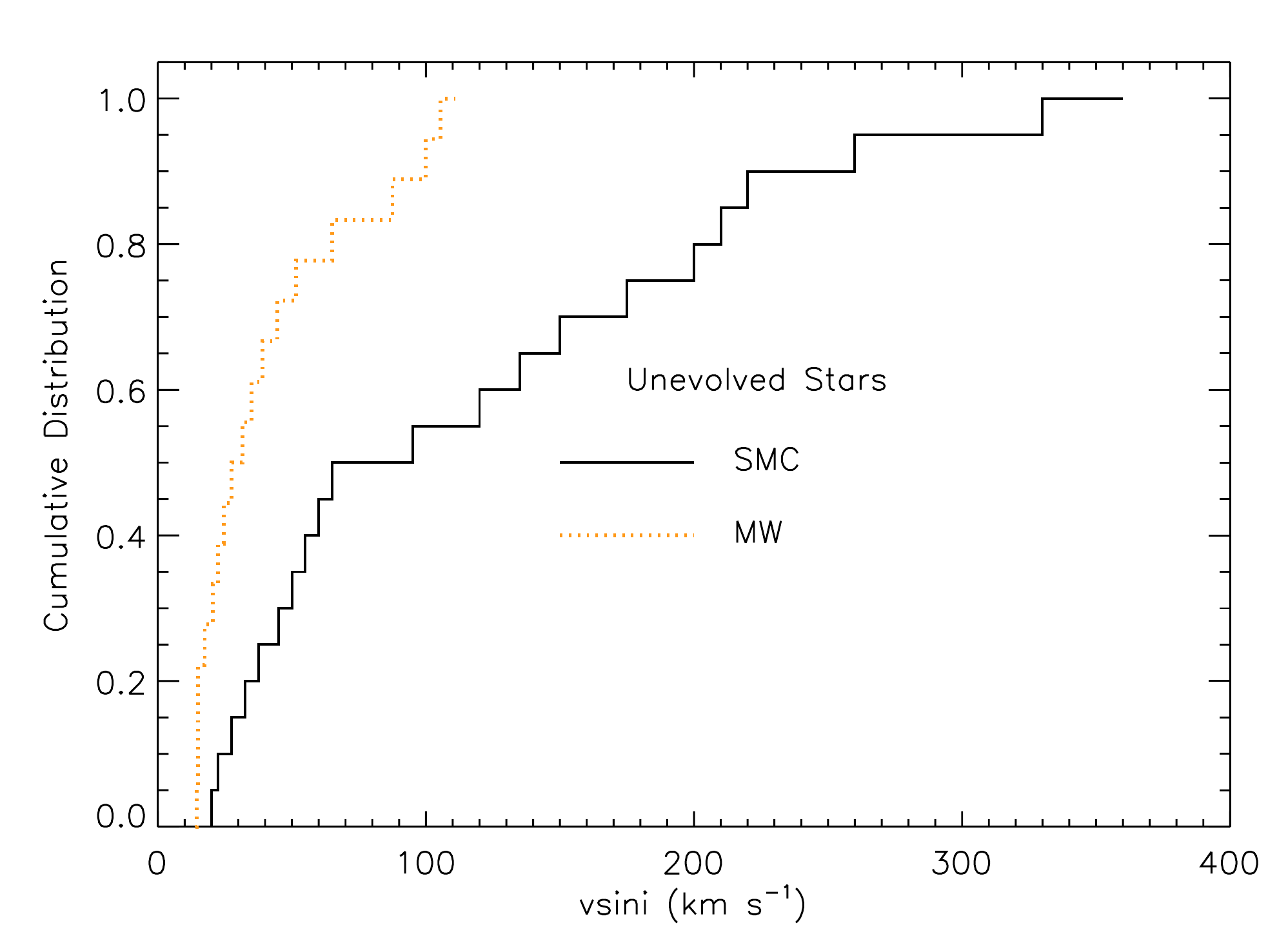}
    \includegraphics[width=9.cm]{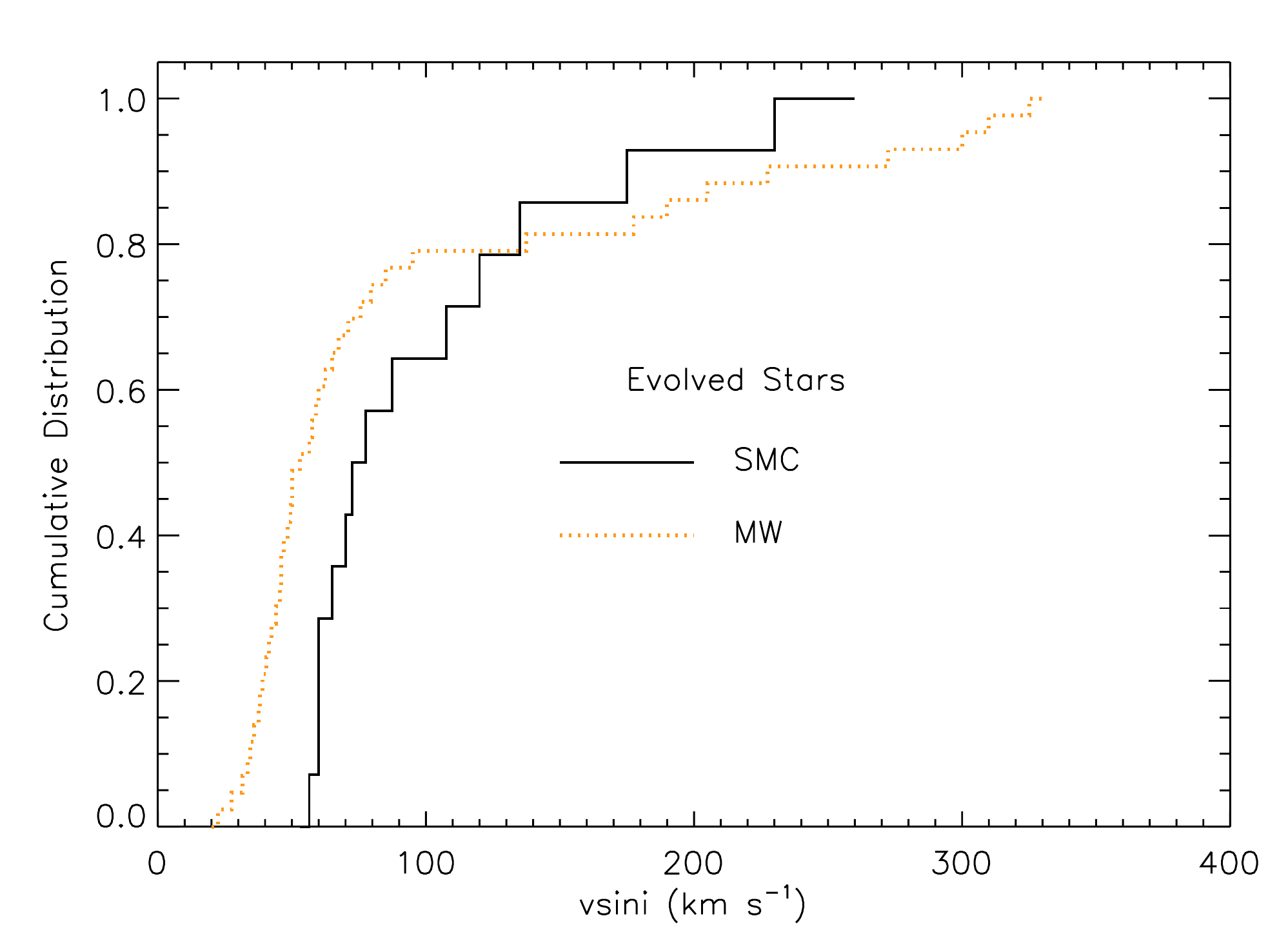}
 \caption{Left: CDFs of \vsini\ values for the sample of unevolved SMC and Galactic stars. Right: Same for evolved stars.}
         \label{fig_vsin_Z}
   \end{figure*}

The two cumulative distribution functions (CDFs) are presented in Fig. \ref{fig_vsin_Z}. 
The shape in the left panel of Fig. \ref{fig_vsin_Z} is striking and arises because
the population of the sample of unevolved Galactic stars in \cite{martins15} is devoid of stars rotating faster than 110 \kms. We considered merging the sample of \cite{martins15} with that of \cite{cazorla17}, for instance, as these authors analysed a sample of Galactic O-type with fast rotation using \cmfgen\ and the same method used here and in \cite{martins15}. 
However, only one star in their sample would be considered unevolved (\logg\ > 3.95), which would not change the shape of the CDF. In the end, we refrained from doing so to avoid biasing the resulting sample towards high rotation.    
The left panel of Fig. \ref{fig_vsin_Z} also shows that 80\% of the sample of unevolved Galactic stars has \vsini\ $<$ 100 \kms, while this fraction is smaller than 60\% for the SMC sample. 

The right panel focuses on evolved stars and shows the opposite: the distribution of \vsini\ extends to higher values for SMC stars than for their Galactic counterparts. Quantitatively, the CDFs in this panel indicate that the fraction of evolved Galactic stars with \vsini\ above 200 \kms\ is 8\% compared
to 14\% for the SMC stars. 

We then performed a KS test on the sample sets of Galactic versus SMC. We first tested for the hypothesis that the
\vsini\ distribution for SMC dwarf stars is shifted towards faster rotators than in the Milky Way (MW) dwarf stars; the null hypothesis for both distributions having the same mean can be rejected at a significance level of 0.025 (p-value = 0.015). In a similar test for the evolved samples in the SMC compared to the Galaxy, the null hypothesis can be rejected at a significance level of 0.01 (p-value =0.006); as for the dwarfs, the \vsini\ distribution for SMC evolved stars are shifted towards faster rotators than the MW stars.
We therefore conclude that there is significant evidence that the \vsini\ distributions for SMC and MW stars are different.

These results should not be viewed as evidence that SMC stars rotate faster than Galactic stars, although \citet{martayan07} have shown that this was the case for B stars \citep[see also][for O stars]{penny96,mokiem06,penny09}. Rather, this indicates that the velocity distributions are different in the Galactic and SMC sample by the construction of these samples. We remind that the sample used by 
\citet{martins15} was biased towards low rotational velocity to maximise magnetic field detection \citep{wade16}. This should be kept in mind when the effects of rotation on stellar evolution are discussed.
In view of these different conclusions, our result must be considered with caution given the small number of stars in each sub-sample as well.

%%%%%%%%%%%%%%%%%%%%%%%%%%%%%%%%%%%%%%%%%
\section{Surface abundances}
\label{sect_abund0}
In this section we first discuss the nature of the surface abundance patterns and their relation to the CNO cycle. We then focus on the metallicity dependence of surface chemical enrichment. Finally, we discuss the effects of rotation and age on the the degree of chemical processing.

\subsection{\mbox{CNO} ratios}
\label{sect_abund}

\begin{figure}[tbp]
   \includegraphics[width=9cm]{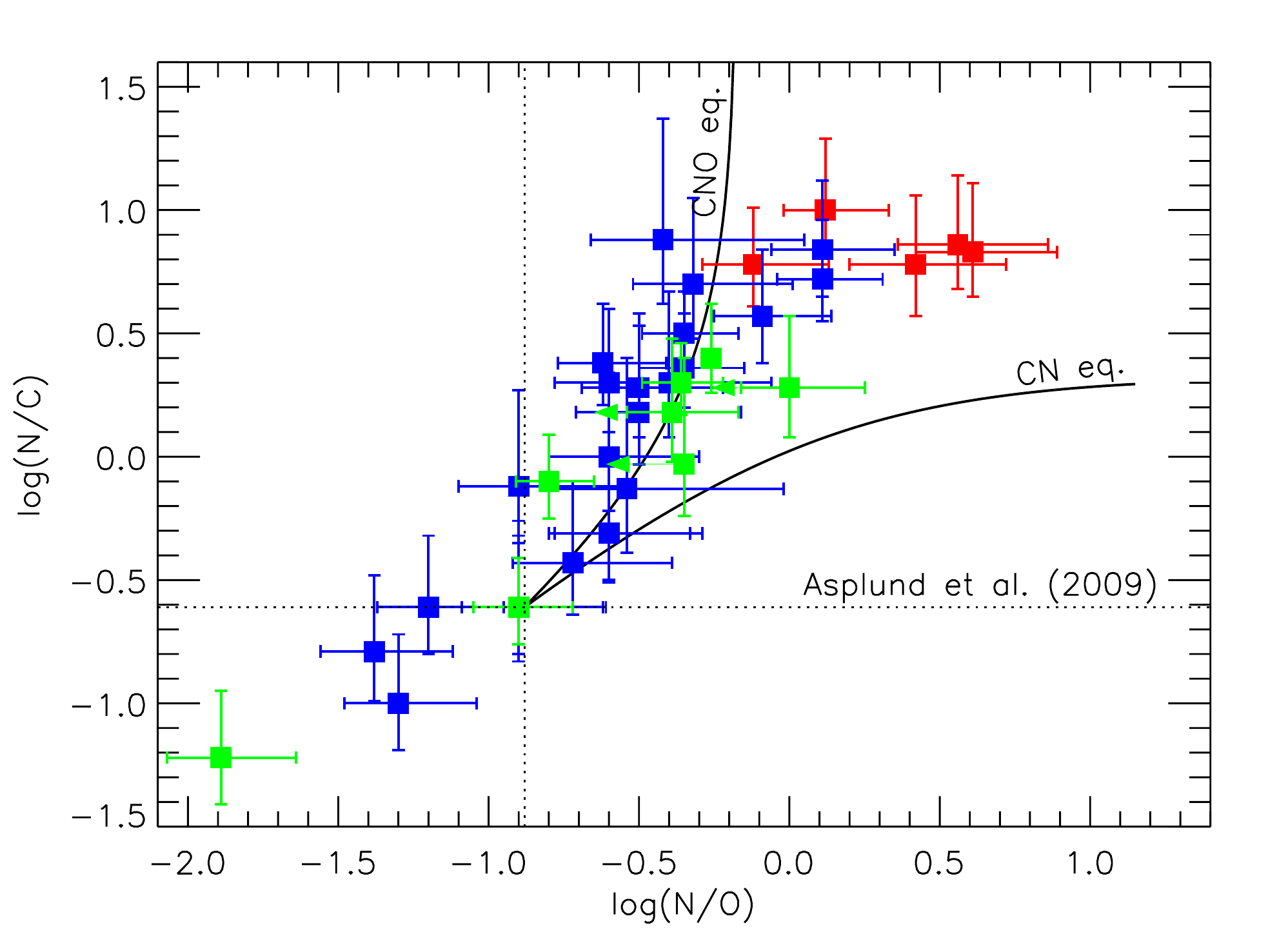}
 \caption{log (N/C) vs. log (N/O) abundances ((by number) for the sample stars. Solid lines indicate the expected trends for the case of the partial \mbox{CN} and complete \mbox{CNO} equilibrium. The dotted lines
 indicate the initial N/O and N/C (here solar). }
         \label{fig_cno}
   \end{figure}

In Fig. \ref{fig_cno} we show the N/C as a function of N/O to probe the chemical evolution of massive stars because these ratios vary substantially throughout the various phases of \mbox{CNO}-cycle \citep[see e.g.][]{maeder14}.
The measured ratios are correlated, showing a trend of higher N/C with higher N/O. This is consistent with the expectation that surface abundances are affected by products of the CNO cycle operating in the interior of massive stars \citep[e.g.][]{maeder14}.
More precisely, a trend for stronger chemical evolution for more evolved objects is observed from dwarfs and giants on one hand to supergiants on the other. However, this trend is not as clearly defined as the trend that was observed for Galactic O-type stars by \cite{martins15}. Most dwarfs and giants populate the same region in the plot in Fig. \ref{fig_cno}, and a few dwarfs exhibits N/C as high as the supergiants. 

Of the seven stars with the highest (log positive) N/Os, five are supergiants and two are dwarfs. Furthermore, \object{AV 15} and \object{AV 75}, the two supergiants  with the lowest value of log(N/O), are also the least evolved.  Of the two dwarfs, one (\object{MPG 355}) has the highest initial mass of the sample, and  the other (\object{AV 177}) is more massive than any of the giants of our sample and is also a fast rotator, with \vsini\ = 220 \kms.

\begin{figure*}[tbp]
  \includegraphics[width=9cm]  {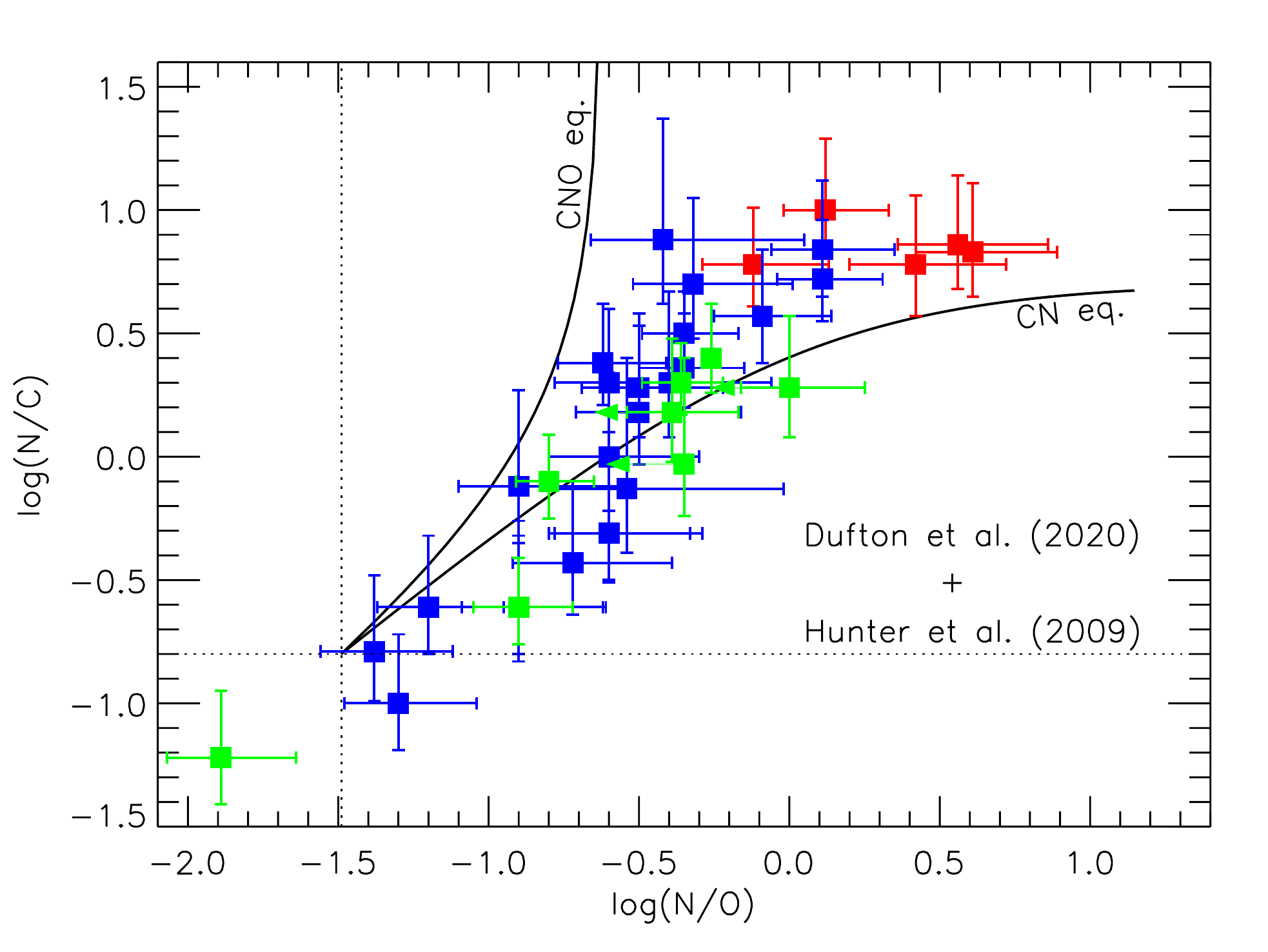} 
   \includegraphics[width=9cm]{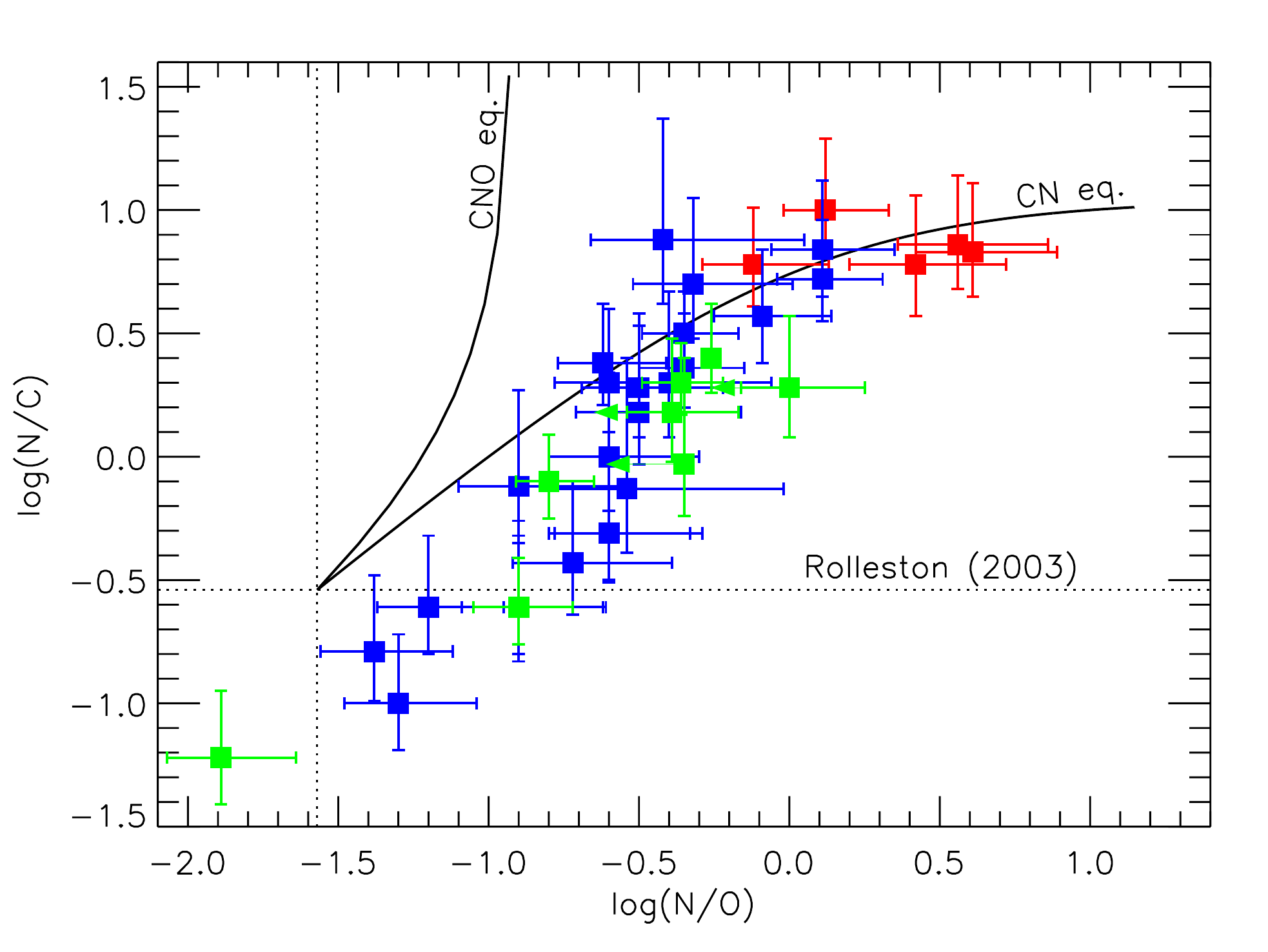}\\
     \includegraphics[width=9cm]{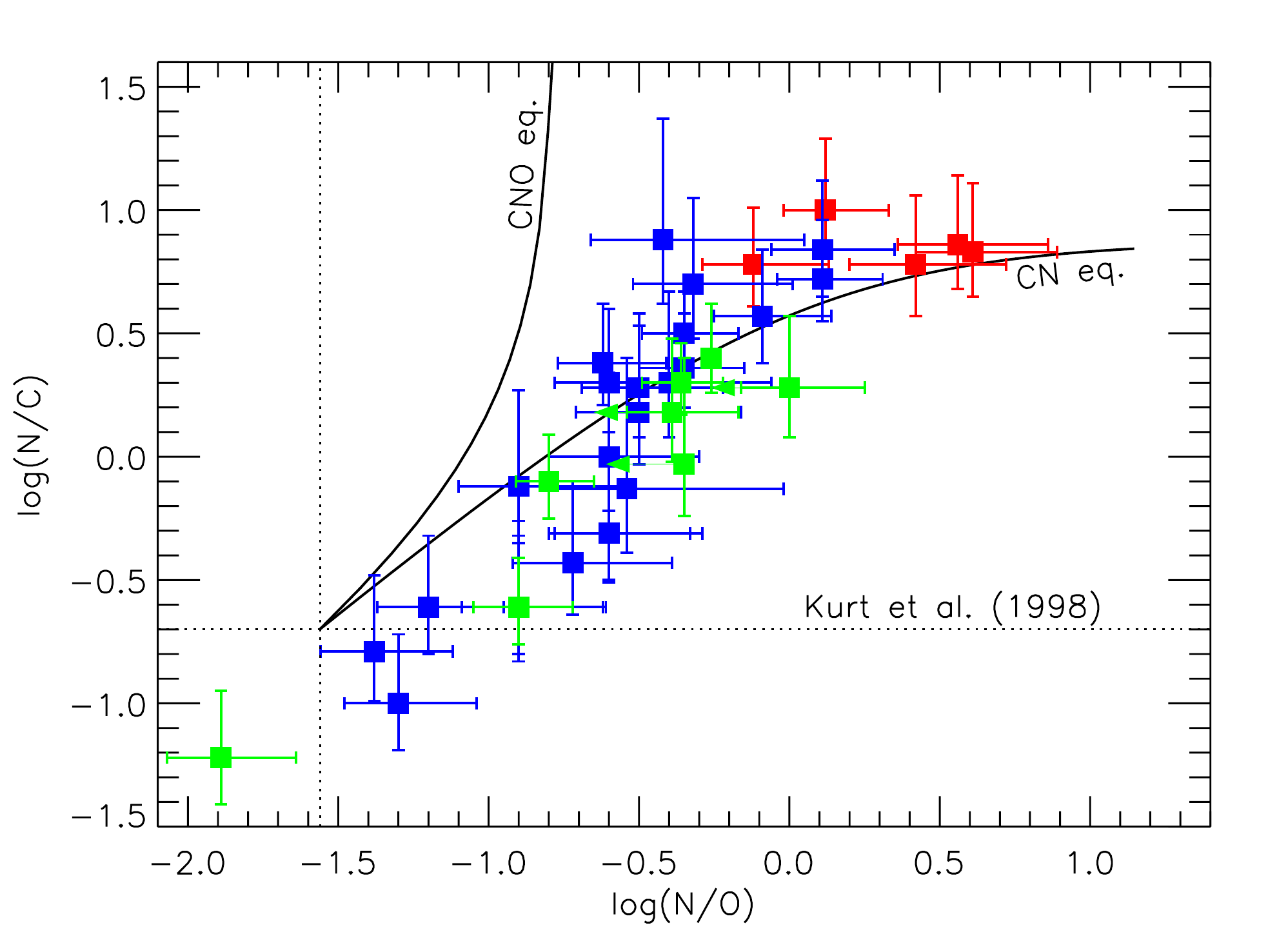}
       \includegraphics[width=9cm]{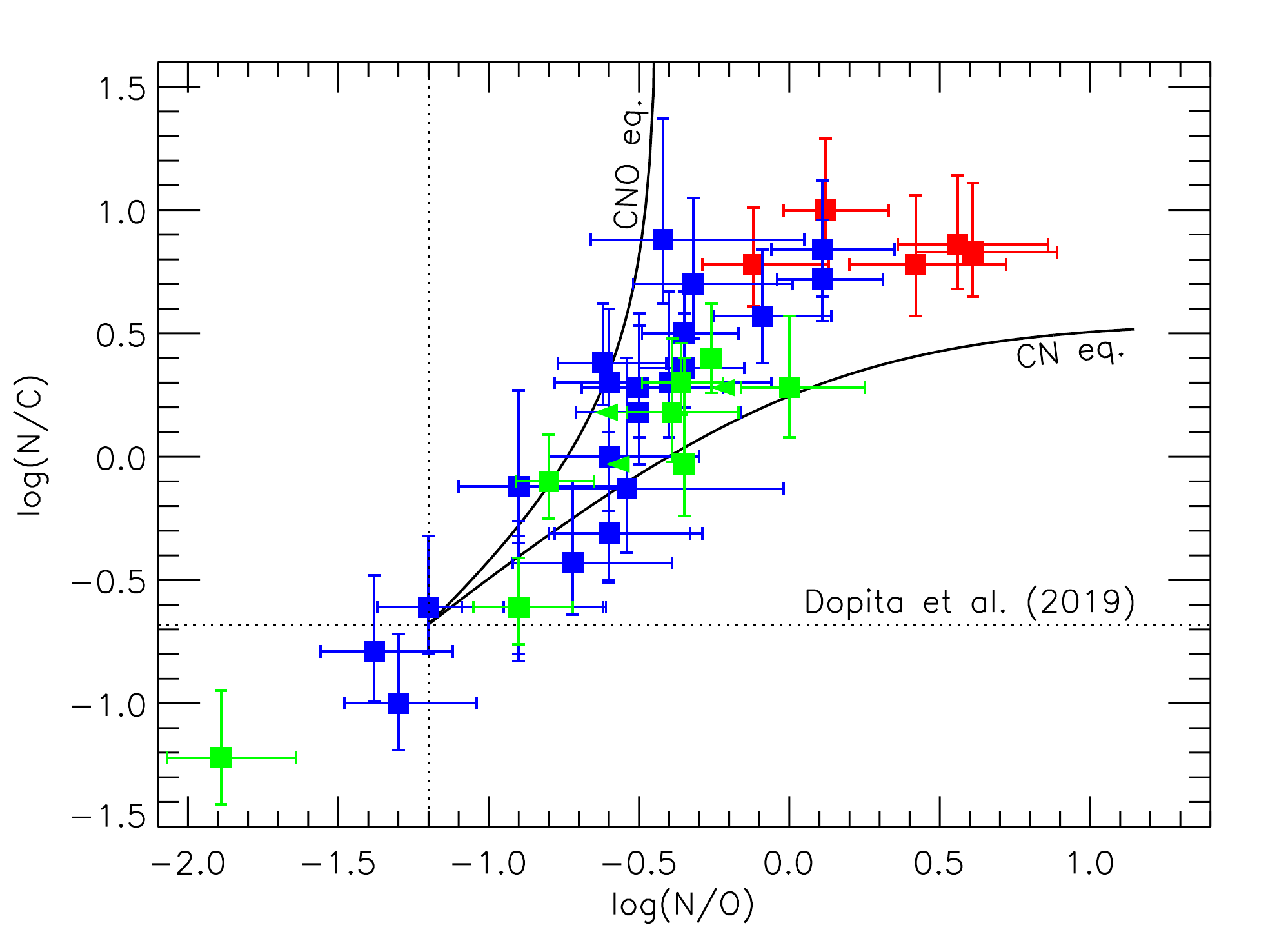}
      \caption{log (N/C) vs. log (N/O) abundances ((by number) for the sample stars. Solid lines indicate the expected trends for the case of the partial \mbox{CN} and complete \mbox{CNO} equilibrium for various initial mixture (see text for details).}
         \label{fig_cno2}
   \end{figure*}

For further information, we overplot in Fig. \ref{fig_cno} the expectations of nucleosynthesis through the \mbox{CNO} cycle, more precisely, the limiting cases of 
the complete \mbox{CNO}, and partial \mbox{CN} cycles derived by \cite{maeder14}. These curves express the degree of dilution of the CNO abundances at equilibrium mixed with the initial \mbox{CNO} 
abundances. These limiting cases above are expected to bracket the (more realistic) solutions obtained from numerical
models. In numerical models, predicted surface N/C and N/O will not only depend on nucleosynthesis, but are also affected by the adopted rotation rates and the assumptions used to describe mixing processes at different depths in the stars.
We adopted a solar initial \mbox{CNO} mixture from \cite{asplund09}, as in the models we use below \citep{georgy13}.

Figure \ref{fig_cno} shows a clear trend for a majority of dwarfs and giants to lie on or even above the \mbox{CNO}-equilibrium line. Only the supergiants all fall between the two limiting case. 
The observed dispersion of N/C for a given N/O is a consequence of the dispersion of the evolutionary stage of each individual object as much as the result of various internal conditions favouring either the \mbox{CN} or the 
\mbox{CNO} cycle. Despite this scatter, Fig.  \ref{fig_cno} provides evidence that internal mixing produces N enrichment at the expense of carbon and oxygen. 

An explanation for some of the N/Cs lying outside the theoretical predictions is that the initial mixture adopted to compute the analytical solutions of \cite{maeder14} is not relevant for SMC stars. To illustrate this, we collected different values for the SMC \mbox{CNO} baseline available in the literature. Figure \ref{fig_cno2} present a comparison of the effect of these initial \mbox{CNO} compositions on the expected trends for the \mbox{CNO} cycle.
The \mbox{CNO} initial abundances have been measured from various sites, from B-type stars \citep[][upper left and right, respectively]{hunter09, rolleston03} to
H~II regions \citep[][lower left]{kurt98} and using supernova remnant radiative shocks \citep{dopita19}.
We list them in Table 3.
All of them lead to N/O and N/Cs that are significantly different from the solar ratios that were adopted for Fig. \ref{fig_cno}. 
The four panels clearly display the effect of the initial N/O and N/Cs on the theoretical expectations for the \mbox{CN} and the \mbox{CNO} equilibrium limiting cases. 
They also emphasise that these initial ratios may lead to different interpretations of the chemical evolution of our sample stars.

\begin{table*}[htbp]
\caption{\mbox{CNO} abundances (12 + log X/H) and ratios for the SMC}
\begin{tabular}{lccccc}
\hline
\hline
Element     & Asplund et al. (2009) &  Hunter et al. (2009)   	& Rolleston (2003)   & Kurt \& Dufour (1998) & Dopita et al. (2019) \\
            & scaled solar (1/5)              &  + Dufton et al. (2020) 	& B-type dwarf    &     H II        & Supernova\\
            & 	               				&   B stars               		&  (AV 304)         &  regions           & remnants\\
            \hline
C           & 7.69         & 7.30               & 7.20        & 7.16           & 7.50 \\
N           & 7.08         & 7.24               & 6.66        & 6.46           & 6.82 \\
O           & 7.96         & 7.99               & 8.23        & 8.02           & 8.02 \\
log(N/C)    & -0.61        & -0.06              & -0.54       & -0.7           & -0.68 \\
log(N/O)    & -0.88        & -0.75              & -1.57       & -1.56           & -1.2 \\

  \hline
\end{tabular}
 \label{tab_abund}
%\tablefoot{
\tablefoot{The scaled \cite{asplund09} abundances for \mbox{CNO} are the baseline that we consider in the rest of the analysis as it is used in the Geneva models (but see text).}
 \end{table*}

The upper left panel, using initial \mbox{CNO} abundances from \cite{hunter09}, 
shows that several stars of our sample have N/O or/and N/Cs that are lower than the baseline ratios. The latter are clearly affected by the high initial value of
the mean nitrogen abundance for the SMC derived by \cite{hunter09}.
A recent study by \cite{dufton20} based on B-type stars in \object{NGC 346,} however, found that $\epsilon_{N}$, the SMC baseline abundance for nitrogen is around 6.5 
(where $\epsilon_{N}$ = log [N/H]+12), that is, about 0.8dex lower than \cite{hunter09}.  
This offset between initial \mbox{CNO} from \cite{hunter09} has been described in \cite{martins13} for Galactic stars. 
On the other hand, a majority of our sample stars fall nicely between these two extreme lines, homogeneously covering the parameter space bracketed by the theoretical \mbox{CN} and \mbox{CNO}
lines. This suggests that they are still evolving towards the full \mbox{CNO}
equilibrium. 

The upper right \citep{rolleston03} and lower left panels \citep{kurt98} present similar trends: a good fraction of our sample stars fall below the \mbox{CN} curve. Although the measured ratios can be reconciled within the error bars with the expectations for the \mbox{CN} and the \mbox{CNO} equilibrium for the adopted \mbox{CNO} baselines, it is possible that these SMC initial abundances do not represent the initial mixture of the O-type stars of our sample. More quantitatively, this is a direct consequence of too high an N/C but too low an N/O in \cite{rolleston03} and \cite{kurt98}.  

\cite{kurt98} measured the \mbox{CNO} baseline abundances from H~II regions, which are expected to best represent the composition of recently formed stars.  However, except for N66, which is located in NGC 346 where several of our sample stars are located, the other H~II regions used by \cite{kurt98} lie well outside the main bar of the SMC, and thus might yield abundances that are not representative of O-type field stars in the SMC.  

The \mbox{CN} and \mbox{CNO} equilibrium trends relying on the SMC baseline \mbox{CNO} composition derived by \cite{dopita19} (lower right panel) best bracket the measured \mbox{CNO} ratios.
These SMC initial abundances therefore likely best represent the \mbox{CNO} mixture out of which our sample stars formed.  In terms of chemical evolution, the location of a large majority of the sample stars between the two extreme lines indicates that they are indeed evolving away from the \mbox{CN}-equilibirum towards the full \mbox{CNO}-equilibrium.
The individual values of \mbox{CNO} elements differ mildly from those used (scaled solar) in the models of \cite{georgy13} that we used in this study. More importantly, the N/O and N/Cs in \cite{dopita19} are lower by a factor of 2 at most than in the Sun. For these abundances the number of outliers is the smallest (only three, and only one when the error bars are taken into account). 
Remarkably, the scaled solar abundances come very close in terms of the smallest number of outliers 
(cf. Table \ref{tab_abund}).

There is now ample evidence that nitrogen in the SMC is depleted relative to the other elements (cf. e.g. the references for the discussion above). Regardless of the true value is, this calls for caution 
when the observed surface abundances of our sample stars are compared to predictions of stellar evolution models built on (scaled) solar abundances. 
Variations in initial abundances induce change in the zero-points of the nuclear 
paths. As a consequence, the relative variations of abundances throughout the evolution must be 
estimated taking the differences in the initial ratios into account. 
\cite{georgy13} noted for instance that when the initial composition measured for the SMC by \cite{hunter09} is used, the nitrogen enrichment for a 15 \msun\ star in the middle and at the end of the MS differs by slightly more than a factor of two with respect to the enrichment obtained at the same stage when scaled solar initial abundances for the heavy elements are used. 

Finally, regardless of the initial mixture we considered, one giant lies well below the baseline N/O and N/Cs. This star is \object{AV 69}, which is classified as an OC7.5 giant \citep{walborn00}. \cite{hillier03} analysed 
\object{AV 69} in detail and found its nitrogen abundance to be very low, more precisely, lower than the SMC baseline.  
The analysis of \cite{hillier03} also indicated that \object{AV 69} is also non-standard in terms of C and O (see Table \ref{tab3}) with respect to the SMC baselines (see Table \ref{tab_abund}). 
\cite{hillier03} further argued that this star is a slow rotator that did not experience significant mixing. Slow rotation is indeed an obvious possibility to explain why the star does not show signatures of strong enough mixing that would take the N/C to higher levels. A direct reading of the \mbox{CNO} abundances suggests that they would have to be be primordial
in \object{AV 69}. 
\cite{martins16} similarly concluded that Galactic OC supergiants likely experienced very little processing through the \mbox{CN} or \mbox{CNO} cycles. 
They further argued that because the OC phenomenon in Galactic stars is related to the chemically unprocessed surface abundances, it is best seen in more evolved objects for which chemical processing is already strong 
in morphologically normal stars. As a consequence, for Galactic stars, the OC phenomenon does
not exist among dwarfs that are only slightly chemically evolved. 
When we extrapolate this to environments with lower metallicities such as the MCs, where stronger and faster chemical processing is expected, the OC phenomenon may appear in earlier evolutionary stages and spectral type such as giants and even dwarfs. 
This is supported by the identification of two LMC early-type stars as OC stars by \cite{evans15}, or by \object{MPG 113}, which is an SMC OC dwarf \citep{bouret03, bouret13}. 

In the following sections, we investigate whether the level of surface chemical processing we measured are consistent with expectations from stellar evolution theory. We focus more specifically on the metallicity, rotation, and the evolutionary status of our sample stars.

%%%%%%%%%%%%%%%%%%%%%%%%%%%%%%%%%%%%%%%%%%%%%%%%%%%%%%%%%%%%%%%%%%%%%%%%%%%%%%%%%%%%%%%%%%%%%%

\subsection{Dependence on metallicity}
\label{sect_metallicity}
\begin{figure}[tbp]

  \includegraphics[width=9.2cm]{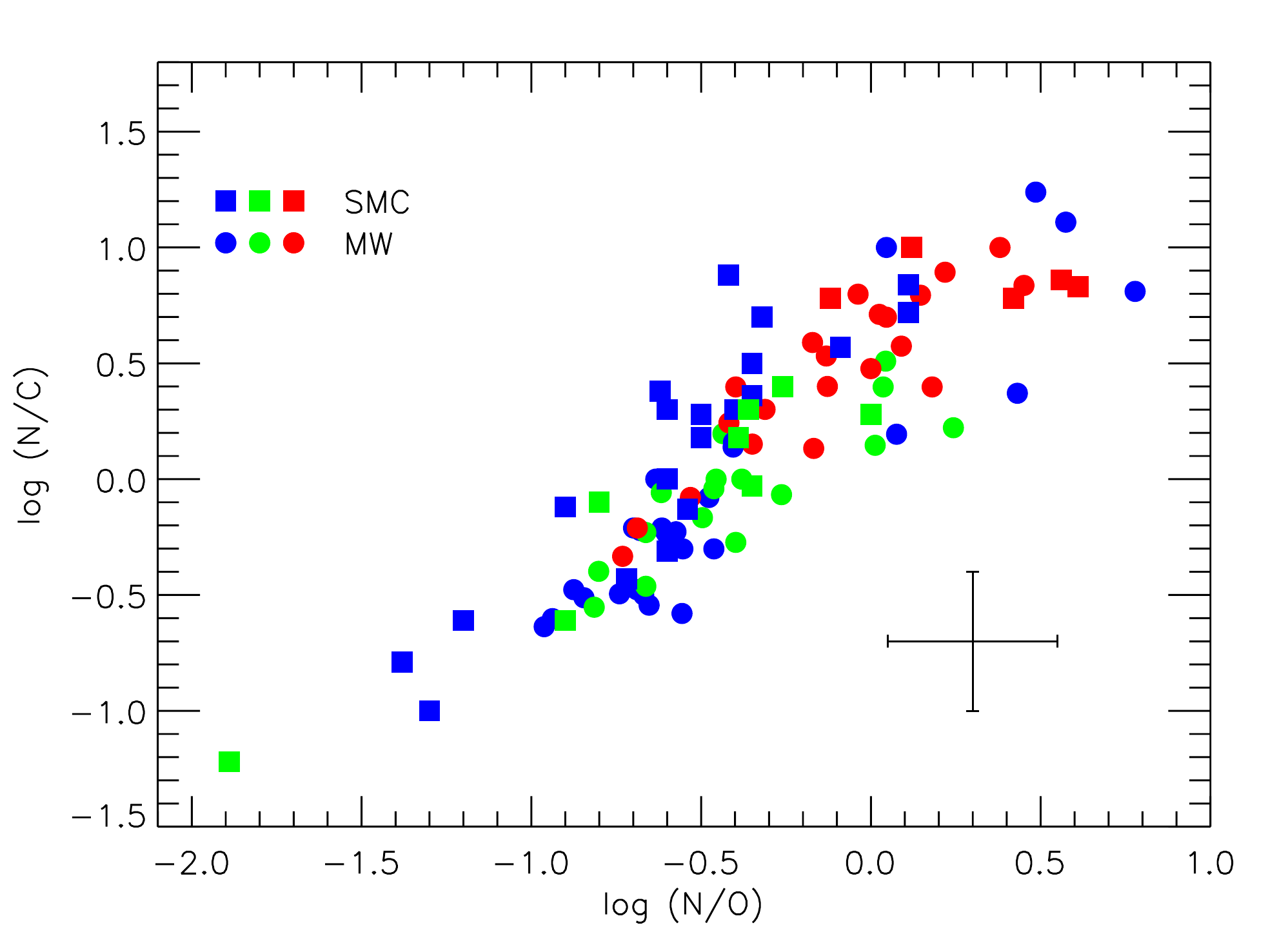}
  \caption{Our sample stars (squares) together with Galactic stars from \cite{martins15} (circles). Same colour-coding as before for the luminosity classes.}
         \label{fig_abund_met}
  \end{figure}

In the past decade, several studies have been published on the surface abundances of massive stars in the Galaxy \citep[e.g.][]{martins15, mahy15, martins16, martins17, cazorla17, markova18, carneiro19} or in the Magellanic Clouds \citep[e.g.][]{rivero12, bouret13, grin17}. Evolutionary computations predict that for stars of similar masses and evolutionary stages, the chemical enrichment is stronger at lower metallicities for identical initial $v_{init}/v_{crit}$ \citep[e.g.][]{maeder09, brott11a, georgy13}.
Hence the relations between the observed N/C and N/Os may differ as a function of mass and evolutionary stage.
Observationally, this prediction still lacks confirmation. Here we compare our results to those of \cite{martins15}, which were obtained with similar tools and methods. Figure \ref{fig_abund_met} presents the two samples in a log(N/O) -- log(N/C) diagram. No clear trend with metallicity is visible. 
Although a few SMC dwarfs appear as evolved as some Galactic O supergiants, the opposite is also true: the O supergiants of our sample appear to be somewhat less evolved than some Galactic O dwarfs. 

\begin{figure*}[]
   \centering
   \includegraphics[width=0.49\textwidth]{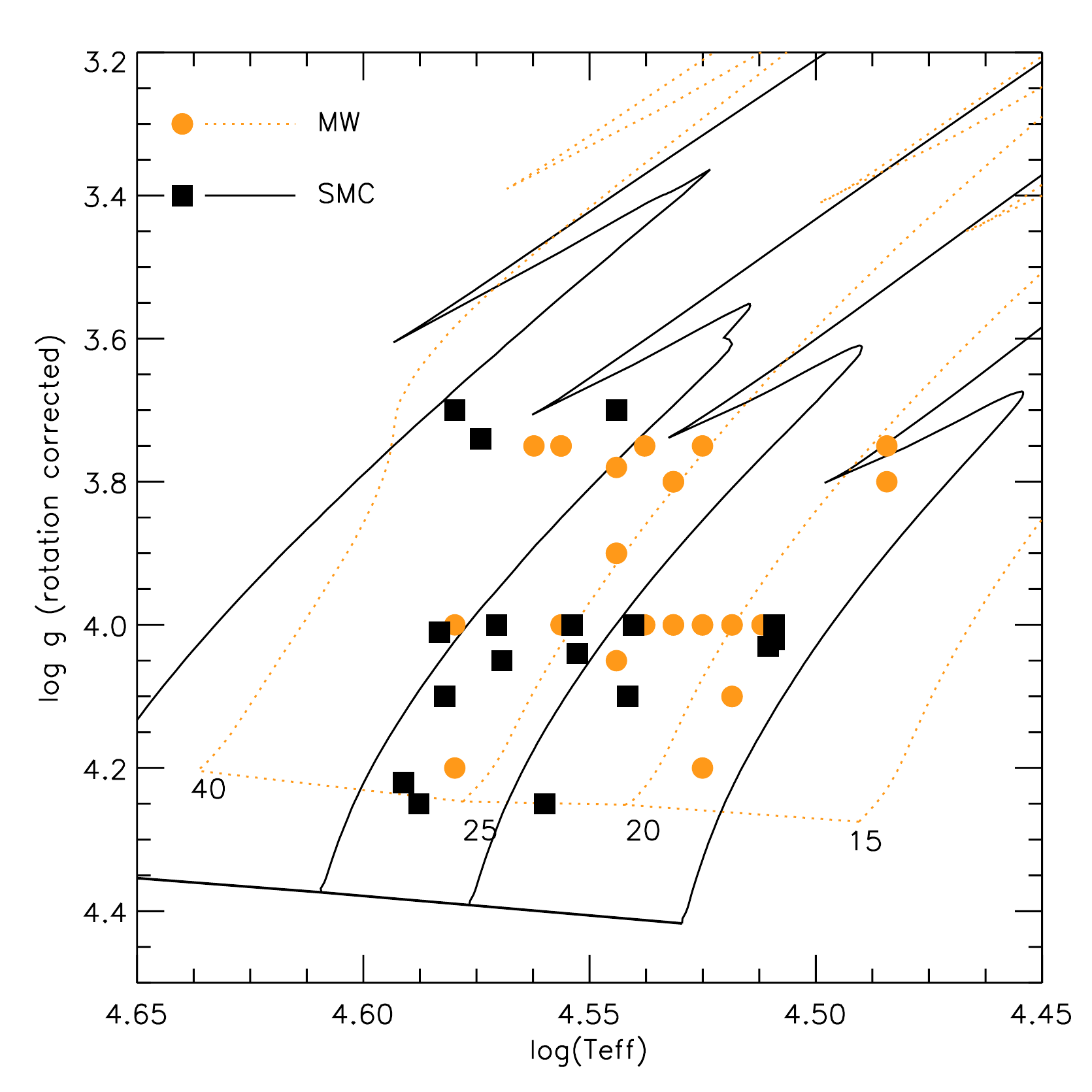}
  \includegraphics[width=0.49\textwidth]{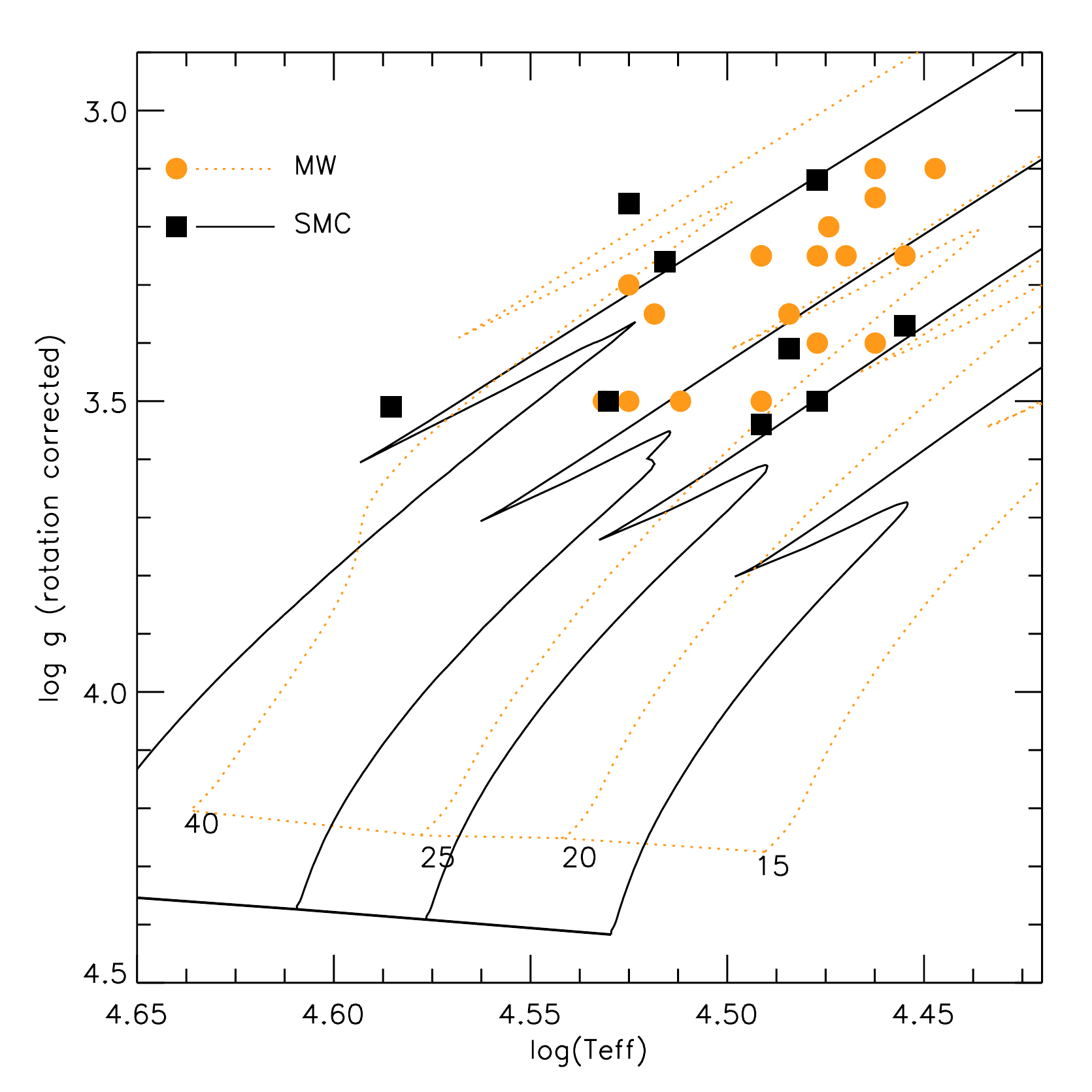} \\
   \includegraphics[width=0.49\textwidth]{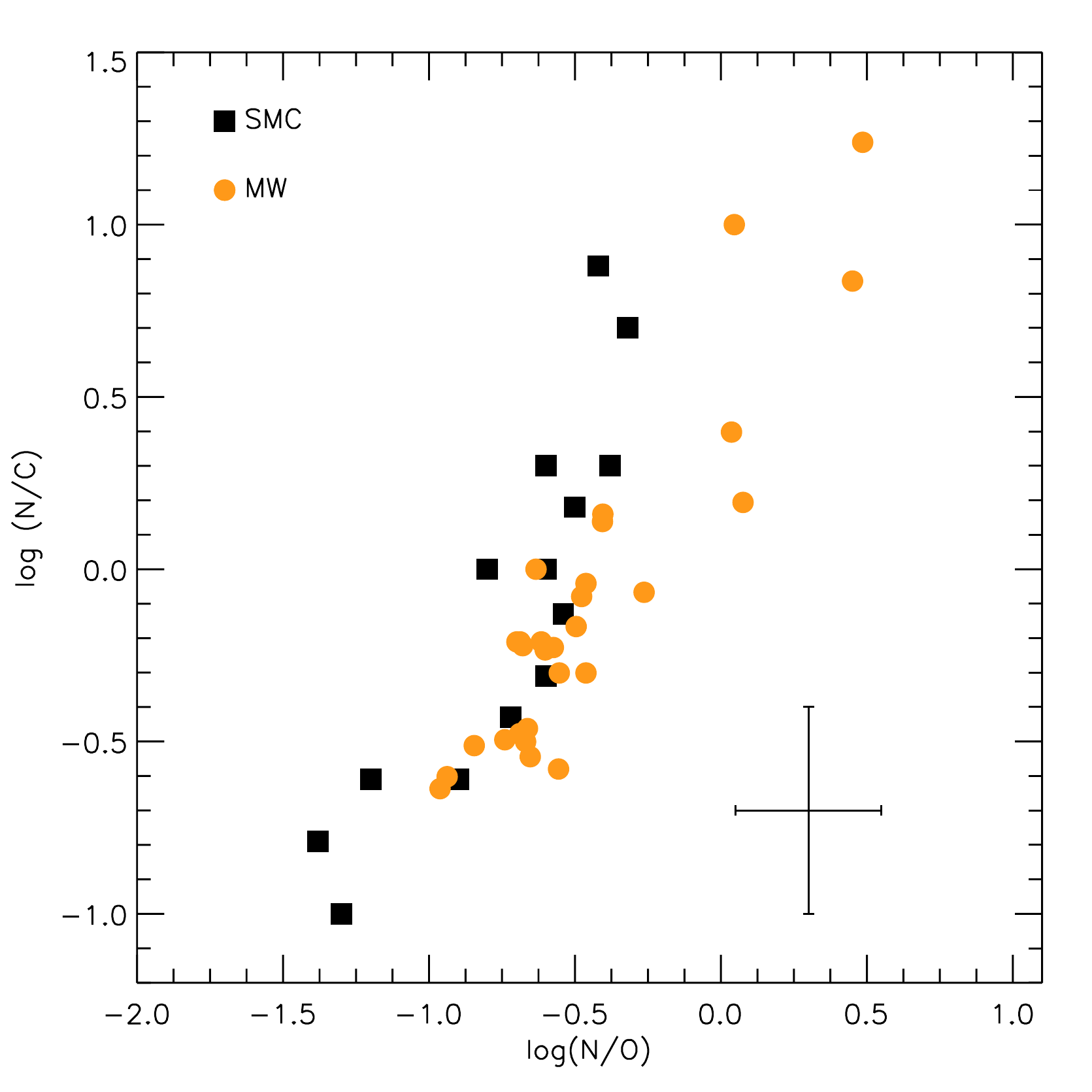}
  \includegraphics[width=0.49\textwidth]{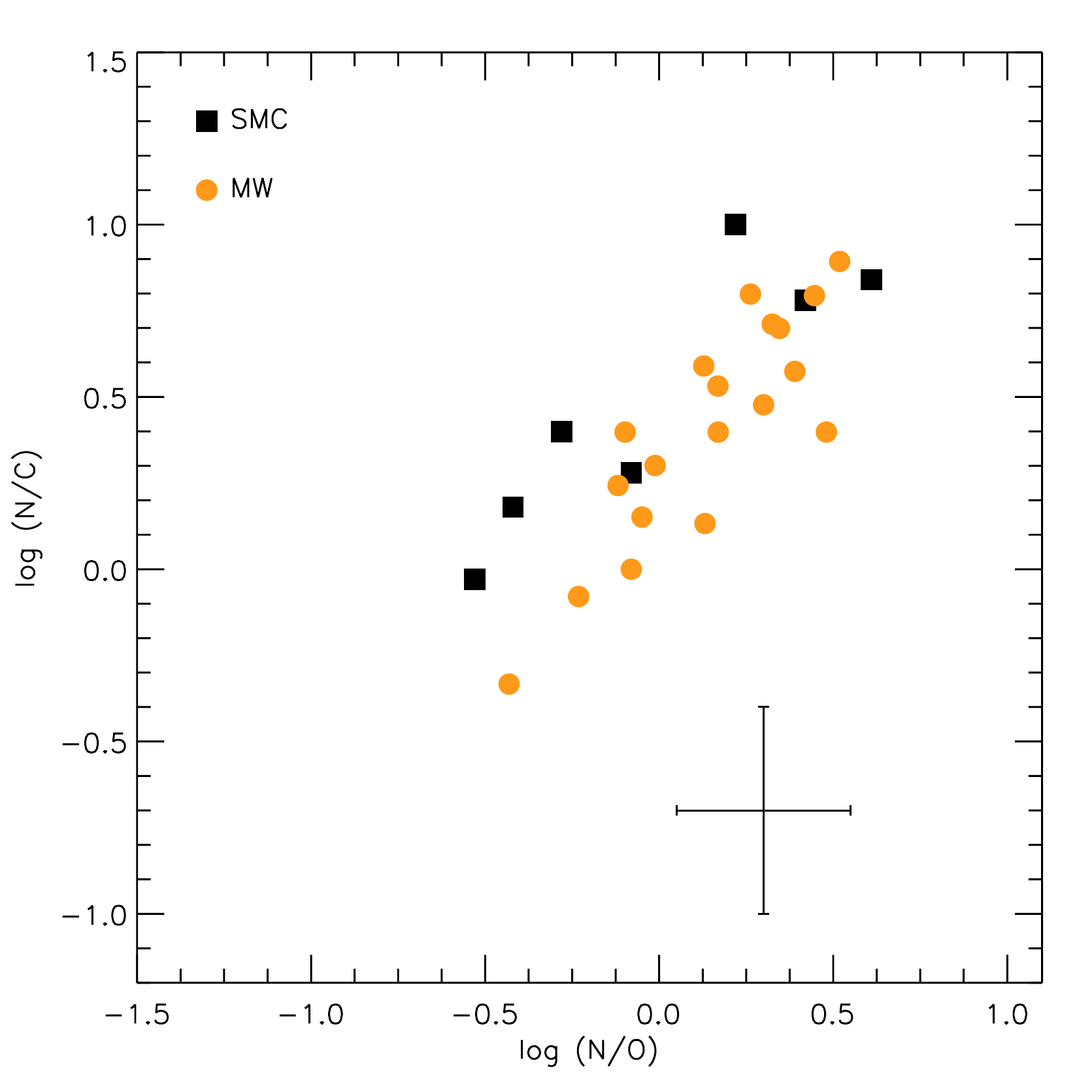}
  \caption{Top: Kiel diagram for SMC stars (black squares and solid lines, this study) and MW stars \citep[orange circles, dotted lines,][]{martins15}. Stars have been selected to overlap significantly in the KD, either on the MS (top left panel) or on the post–MS (top right panel). Bottom: log (N/C) vs. log (N/O) diagram for the same sub-sample of stars. Evolutionary tracks are from \citet{ekstrom12} and \citet{georgy13}.} 
         \label{fig_CNO_Z}
   \end{figure*}

To better isolate any potential metallicity effect between the Galactic and SMC stars, we restricted the samples to stars that cover the same region of the KD (see Fig.~\ref{fig_CNO_Z}, top panels). Their surface abundances are shown in the bottom panels of Fig.~\ref{fig_CNO_Z}. No conspicuous trend is observed, but a higher N/C (for a given N/O) in SMC stars compared to Galactic stars may be spotted. 

A potential pitfall here is that this visual indication might be biased by initial $v_{init}/v_{crit}$ being higher in low metallicity stars 
than in Galactic stars. In this case,  the (potentially) stronger surface enrichment in metal-poor stars might indeed be the consequence
of their higher initial rotational velocities rather than be due to steeper internal gradients of angular velocities. 
This word of caution is supported by the shape of the cumulative distribution functions of \vsini\ for both populations \citep[cf. Sect. \ref{sect_vsini}, see also the discussion in ][of this topic]{penny09}.
 
For a more robust view, we applied a principal component analysis (PCA) to the log(N/O) -- log(N/C) data to verify whether the SMC stars are on average 
more enriched in log(N/C) than their MW counterparts for a given log(N/O) value.
For a two-dimensional data sample, the PCA analysis is equal to a single rotation of the axes, yielding two new components, PC1 and PC2, the first of which points to the direction of maximum data variance, whereas the second is orthogonal to that direction. In our case, PC1 should in particular coincide with the ordinary least-squares regression line, and PC2 would correspond to orthogonal deviations from this regression. 

To define the principal components, we joined the two samples (MW + SMC) because we wish to find the directions of maximum variance in the abundance space \citep[see e.g.][]{boesso18}. The coefficients of the linear combination of the initial variables from which the principal components are constructed are
\begin{equation*}
{\rm PC1  =   0.664 log(N/O)  +  0.748 log(N/C)}
\end{equation*}
\begin{equation*}
{\rm PC2  =   -0.748 log(N/O)  +  0.664 log(N/C).}
\end{equation*}

\begin{figure}[tbp]
   \centering
   \includegraphics[width=9.cm]{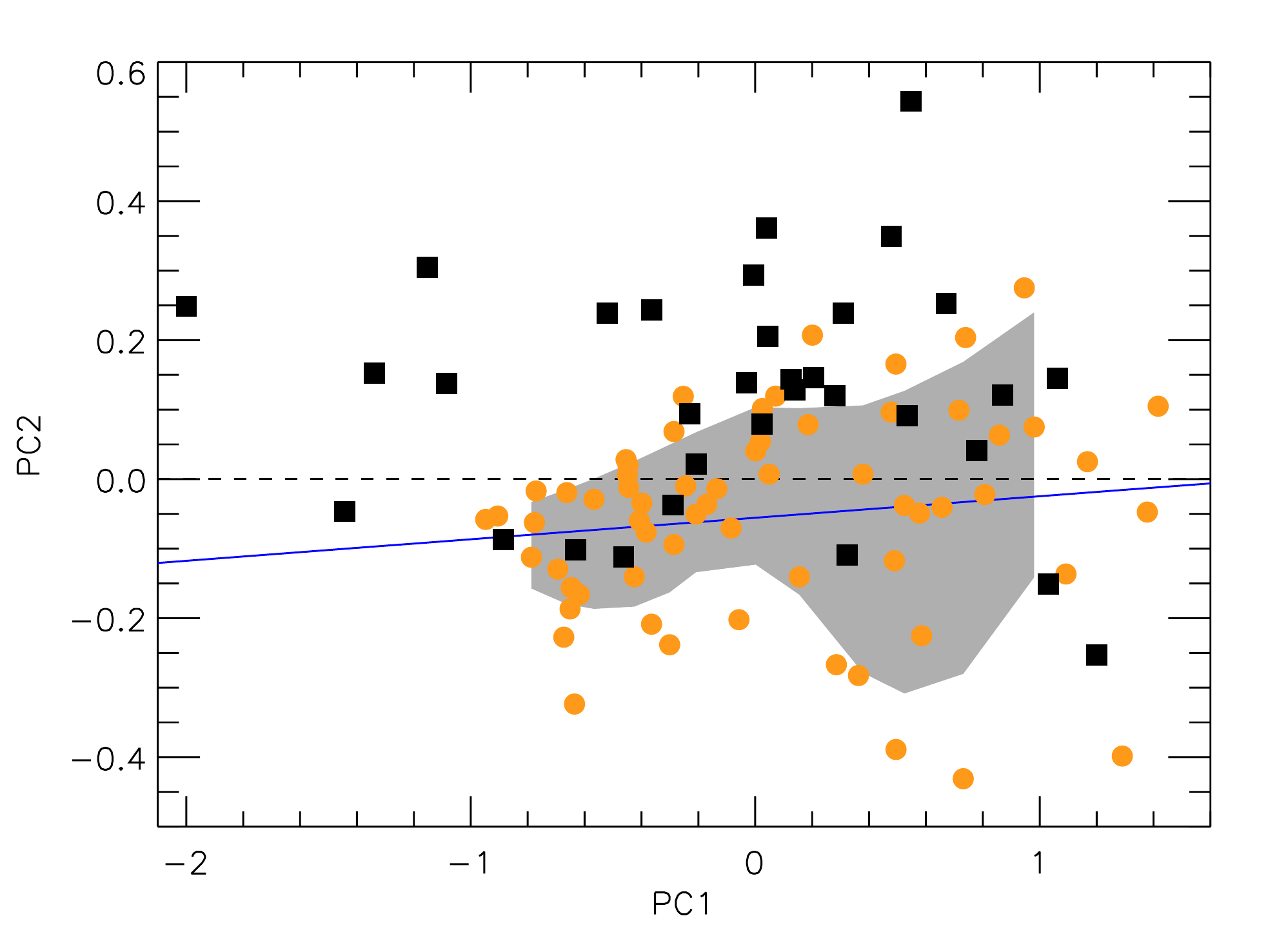}
  \caption{PCA analysis of the full merged SMC + Galactic sample. Squares show SMC stars (this study), and circles  correspond to Galactic stars \citep{martins15}. See text for details.}
         \label{fig_pca}
   \end{figure}

Figure \ref{fig_pca} shows the abundance space log (N/O) -- log(N/C) in this new coordinate system. The dashed black line is the ordinary least-squares regression line when the two data samples are joined. A quick glance at this plot shows that the SMC data have higher PC2 scores than MW data. This is reinforced by the disposition of the SMC points with respect to the grey area, which corresponds to the 2$\sigma$ confidence band for a locally estimated scatterplot smoothing (LOESS)
%\LEt{please introduce} 
regression \citep[e.g.][]{cappellari13} of the Milky Way log(N/O) -- log(N/C) relation. The LOESS curve is found by a non-parametric regression that fits simple models to localised subsets of the data and smooths the relations. It can be compared with the blue line, which corresponds to the ordinary least-squares regression line for the Milky Way data alone.

%{DHG}

We used a two-sample Kolmogorov-Smirnoff test to confirm the null hypothesis that the distribution of PC1 and PC2 scores for the MW and SMC comes from the same distribution. While the data samples do not differ statistically in the first principal component, the null hypothesis can be discarded for the PC2 scores with a probability (p-value) $\sim 4.9\times 10^{-5}$. 

By joining the two samples before applying the PCA, our analysis took a conservative approach. Had we defined the first principal component by using only the MW log(N/O) -- log(N/C) data, the PC1 axis would have coincided with the blue line in Fig. \ref{fig_pca}, and the difference between the PC2 scores for the MW and SMC stars would have been larger.
From this we conclude that the SMC stars are on average more enriched in log(N/C) than the MW for a given log(N/O) value.

The PCA of our sample and \cite{martins15} supports stellar evolution predictions that low metallicity indeed favours stronger chemical processing. 
We acknowledge that a more definitive conclusion still requires the analysis of a larger sample of SMC stars (at least commensurate to the available sample of Galactic stars) to define more statistically significant bins of the same initial masses. These should be analysed as homogeneously as possible using a larger variety of tools (atmosphere codes) and methods (e.g. line fitting, equivalent widths, and curve of growth). 

%%%%%%%%%%%%%%%%%%%%%%%%%%%%%%%%%%%%%%%%%%%%%%%%%%%%%%%%%%%%%%%%%%%%%%%%%%%%%%%%%%%%%%%%%%%%%%%%%%%%

\subsection{Surface abundances and evolutionary status}
\label{sect_age}

\begin{figure}[tbp]
   \centering
   \includegraphics[width=9.2cm]{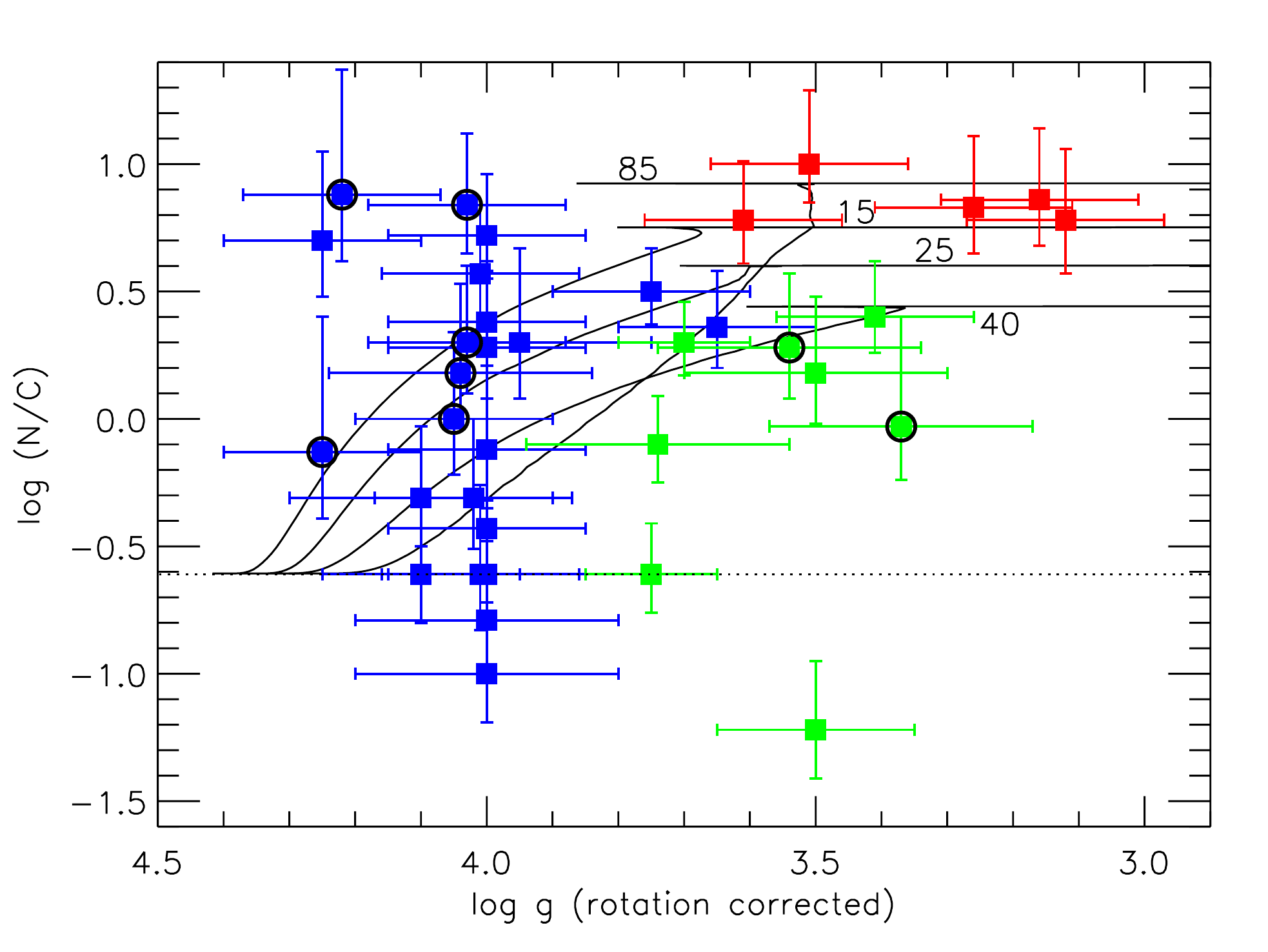}
  \caption{log (N/C) vs. \logg\ for the sample stars. The evolutionary tracks are labelled by initial mass (initial rotation $v_{init}$ = 0.4$v_{crit}$). The horizontal dotted line indicates the baseline N/C for the SMC used in the models. Stars rotating faster than 200 \kms\ are highlighted by thick black circles.}
         \label{fig_abund3}
   \end{figure}

Figure \ref{fig_abund3} shows log (N/C) plotted against the (log of) surface gravity. The N/C indicates the evolutionary level of a star in terms of nucleosynthesis, while the surface gravity is a proxy for the evolutionary point reached by a star on its mass track.   
While there is clear separation in \logg\ between dwarfs on one hand and the giants and supergiants sample on the other, there is a significant overlap of these two categories of evolved stars that correspond to stars in similar evolutionary states at or just past the TAMS (cf. our comments on Sect.~\ref{sect_evol}, see also Fig.~\ref{fig_tefflogg}). 
Dwarfs display a wide range of N/C values, while supergiants show a remarkably homogeneous set of N/C values. Giants have a somewhat intermediate behaviour.
The two dwarfs with the lowest \logg, typical of what we find for some giants, display log(N/C) values similar to those of their more evolved (giants) counterparts. This property further suggests that these two stars are indeed more evolved than the remaining stars in the dwarf sample.
We also note that while high N/Cs are not systematically associated with fast rotation, all the fast rotators in our sample display  significant signs of chemical processing (cf. Fig. \ref{fig_abund3}). This is further discussed in the next section.

\begin{figure*}[tbp]
   \centering
      \includegraphics[width=9.cm]{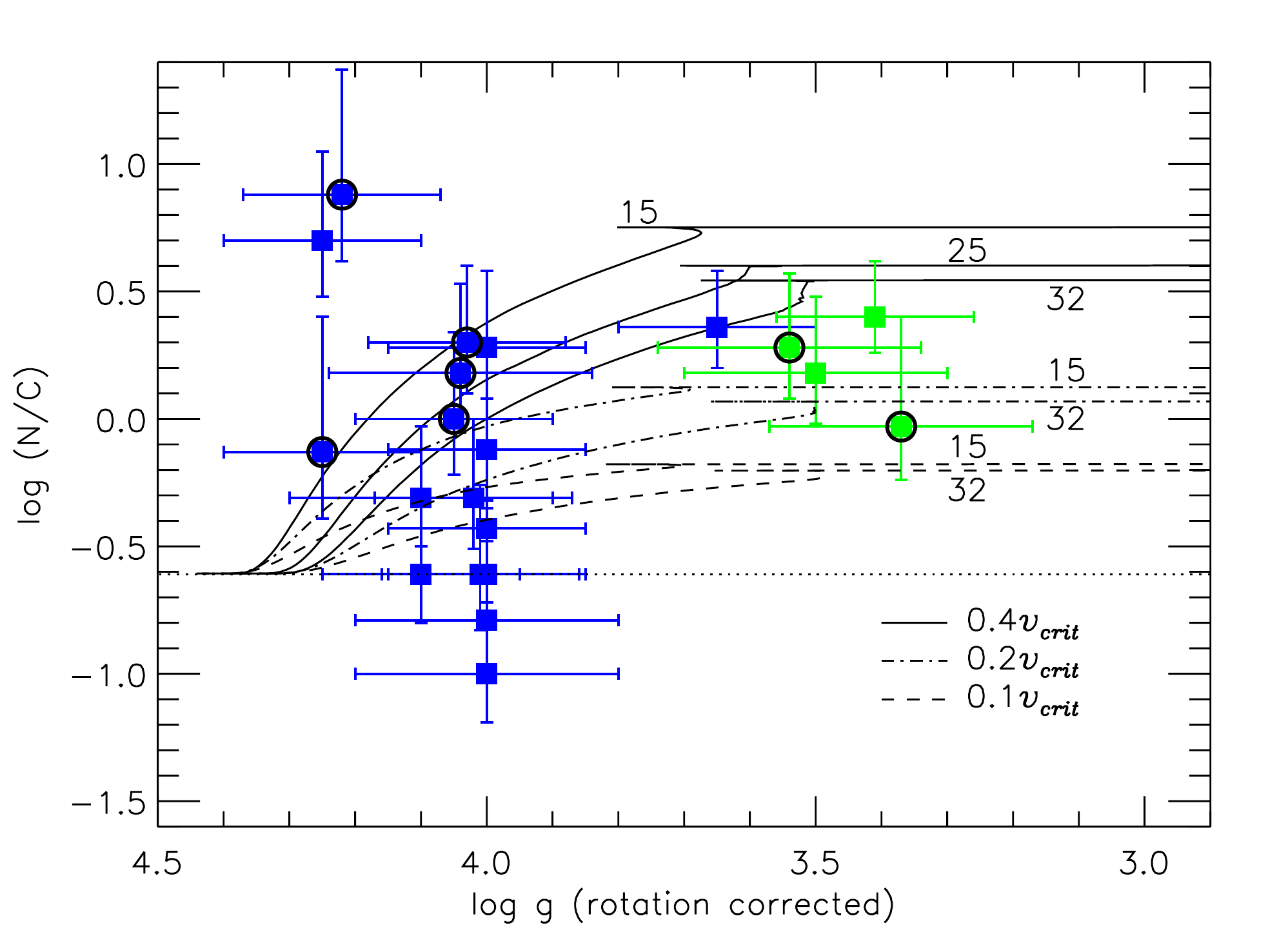}
       \includegraphics[width=9.cm]{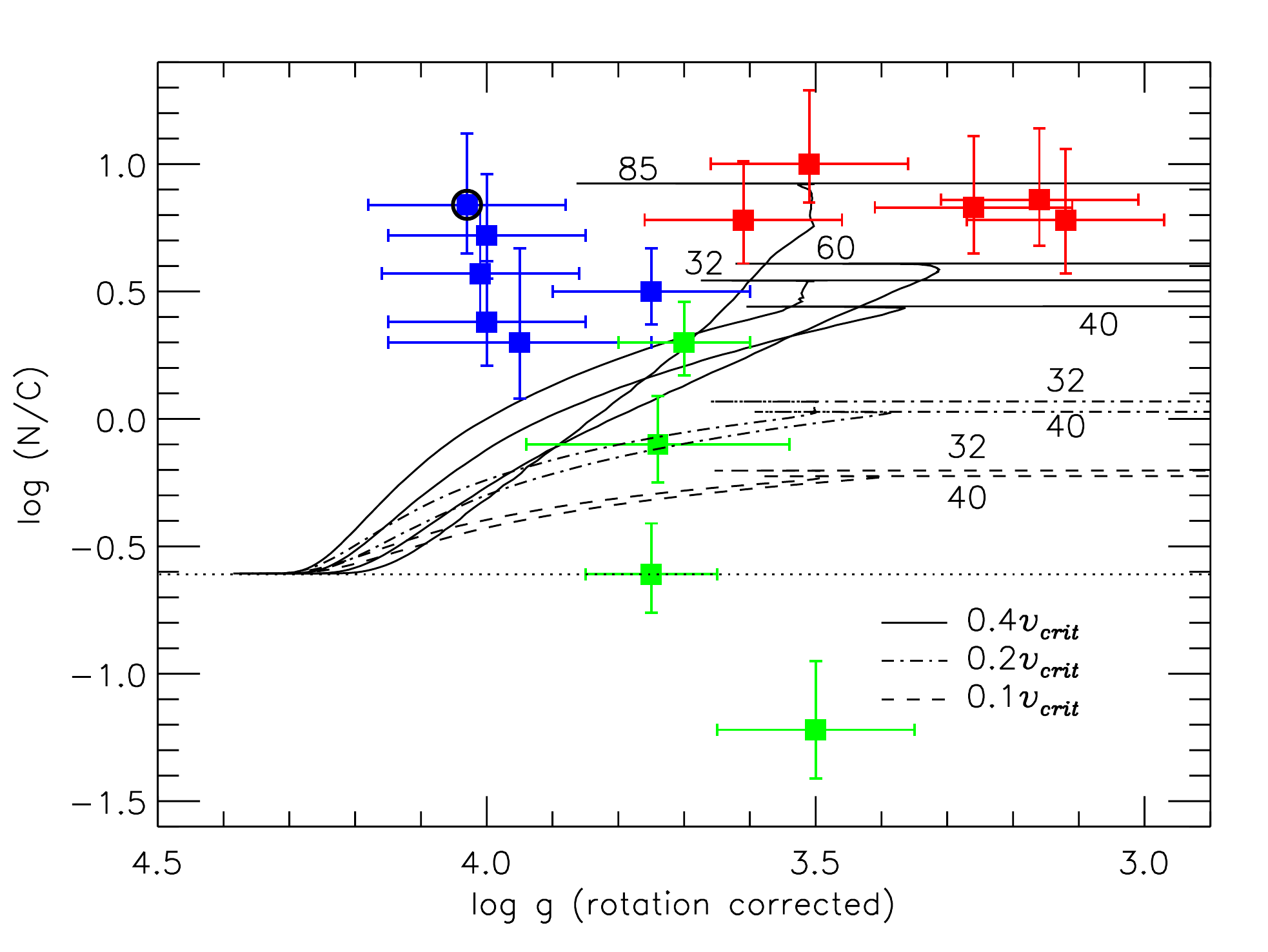}
  \caption{log (N/C) vs. $\log g$ for the sample stars. The evolutionary tracks are labelled by initial mass. 
  Full lines show models with initial rotation $v_{init}$ = 0.4$v_{crit}$, dot-dashed and dashed lines show models with $v_{init}$ = 0.1$v_{crit}$ and 0.2$v_{crit}$, respectively. Left: Stars less massive than the average mass of the sample, i.e. 30 \msun. Stars rotating faster than 200 \kms\ are highlighted by thick black circles. Right: Stars with initial masses above the average for the whole sample. See text for more details}
         \label{fig_abund4}
   \end{figure*}

To better isolate the effect of evolution, we again plot the N/Cs as a function of \logg, but we sort the stars by bins of (initial) masses in the two panels of Fig. \ref{fig_abund4}. We used the average mass calculated for the whole sample, that is, $\approx$ 30 \msun, to separate the sample. Also illustrated are the evolutionary tracks for 15 \msun\ and 40 \msun\ with $v_{init}$ = 0.1$v_{crit}$,  $v_{init}$ = 0.2$v_{crit}$, and  $v_{init}$ = 0.4$v_{crit}$. % (cf. Fig. \ref{fig_abund5}). 
 
The models of \cite{georgy13} that we used for SMC stars predict 
a non-monotonic behaviour of the maximum surface enrichment with mass, unlike the models of \cite{ekstrom12} for Galactic stars \citep[see e.g.][]{martins15}: from 15 to 40 \msun\  , this maximum value decreases before it increases for masses between 40 and 85 \msun.
The reason for this behaviour in the low-metallicity models of \cite{georgy13} is manyfold and discussed in Appendix A.

In the mass range 15 \msun\ to 30 \msun\ (left panel), all but three stars have log(N/C) falling in the part of the diagram covered by the 15\msun\ and 32 \msun\ tracks for various initial rotation rates, demonstrating the ability of models with rotation to account for the surface abundances of a large majority of stars in that mass range. 

Thirteen out of the sixteen dwarfs are reasonably well reproduced by the models with various degrees of rotation. Most of these objects are little evolved in terms of surface abundances. This is expected given their position in the HRD (or KD), where they lie in the first half of the MS.
The N/Cs of the giants in the 15 -- 30 \msun\ mass bin are correctly reproduced by the models. They are also higher than those of the dwarfs in the same mass bin, supporting the prediction from theoretical models that surface chemical processing increases with time. 

The three outliers not accounted for by any model are all dwarfs. 
Two have N/Cs similar to those we find for supergiants, as shown in Fig. \ref{fig_abund3}. 
The location of these stars in the diagram argues for a much earlier and stronger processing and mixing than displayed in the tracks available in the public grid of models\footnote{https://www.unige.ch/sciences/astro/evolution/fr/base-de-donnees/syclist/}.
The third outlier has log(N/C) = -1, which is 0.4 dex below the (solar-like) baseline adopted for the models of \cite{georgy13}. The measured log(N/C) for \object{AV 468} could be reconciled with a baseline for the SMC if values such as \cite{kurt98} were adopted instead, so that at this point we cannot consider this object incompatible with model predictions.  

The evolved status of the dwarf with the lowest surface gravity (\object{MPG 12}, \logg\ = 3.65) is suggested by the luminosity class IV ascribed to this star. It is further supported by a high enhancement in surface nitrogen \citep[see also][and references therein for more comments]{bouret13}. The location of this star in the diagram is closer to the location of the giants of this mass sub-sample, both in terms of \logg\ and N/Cs. The initial mass of \object{ MPG 12} is also very similar (if not identical within the error bars) with those
of these four giants. 

Stars with masses above 30 \msun, that is six dwarfs, four giants and the five supergiants of the original sample, are shown in the right panel of Fig.~\ref{fig_abund4}.
All but one star in this mass bin have M < 60 \msun, the exception being \object{MPG 355} with M = 85 \msun. 
The five supergiants show the highest N/C of the whole sample, in agreement with qualitative expectations for their evolutionary status. The average log(N/C) for these five stars is 0.85, with a small dispersion around this value (0.1). Similarly, dwarfs with \mstar\ $\geq$ 30 \msun\ have an average log(N/C) $\approx$ 0.55, again with a relatively small dispersion, which is far smaller that what we observed for the lower mass bin (left panel).

For the dwarfs and supergiants the N/Cs are typically higher than predicted by models for the three values of $v_{init}$ we considered. 
Interpreting the behaviour of our measurements with respect to model predictions is made difficult by the non-monotonic evolution of N/C with mass, that is, the reversal above 40 \msun. Nevertheless, the systematic increase of the N/C with the initial rotation rate in the theoretical models suggests that the enhanced abundances more in line with the observed abundance might be obtained if a larger $v_{init}/v_{crit}$ were used in the evolutionary models.

The four giants in the right panel of Fig. \ref{fig_abund4} display very different N/Cs, from a significantly processed (\object{AV 95} with log(N/C) = 0.3)
to an essentially unprocessed ratio (\object{AV 47} with log(N/C) = -0.61). 
These three stars have very similar physical parameters and masses ($\approx$ 32 $\pm$ 2 \msun), and the HRD (Fig. \ref{fig_tefflogg}) indicates that they are located in the second half of the MS, before the TAMS, hence they have similar evolutionary status. The conspicuous outlier among giants is \object{AV 69} (see Sect. \ref{sect_abund}). 

In the high-mass bin, above 30~\msun, the models therefore quantitatively account less well for the observed surface abundances. Qualitatively, observations confirm that more evolved objects show on average more surface chemical processing, in agreement with expectations from mixing due to rotation.
For each luminosity class, there is of course a dispersion in each mass bin due to different rotational velocities (cf. next section), but the average trend is relatively clear: in our sample, more massive O-type stars are on average more chemically enriched.

We emphasise that the same conclusion holds qualitatively when initial masses from the KD are used instead of those from the H-R diagram. Fewer than a handful of stars shift from one mass bin to another, and this does not affect the general conclusion we just reached. For a more quantitative test of the ability of theoretical models to reproduce the surface chemistry of SMC O stars, it is obvious that 1) a larger sample and 2) dedicated population synthesis models are necessary.

%%%%%%%%%%%%%%%%%%%%%%%%%%%%%%%%%%%%%%%%%%%%%%%%%%%%%%%%%%%%%%%%%%%%%%%%%%%%%%%%%%%%%%%%%%%%%%%%%
%%%%%%%%%%%%%%%%%%%%%%%%%%%%%%%%%%%%%%%%%%%%%%%%%%%%%%%%%%%%%%%%%%%%%%%%%%%%%%%%%%%%%%%%%%%%%%%%%

\subsection{Surface abundances and rotation}
\label{sect_rotation}

In the specific case of the SMC, and to the best of our knowledge, systematic studies on the effect of rotation on surface abundances have been limited to B-type stars \citep{hunter07, hunter08, hunter09, dufton18, dufton20}.
These studies showed that evolutionary models based on the code of \cite{yoon06} \citep[see also][]{brott11a} failed to reproduce the position of up to 40\% 
of core-hydrogen burning stars in a log (N/H) -- \vsini\ diagram (e.g. a population of slowly rotating nitrogen-rich stars).
Our analysis extends the study of \cite{bouret13} towards more evolved evolutionary phases, and allows a full comparison to the studies above.  

%----------------------------------------
\subsubsection{O versus B stars}
\label{sect_rot_obs}

\begin{figure*}[tbp]
   \centering
   \includegraphics[width=0.49\textwidth]{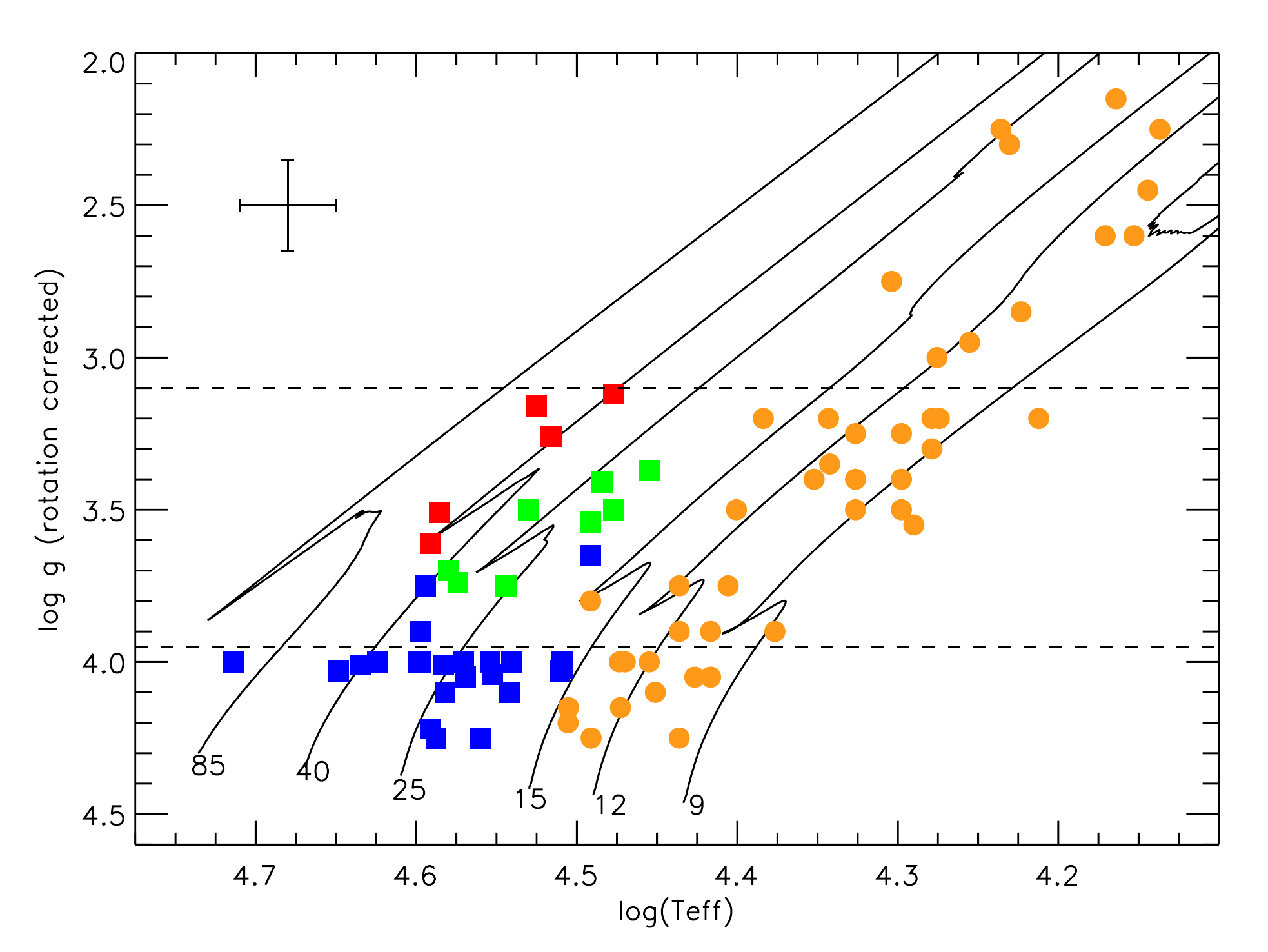}\\
   \includegraphics[width=0.49\textwidth]{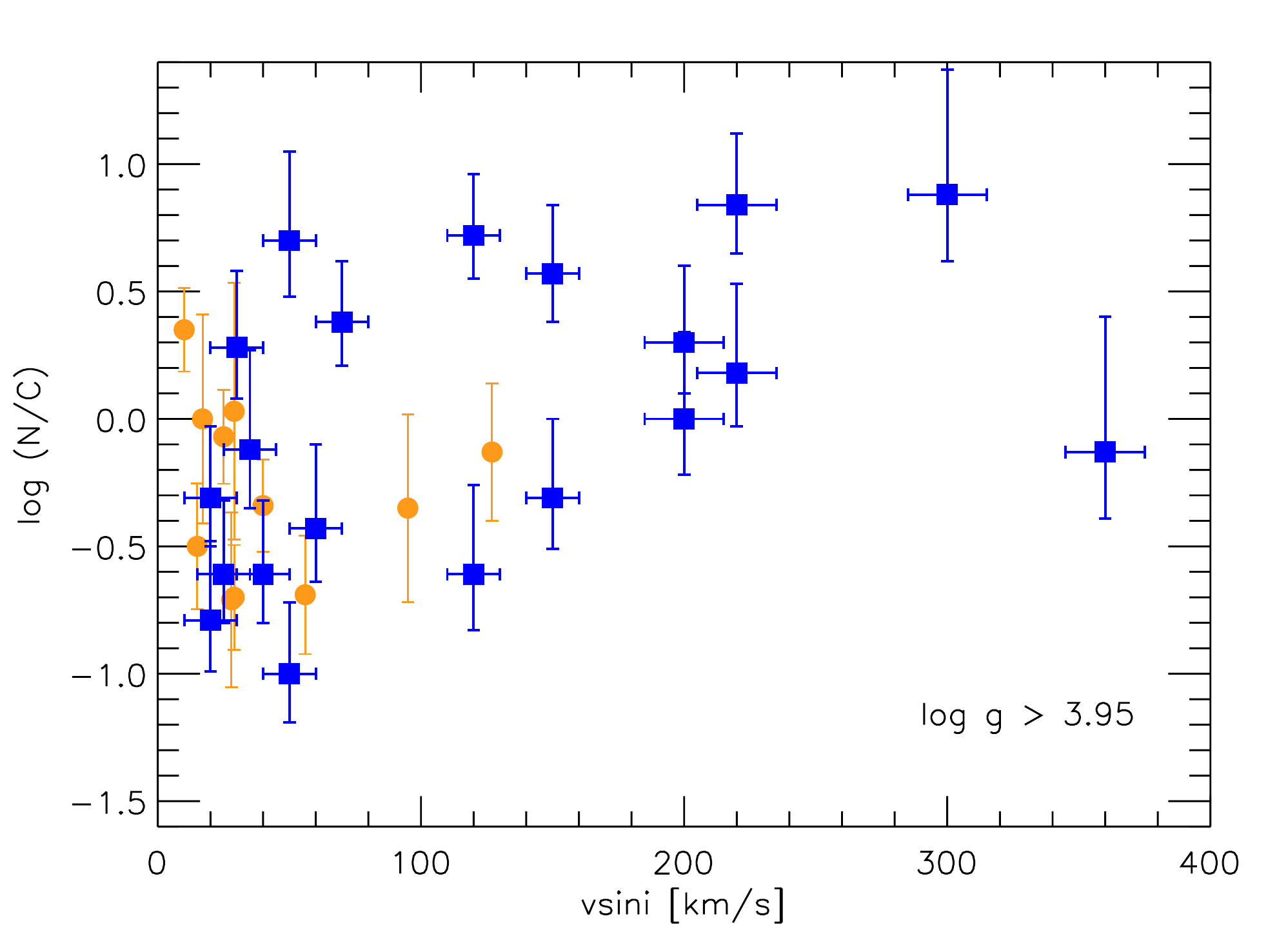}
   \includegraphics[width=0.49\textwidth]{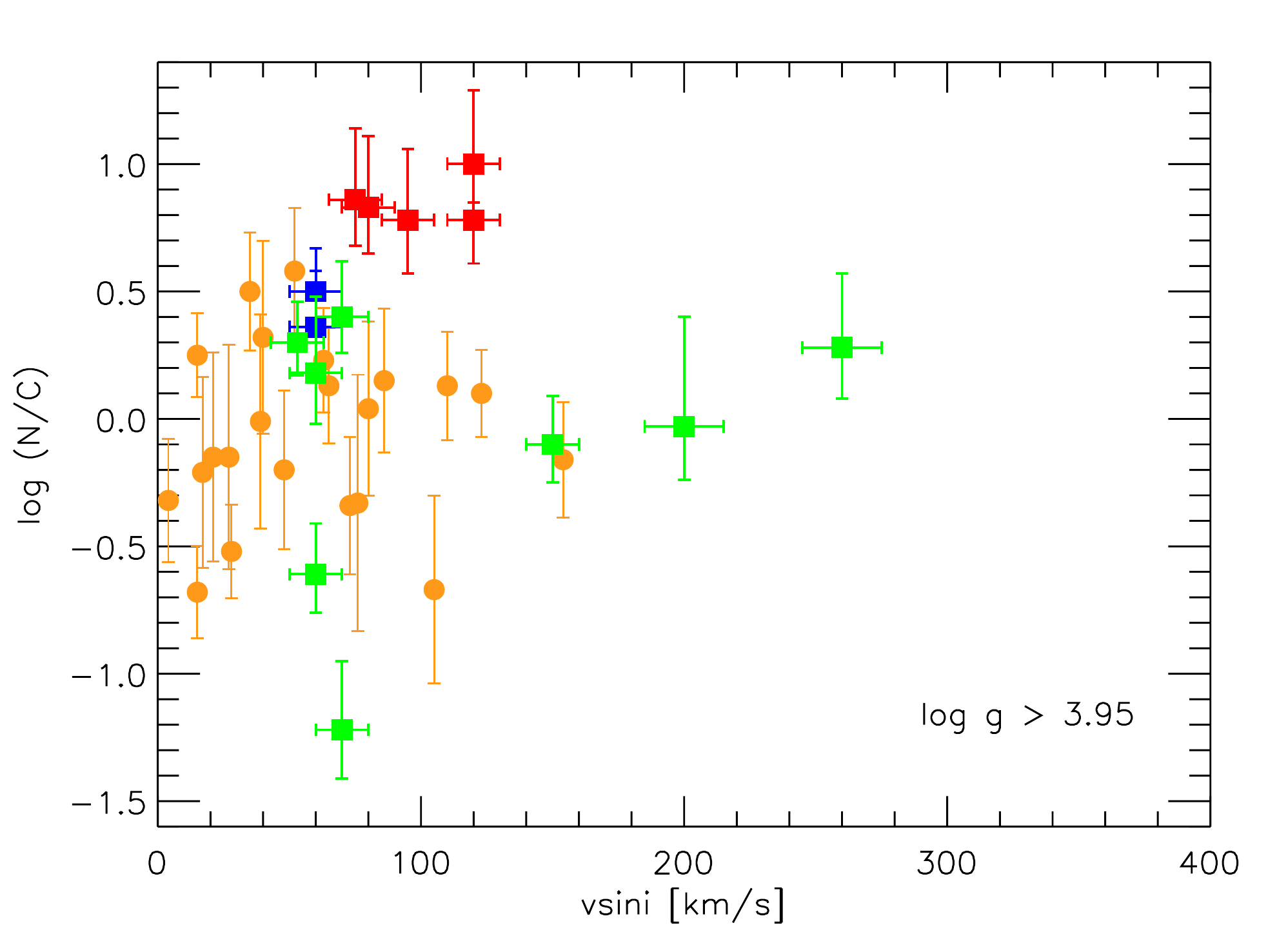}   
  \caption{\textit{Upper panel}: \logg\ -- \teff\ diagram for the SMC O stars (squares) and the SMC B stars (orange circles) of \citet{hunter09}. The evolutionary tracks are from \citet{georgy13}. The dashed horizontal lines mark the \logg\ limits used in the lower panels. \textit{Lower panels}: log (N/C) as a function of  projected rotational velocity (\vsini). The left panel corresponds to stars with \logg\ $>$ 3.95, the right panel to those with 3.1 $<$ \logg\ $<$ 3.95. We have omitted stars from the Hunter et al. sample that have only upper limits on the nitrogen abundance.}
         \label{fig_abFM}
   \end{figure*}

In Fig. \ref{fig_abFM} we show the N/Cs of our sample as a function of the (projected) rotation velocity, compared to the location of the B stars from \cite{hunter09}. With this comparison we determine whether there is a trend of more chemical processing for more massive stars, as predicted by models including rotation \citep{brott11a,ekstrom12} and observed in the Galaxy \citep{martins17}.

We separated the stars according to their surface gravity. In a first group we gathered stars located on the MS, while in the second group we collected stars near the TAMS and the early post–MS phases. In doing so, we tried to reduce the effect of age on surface abundances. In each sub-sample, N/C should thus depend mainly on rotation and initial mass. The bottom panels of Fig.~\ref{fig_abFM} display the log(N/C) -- \vsini\ diagrams for both sub-samples. In the MS sample, O stars with small \vsini\ have N/C surface values that cover a wide range. At high \vsini\ the maximum N/C values seem to be the same as in the low \vsini\ sample, but fast-rotating O stars with low N/C are not observed. This is an indication that high \vsini\ implies on average high N/C. The opposite is not true because there are slow rotators with N/C as high as fast rotators. 

The comparison B stars are all slow rotators and thus no conclusion regarding the effect of \vsini\ on surface abundance can be drawn for this sub-sample. At low \vsini\ B stars show a range of N/C values that is almost comparable to that of O stars. 
In the sub-sample covering the TAMS and early post–MS phases, only a few stars have \vsini\ $>$ 200 \kms. It is thus impossible to discuss any trend of chemical enrichment with rotation. At low \vsini\ (i.e. up to $\sim$150 \kms), the few O stars have a range of N/C values similar to that of the MS O stars. The B stars seem to have N/C in the same range, although only two objects reach the N/C of O supergiants. 
The top panel of Fig.~\ref{fig_abFM} shows in that sub-sample B stars, O giants and O supergiants correspond to groups of stars having increasing masses. The supergiants are significantly more enriched than both O giants and B stars, which share the same range of N/Cs. Surface chemical processing thus appears to increase with mass above $\sim$30~\msun, while it is essentially constant below that limit. This qualitative behaviour is reminiscent of what models predict (see Appendix~\ref{ap_mod}).

We stressed in the previous section that for the most massive stars the maximum \mbox{N/C} observed value seemed to correlate with mass but does not do so for the least massive stars, although this last point is not robustly demonstrated. This may be a first observational  evidence of the theoretical effects just described. 

We acknowledge, however, that this needs to be further investigated with larger samples because the number of stars in the current study remains small (about 10 to 20 objects at most). If true, and given our previous conclusion that among O stars the enrichment is stronger for more massive stars, this would indicate that surface abundances depend on mass only for the most massive stars.

%----------------------------------------
\subsubsection{Comparison to theoretical predictions}
\label{sect_rot_pred}

\begin{figure*}[tbp]
   \centering
   \includegraphics[width=9.cm]{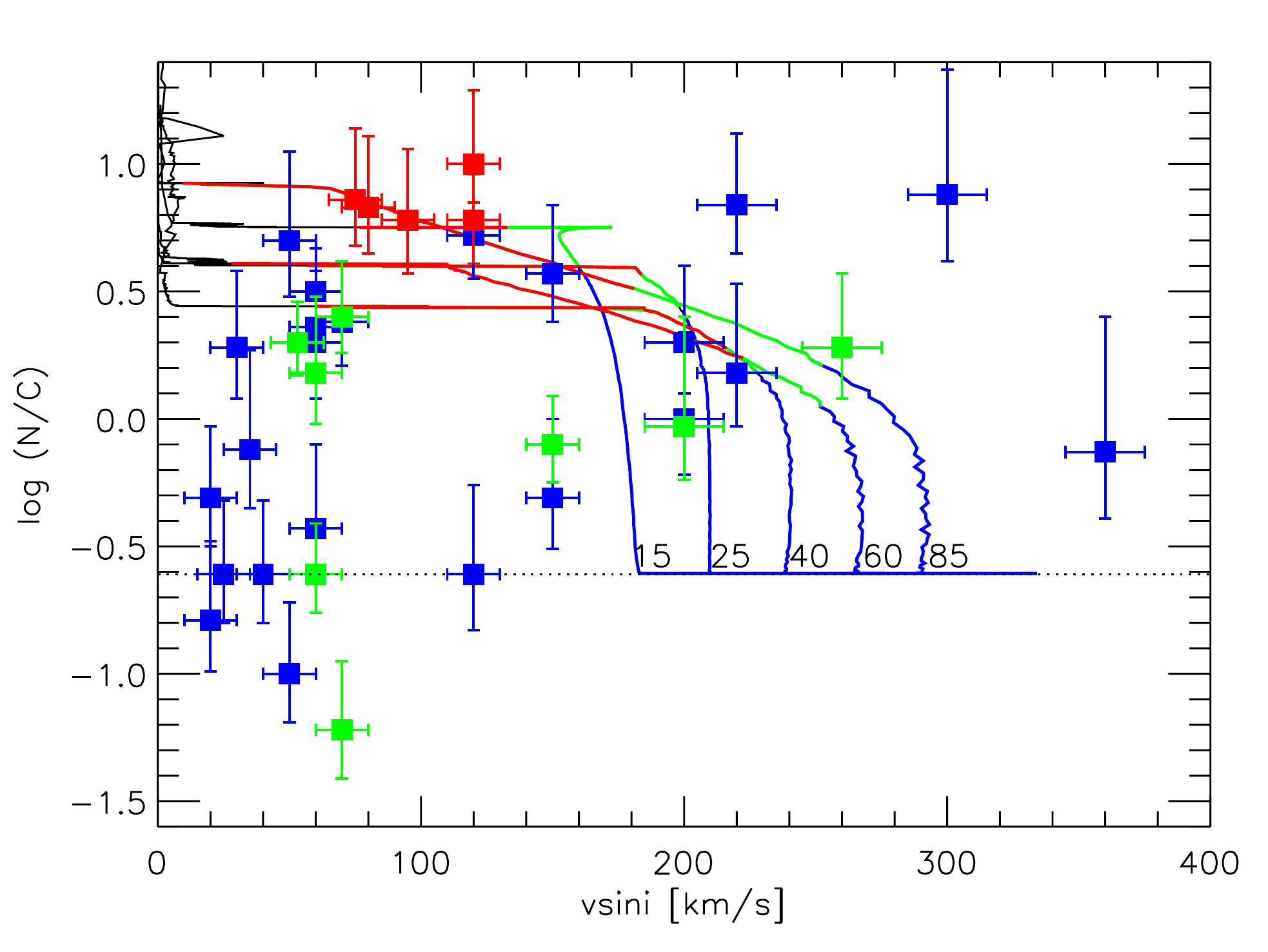}
     \includegraphics[width=9.cm]{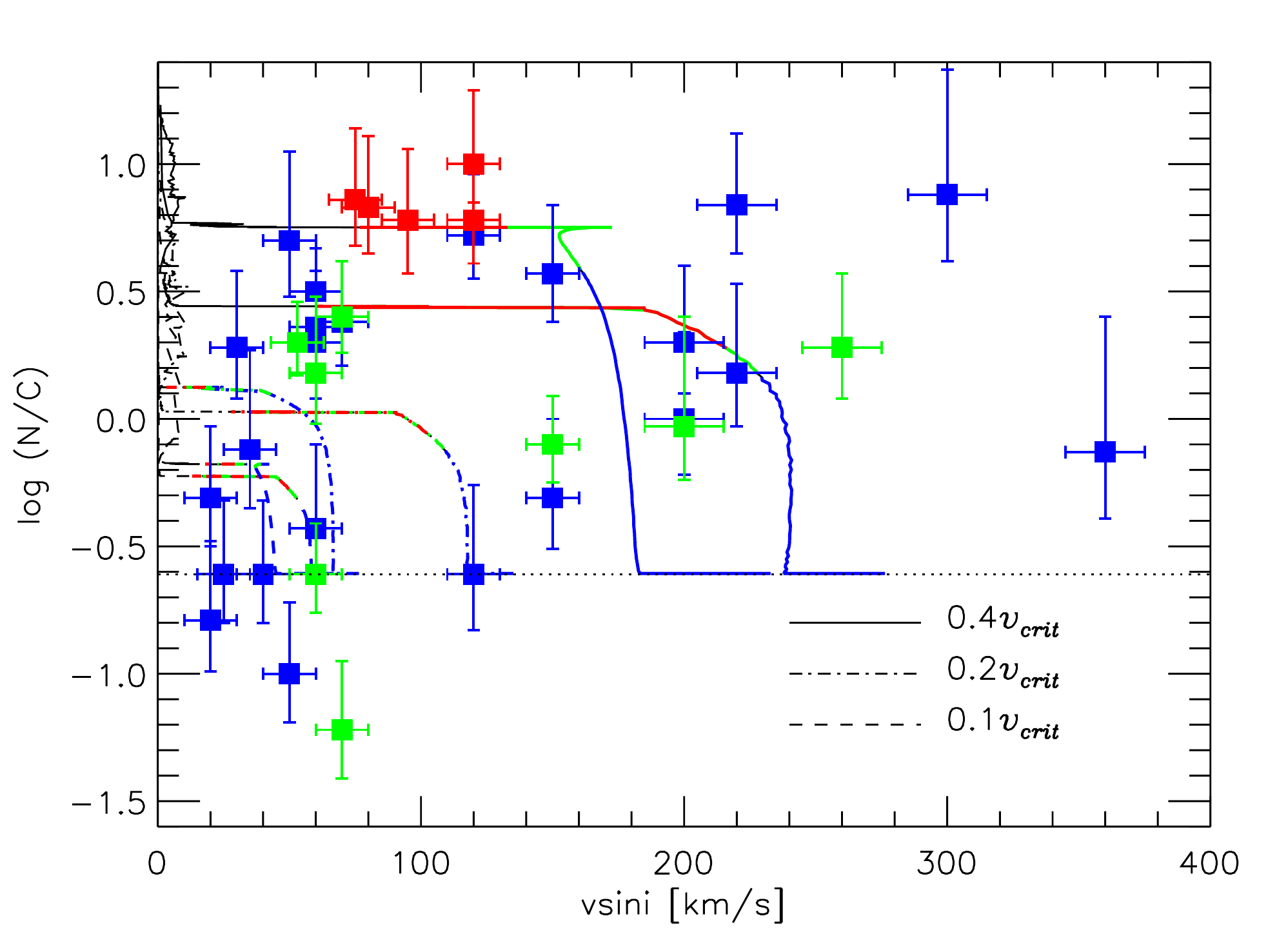}
  \caption{log (N/C) as a function of  projected rotational velocity (\vsini) for the sample stars. Left: The evolutionary tracks are labelled by initial mass (initial rotation $v_{init}$ = 0.4$v_{crit}$). We overplot in blue, green, and red the parts of the tracks that correspond to the minimum and maximum value of \logg\ as measured for the dwarfs, giants, and supergiants, respectively. Right: Models with 
  a range of initial rotation $v_{init}$ = 0.1$v_{crit}$, 0.2$v_{crit}$, and 0.4$v_{crit}$, and initial masses of 15\,\msun\ and 40\,\msun\ are shown.  In both panels, a horizontal scaling corresponding to a factor $\pi$/4 to account for an average projection effect has been applied to the theoretical curves. The horizontal dotted line indicates the baseline N/C for the SMC used in the models.}
         \label{fig_abund5}
   \end{figure*}

In Fig. \ref{fig_abund5} we compare the N/Cs as a function of \vsini\ for the sample as a whole.
We further compare the positions of the stars to the different parts of the tracks
that correspond to their range of surface gravities as deduced from the spectroscopic analysis. 
In the left panel, the models are those we adopted throughout this work, that is, they have an initial rotation $v_{init}$ = 0.4$v_{crit}$ \citep{georgy13}. The non-monotonic behaviour of the maximum surface enrichment with mass predicted by the models of \cite{georgy13}, discussed in Appendix A, is also visible in Fig. \ref{fig_abund5}. 

For most of the dwarfs located in the area covered by the evolutionary tracks, the measured N/C (around 0.5 $\pm$ 0.2 for values of \vsini\ ($\simeq$ 160 \kms) is predicted to be reached at a later evolutionary stage, regardless of the initial value of the mass in the models. 
Similarly, all the supergiants fall close to the tracks at the correct evolutionary stage, although for masses different from those measured from the evolutionary diagrams. 
For both cases, the location of the stars on the tracks qualitatively supports the assumptions of rotational mixing in the models, although quantitatively, the measured \mbox{N/C} appear to require stronger mixing than currently implemented in the models. 

Another conspicuous feature of Fig. \ref{fig_abund5} is the population of dwarfs and giants with \vsini\ $\leq$ 160 \kms, and log (N/C) < 0, which fall in a region  of the log (N/C) -- \vsini\ plot that no tracks with $v_{init}$ = 0.4$v_{crit}$ reach.  
This may indicate that the initial rotation rates selected for the models are not relevant for a fraction of our sample.
To further explore this assumption, the right panel of Fig. \ref{fig_abund5} shows that models with a range of initial rotation ($v_{init}$ = 0.1$v_{crit}$, and $v_{init}$ = 0.2$v_{crit}$) can indeed reproduce the location of the stars of our sample, although they too seem to struggle to do so for the correct 
evolutionary stage or mass for several objects. The same comments likely hold for stars with \vsini\ larger than 250 \kms, but for $v_{init}$ that would be larger than 0.4$v_{crit}$. 

Again, we acknowledge that this discussion is limited by the size of the sample. Nevertheless, the analysis of the properties of our sample stars provides an indication that most objects can be accounted for by models including rotation, although a small fraction clearly escape these predictions. For these objects, faster braking or more efficient mixing may be needed.

%%%%%%%%%%%%%%%%%%%%%%%%%%%%%%%%%%%%%%%%%%%%%%%%%%%%%%%%%%%%%%%%%%%%%%%%%%%%%%%%%%%

\section{Summary}
\label{sect_summary}

We have analysed the UV and optical spectra of 13 giants  (luminosity class III) and supergiants (class I)  O stars in the SMC. Using atmosphere models computed with the code \cmfgen, we derived their physical and evolutionary properties and their surface abundances. 
These properties and abundances
were discussed together with those of the 23 SMC dwarf stars (luminosity class V) presented by \cite{bouret13} and analysed with the same method. The properties of the full sample as a whole were  compared to predictions of stellar evolution models in \cite{georgy13}.  
A discussion of the wind properties of this (full) sample will be presented in a coming paper. 

Evolutionary masses determined from the HRD and the KD usually agree relatively well given the respective uncertainties. Spectroscopic and evolutionary masses do not show a clear systematic difference as a function of mass, as reported for instance by \citet{markova18} for Galactic O stars. 
The high-resolution spectra used in this work to derive surface gravities are echelle spectra. The rectification of these spectra is notoriously difficult and often uncertain in the vicinity of the broad H lines that are the primary diagnostics of \logg. This is even more severe when the signal-to-noise ratios are limited. Because improving \logg\ determinations is crucial for determining stellar masses by spectroscopic methods, a major undertaking in the near future would be to design reliable and accurate spectrum rectification methods.

The surface C, N, and O abundances are consistent with nucleosynthesis from the hydrogen-burning CNO cycle. Comparisons to theoretical predictions reveal the importance of the initial mixture for reproducing the observed trends in the N/C versus N/O diagram.

A PCA analysis of the abundance ratios for the whole (dwarfs and evolved stars) sample supports the theoretical prediction of models including rotation that massive stars at low metallicity are more chemically processed than their Galactic counterparts. 

A trend for stronger chemical evolution for more evolved objects is observed, from dwarfs to giants and supergiants. Similarly, for a given evolutionary stage, there is evidence that above $\sim$30~\msun, more massive O stars are more chemically processed at their surface. However, comparison with literature data indicates that this may not be the case below $\sim$30~\msun when B stars are considered. This pattern is expected from stellar evolution at low metallicity, although the limit above which the mass dependence kicks in is quantitatively different from what we observe. 
The N/C as a function of the projected rotation is overall well accounted for by models when different initial rotation rates are adopted for the models. 
However, for the most massive stars of our sample, a stronger mixing efficiency is required to reproduce the N/Cs. In addition, a handful of slowly rotating O stars with high N/Cs is not reproduced by evolutionary models.
A similar result has been found for slowly rotating B stars \citep{hunter08,grin17,dufton20}. For these objects, faster braking may be required.

Extending our work to much larger samples of O stars is needed if we are to fully understand the factors controlling surface abundances and the interplay between surface abundances and rotation. 
This is the goal of ambitious programmes such as \ullyses\ and \xshootu, which will target more than 200 OB stars in the MCs, and provide full spectral coverage from the UV to the optical and NIR.
The enabled science supported by the \ullyses\ -- \xshootu\ data spans from constraining the properties of massive stars at low metallicity and their evolution to building spectral templates for stellar population synthesis, and using them to study stellar populations at low metallicity.
 %

%%%%%%%%%%%%%%%%%%%%%%%%%%%%%%%%%%%%%%%%%%%%%%%%%%%%%%%%%%%%%%%%%%%%%%%%%%%%%%%%%%%%%%%%%%%%%%%%%%%%%%%%%%%%
\begin{table*}[htbp]
\caption{Surface abundances}
\centering
\begin{tabular}{llccccccc}
\hline
\hline  
Star             & Sp. Type      &      \vsini\ & $Y_{He} $     & $\epsilon$(C)                    & $\epsilon$(N)                      & $\epsilon$(O)                 & log(N/C)                               & log(N/O)                       \\
        \hline
\object{AV 75}   &     O5.5I(f)  &  120.00  & 0.12      &   7.00$^{0.18}_{-0.12}$  &    8.00$^{0.15}_{-0.13}$  &    7.88$^{0.10}_{-0.10}$  &    1.00$^{0.29}_{ 0.15}$  &    0.12$^{0.21}_{ 0.14}$  \\
\object{AV 15}   &     O6.5I(f)  &  120.00  & 0.10      &   7.00$^{0.11}_{-0.11}$  &    7.78$^{0.16}_{-0.16}$  &    7.90$^{0.13}_{-0.13}$  &    0.78$^{0.23}_{ 0.17}$  &   -0.12$^{0.25}_{ 0.17}$  \\
\object{AV 232}  &       O7Iaf+  &   75.00  & 0.20      &   7.50$^{0.10}_{-0.10}$  &    8.36$^{0.20}_{-0.20}$  &    7.80$^{0.14}_{-0.14}$  &    0.86$^{0.28}_{ 0.18}$  &    0.56$^{0.30}_{ 0.20}$  \\
\object{AV 83}   &       O7Iaf+  &   80.00  & 0.20      &   7.58$^{0.10}_{-0.10}$  &    8.41$^{0.20}_{-0.20}$  &    7.80$^{0.11}_{-0.11}$  &    0.83$^{0.28}_{ 0.18}$  &    0.61$^{0.28}_{ 0.19}$  \\
\object{AV 327}  &        O9.7I  &   95.00  & 0.15      &   7.30$^{0.11}_{-0.11}$  &    8.08$^{0.20}_{-0.25}$  &    7.66$^{0.13}_{-0.13}$  &    0.78$^{0.28}_{ 0.21}$  &    0.42$^{0.30}_{ 0.22}$  \\
\object{AV 77}   &        O7III  &  150.00  & 0.15      &   7.18$^{0.12}_{-0.18}$  &    7.08$^{0.11}_{-0.05}$  &    7.88$^{0.08}_{-0.12}$  &   -0.10$^{0.19}_{ 0.15}$  &   -0.80$^{0.15}_{ 0.11}$  \\
\object{AV 95}   &   O7III((f))  &   53.00  & 0.10      &   7.30$^{0.09}_{-0.10}$  &    7.60$^{0.11}_{-0.11}$  &    7.96$^{0.06}_{-0.10}$  &    0.30$^{0.16}_{ 0.13}$  &   -0.36$^{0.14}_{ 0.13}$  \\
\object{AV 69}   & OC7.5III((f)  &   70.00  & 0.10      &   7.56$^{0.15}_{-0.15}$  &    6.34$^{0.17}_{-0.17}$  &    8.23$^{0.12}_{-0.12}$  &   -1.22$^{0.27}_{ 0.19}$  &   -1.89$^{0.25}_{ 0.18}$  \\
\object{AV 47}   &   O8III((f))  &   60.00  & 0.10      &   7.69$^{0.08}_{-0.09}$  &    7.08$^{0.15}_{-0.15}$  &    7.98$^{0.04}_{-0.08}$  &   -0.61$^{0.20}_{ 0.15}$  &   -0.90$^{0.18}_{ 0.15}$  \\
\object{AV 307}  &        O9III  &   60.00  & 0.12      &   7.38$^{0.17}_{-0.17}$  &    7.56$^{0.18}_{-0.18}$  &    > 7.95                 &    0.18$^{0.30}_{ 0.20}$  &   < -0.39  \\
\object{AV 439}  &      O9.5III  &  260.00  & 0.11      &   7.62$^{0.12}_{-0.14}$  &    7.90$^{0.20}_{-0.20}$  &    > 7.90                 &    0.28$^{0.29}_{ 0.20}$  &   <  0.00  \\
\object{AV 170}  &      O9.7III  &   70.00  & 0.12      &   7.30$^{0.10}_{-0.12}$  &    7.70$^{0.16}_{-0.10}$  &    7.96$^{0.06}_{-0.08}$  &    0.40$^{0.22}_{ 0.14}$  &   -0.26$^{0.20}_{ 0.12}$  \\
\object{AV 43}   &      B0.5III  &  200.00  & 0.15      &   7.48$^{0.21}_{-0.17}$  &    7.45$^{0.25}_{-0.20}$  &    > 7.80                 &   -0.03$^{0.43}_{ 0.21}$  &   < -0.35 \\

\object{MPG 355} &    ON2III(f)  &  120.00  & 0.09      &   7.39$^{0.14}_{-0.14}$  &    8.11$^{0.15}_{-0.15}$  &    8.00$^{0.08}_{-0.08}$  &    0.72$^{0.24}_{ 0.17}$  &    0.11$^{0.20}_{ 0.15}$  \\
\object{AV 177}  &     O4V((f))  &  220.00  & 0.09      &   7.17$^{0.16}_{-0.16}$  &    8.01$^{0.17}_{-0.17}$  &    7.90$^{0.10}_{-0.10}$  &    0.84$^{0.28}_{ 0.19}$  &    0.11$^{0.24}_{ 0.17}$  \\
\object{AV 388}  &          O4V  &  150.00  & 0.09      &   7.39$^{0.15}_{-0.15}$  &    7.96$^{0.17}_{-0.17}$  &    8.05$^{0.09}_{-0.09}$  &    0.57$^{0.27}_{ 0.19}$  &   -0.09$^{0.23}_{ 0.16}$  \\
\object{MPG 324} &          O4V  &   70.00  & 0.09      &   7.00$^{0.14}_{-0.14}$  &    7.38$^{0.15}_{-0.15}$  &    8.00$^{0.09}_{-0.09}$  &    0.38$^{0.24}_{ 0.17}$  &   -0.62$^{0.21}_{ 0.15}$  \\
\object{MPG 368} &          O6V  &   60.00  & 0.09      &   7.20$^{0.11}_{-0.11}$  &    7.70$^{0.10}_{-0.10}$  &    8.05$^{0.12}_{-0.12}$  &    0.50$^{0.17}_{ 0.13}$  &   -0.35$^{0.18}_{ 0.14}$  \\
\object{MPG 113} &        OC6Vz  &   35.00  & 0.09      &   7.52$^{0.16}_{-0.16}$  &    7.40$^{0.25}_{-0.25}$  &    8.30$^{0.08}_{-0.08}$  &   -0.12$^{0.39}_{ 0.23}$  &   -0.90$^{0.35}_{ 0.20}$  \\
\object{AV 243}  &          O6V  &   60.00  & 0.09      &   7.20$^{0.15}_{-0.15}$  &    7.50$^{0.24}_{-0.24}$  &    7.90$^{0.10}_{-0.10}$  &    0.30$^{0.37}_{ 0.22}$  &   -0.40$^{0.34}_{ 0.20}$  \\
\object{AV 446}  &        O6.5V  &   30.00  & 0.09      &   7.20$^{0.13}_{-0.13}$  &    7.48$^{0.20}_{-0.20}$  &    7.98$^{0.11}_{-0.11}$  &    0.28$^{0.30}_{ 0.20}$  &   -0.50$^{0.28}_{ 0.19}$  \\
\object{MPG 356} &        O6.5V  &   20.00  & 0.09      &   7.69$^{0.11}_{-0.11}$  &    7.38$^{0.20}_{-0.20}$  &    7.98$^{0.08}_{-0.08}$  &   -0.31$^{0.28}_{ 0.19}$  &   -0.60$^{0.27}_{ 0.18}$  \\
\object{AV 429}  &          O7V  &  120.00  & 0.09      &   7.69$^{0.19}_{-0.19}$  &    7.08$^{0.20}_{-0.20}$  &    7.98$^{0.12}_{-0.12}$  &   -0.61$^{0.35}_{ 0.22}$  &   -0.90$^{0.29}_{ 0.19}$  \\
\object{MPG 523} &         O7Vz  &   50.00  & 0.09      &   6.38$^{0.15}_{-0.15}$  &    7.08$^{0.23}_{-0.23}$  &    7.40$^{0.11}_{-0.11}$  &    0.70$^{0.35}_{ 0.22}$  &   -0.32$^{0.33}_{ 0.20}$  \\
\object{ELS 046} &         O7Vn  &  300.00  & 0.09      &   6.68$^{0.15}_{-0.15}$  &    7.56$^{0.31}_{-0.31}$  &    7.98$^{0.12}_{-0.12}$  &    0.88$^{0.49}_{ 0.26}$  &   -0.42$^{0.47}_{ 0.24}$  \\
\object{ELS 031} &         O8Vz  &   25.00  & 0.09      &   7.69$^{0.14}_{-0.14}$  &    7.08$^{0.19}_{-0.19}$  &    7.98$^{0.12}_{-0.12}$  &   -0.61$^{0.29}_{ 0.19}$  &   -0.90$^{0.28}_{ 0.19}$  \\
\object{AV 461}  &          O8V  &  200.00  & 0.09      &   7.38$^{0.17}_{-0.17}$  &    7.38$^{0.21}_{-0.21}$  &    7.98$^{0.12}_{-0.12}$  &    0.00$^{0.34}_{ 0.22}$  &   -0.60$^{0.30}_{ 0.20}$  \\
\object{MPG 299} &         O8Vn  &  360.00  & 0.13      &   7.69$^{0.15}_{-0.15}$  &    7.56$^{0.33}_{-0.33}$  &    8.10$^{0.14}_{-0.14}$  &   -0.13$^{0.53}_{ 0.26}$  &   -0.54$^{0.52}_{ 0.26}$  \\
\object{MPG 487} &          O8V  &   20.00  & 0.15      &   7.39$^{0.16}_{-0.16}$  &    6.60$^{0.19}_{-0.19}$  &    7.98$^{0.10}_{-0.10}$  &   -0.79$^{0.31}_{ 0.20}$  &   -1.38$^{0.26}_{ 0.18}$  \\
\object{AV 267}  &          O8V  &  220.00  & 0.09      &   7.30$^{0.14}_{-0.14}$  &    7.48$^{0.23}_{-0.23}$  &    7.98$^{0.13}_{-0.13}$  &    0.18$^{0.35}_{ 0.21}$  &   -0.50$^{0.34}_{ 0.21}$  \\
\object{AV 468}  &        O8.5V  &   50.00  & 0.09      &   7.30$^{0.13}_{-0.13}$  &    6.30$^{0.19}_{-0.19}$  &    7.60$^{0.10}_{-0.10}$  &   -1.00$^{0.28}_{ 0.19}$  &   -1.30$^{0.26}_{ 0.18}$  \\
\object{AV 148}  &        O8.5V  &   60.00  & 0.09      &   7.69$^{0.12}_{-0.12}$  &    7.26$^{0.23}_{-0.23}$  &    7.98$^{0.11}_{-0.11}$  &   -0.43$^{0.33}_{ 0.21}$  &   -0.72$^{0.33}_{ 0.20}$  \\
\object{MPG 682} &          O9V  &   40.00  & 0.09      &   7.39$^{0.15}_{-0.15}$  &    6.78$^{0.18}_{-0.18}$  &    7.98$^{0.10}_{-0.10}$  &   -0.61$^{0.29}_{ 0.19}$  &   -1.20$^{0.25}_{ 0.17}$  \\
\object{AV 326}  &          O9V  &  200.00  & 0.09      &   7.08$^{0.14}_{-0.14}$  &    7.38$^{0.20}_{-0.20}$  &    7.98$^{0.08}_{-0.08}$  &    0.30$^{0.30}_{ 0.20}$  &   -0.60$^{0.27}_{ 0.18}$  \\
\object{AV 189}  &          O9V  &  150.00  & 0.09      &   7.69$^{0.15}_{-0.15}$  &    7.38$^{0.20}_{-0.20}$  &    7.98$^{0.15}_{-0.15}$  &   -0.31$^{0.31}_{ 0.20}$  &   -0.60$^{0.31}_{ 0.20}$  \\
\object{MPG 12}  &         B0IV  &   60.00  & 0.09      &   7.34$^{0.12}_{-0.12}$  &    7.70$^{0.14}_{-0.14}$  &    8.05$^{0.10}_{-0.10}$  &    0.36$^{0.22}_{ 0.16}$  &   -0.35$^{0.20}_{ 0.15}$  \\

\\
Scaled Solar    &                       &               & 0.09          &                7.69                    & 7.08                          & 7.96                          & -0.61                           &-0.88\\
  \hline
\end{tabular}
 \label{tab_abund_full}
   \begin{list}{}{}
\item[Note :]  For \mbox{CNO} abundances, $\epsilon_{\rm X}  =$ 12 +  log (X /H ). For comparison, we give the abundances for He, C, N, and O that we consider as the baseline in our the analysis \citep[scaled solar,][]{asplund09}.
\end{list}

 \label{tab3}
 \end{table*}

\begin{acknowledgements}
We are grateful to P. Crowther, C. Evans and P. Massey who provided several optical spectra most useful to this work. 
CG has received funding from the European Research Council (ERC) under the European Union’s Horizon 2020 research and innovation programme (Grant Agreement No. 833925).
DJH was provided funding by STScI grants {\sl HST\/}-GO-14683.002-A and {\sl HST\/}-AR-14568.001-A. This research has made use of the SIMBAD database, operated at CDS, Strasbourg, France.  STScI is operated by the Association of Universities for Research in Astronomy, Inc., under NASA contract NAS5-26555. 

\end{acknowledgements}
 
\bibliographystyle{aa}
\bibliography{article2}

\begin{appendix}
\section{Evolution models at SMC metallicity}
\label{ap_mod} 
\begin{figure*}[tbp]
  \centering
  \includegraphics[width=9.cm]{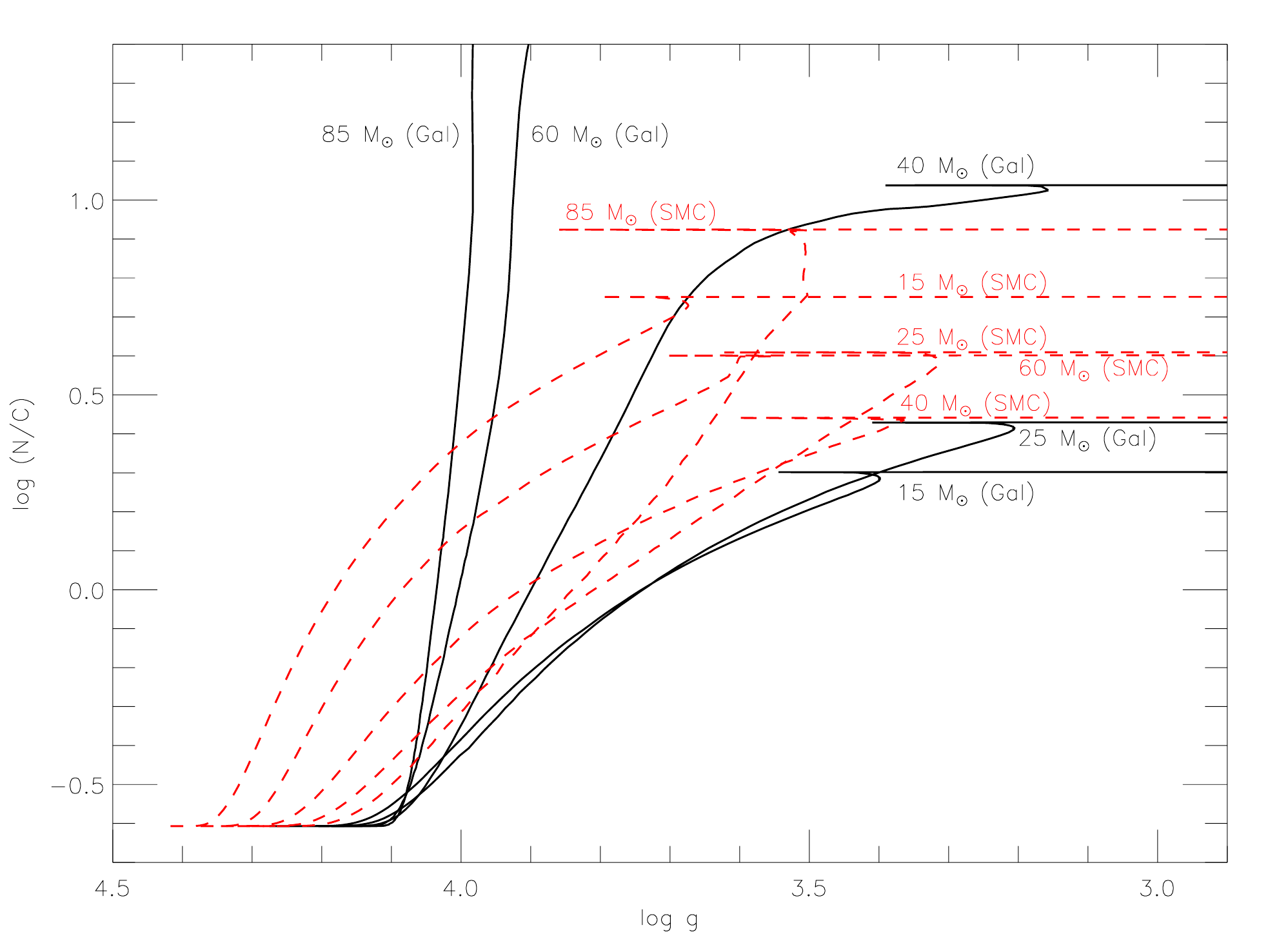}
    \includegraphics[width=9.cm]{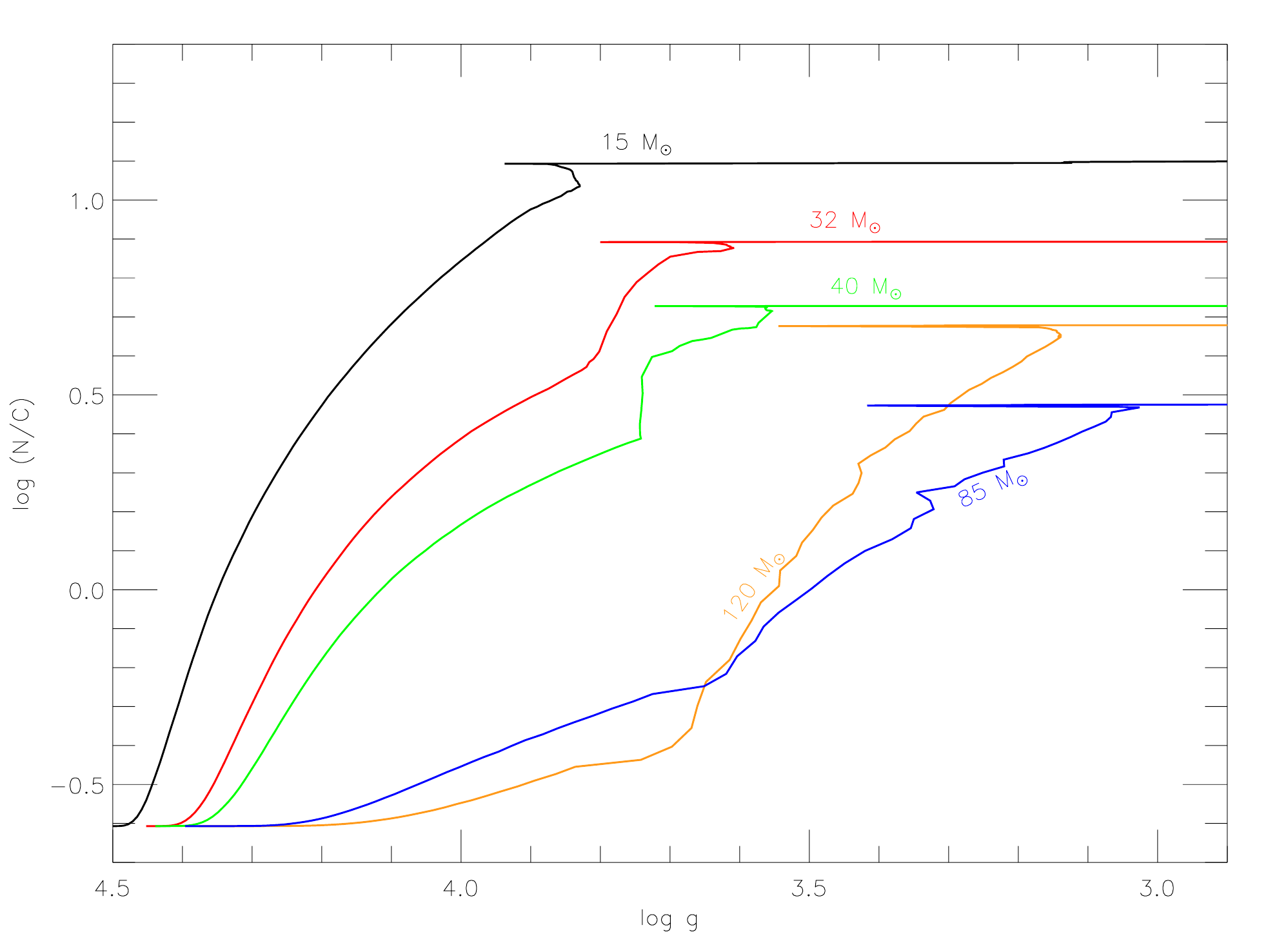}  
  \caption{Left: log(N/C) as a function of log$g$ predicted by models for several masses and two metallicities: Galactic (black) vs. SMC (red) metallicity. Right: log(N/C) vs. \logg\ for various masses and Z = 0.0004. See text for comment.}
         \label{fig_abund_Ap1}
  \end{figure*}

Evolution models for Galactic stars \citep[e.g.][]{brott11a, ekstrom12} predict that the N/C should increase as stellar mass increases. This has been interpreted as a consequence of stronger effects of rotation in models for more massive stars, and it has been found to be consistent with observations \citep[e.g.][]{martins15, martins17, markova18}. On the other hand, the left panel of Fig. \ref{fig_abund_Ap1} shows that the predicted surface abundance ratios are not a monotonic function of the stellar mass for models at SMC metallicity: N/C decreases for higher masses, then increases again above a certain threshold (40 \msun). There are several reasons for this behaviour in the low-metallicity models \citep{georgy13}.

The evolution of surface abundances depends on the mixing efficiency (hence on rotation), and  also on the amount of time each star has to undergo this mixing and hence bring the processed material to the surface. Mixing efficiency increases with mass \cite[see][and references therein]{ekstrom12}, but this might be offset because the MS lifetime decreases with increasing mass. Two other factors also play an important role in mixing efficiency: at low metallicity, stars are more compact because of lower opacities, and thus they have a shorter diffusion timescale. Consequently, their surfaces are also more efficiently enriched at a given time of their evolution. Secondly, mass loss has a decisive effect on the surface abundances. For O-type stars at solar metallicity, mass loss is strong enough for deeper layers, showing stronger effects of chemical processing, to be visible. The net product of the combination of these effects is that a monotonic dependence of the surface chemical enrichment on mass is predicted at solar metallicity. However, the mass-loss dependence on metallicity is such that mass-loss rates in lower metallicity galaxies may not be high enough to compensate for the shorter evolution lifetime above a mass threshold, leading to a non-monotonic evolution of the surface chemical enrichment with mass. For the SMC, this occurs at roughly 40 \msun\ (see the left panel of Fig \ref{fig_abund_Ap1}). 
In the right panel of Fig. \ref{fig_abund_Ap1} we plot log (N/C) as a function of \logg\ for a metallicity Z = 0.0004, that is, 35 times lower than ${\rm Z}_{\rm SMC}$. The chemical processing observed at the stellar surface decreases for lower masses, as expected, but this trend starts to reverse at higher mass than for the SMC, that is, at about 85 \msun, which confirms this interpretation.

\begin{figure}[tbp]
  \centering
 \includegraphics[width=9.2cm]{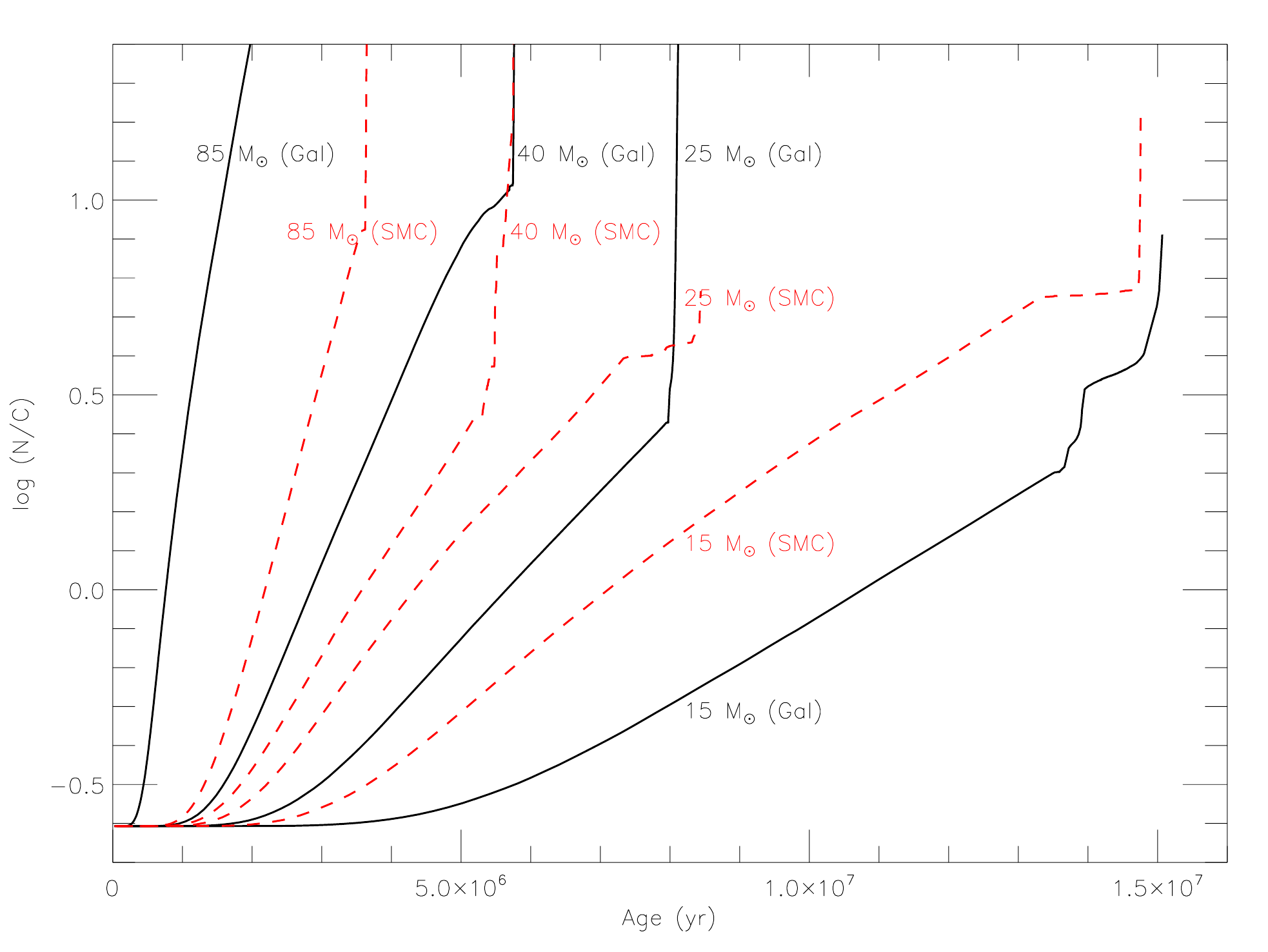}
   \caption{log(N/C) as a function of time predicted by models for several masses and two metallicities: Galactic (black) vs. SMC (red) metallicity}
         \label{fig_abund_Ap2}
  \end{figure}

Another way to understand the model outputs is also to look at the predicted N/C as a function of time. Figure \ref{fig_abund_Ap2} shows that (for a given metallicity), the N/C monotonically increases with mass when plotted against time (age). This plot explicitly includes the intricate effects of mass, rotation (and induced mixing), and mass loss. Figure \ref{fig_abund_Ap2} and the left panel of Fig. \ref{fig_abund_Ap1} show that up to approximately 25 \msun, the chemical enrichment  is stronger for the SMC metallicity than for the Galaxy, which is due to more efficient mixing in the more compact stars at lower Z. This effect has previously been commented on by \cite{georgy13}.

\section{Best fits}
In this appendix we present our best-fit models to the UV and optical spectra.
Absolute fluxes in UV spectra are shown after correction for reddening (see Sect. \ref{sect_uv_opt_spec} and Table \ref{tabtwo}) and expressed in ergs\,.\,\AA$^{-1}$.\,s$^{-1}$.\,cm$^{-2}$. The fits to \object{AV 83} and \object{AV 69} can be found in \cite{hillier03} and are not reproduced here.

\begin{figure*}[tbp]
\includegraphics[scale=0.51, angle=0]{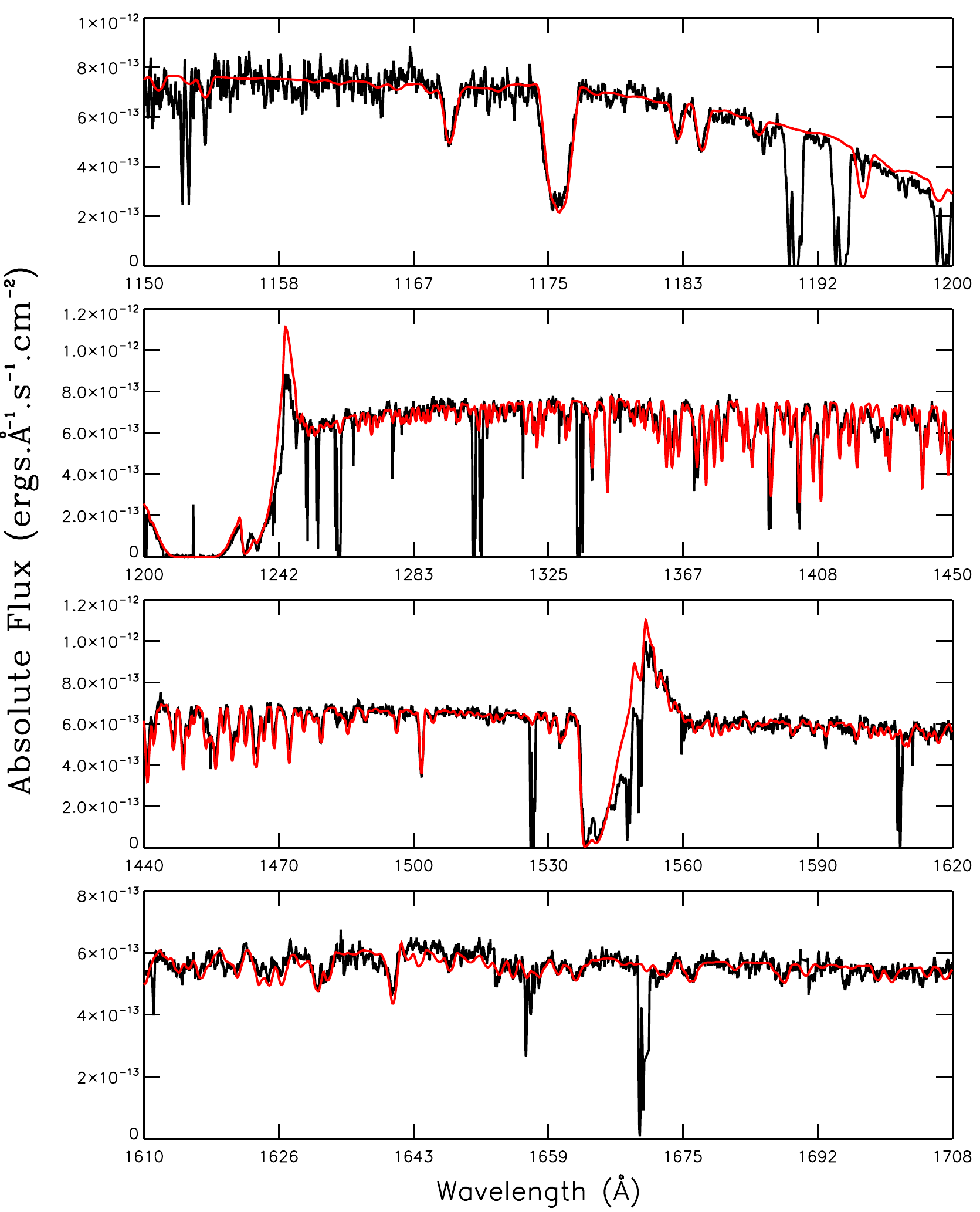}
\includegraphics[scale=0.51, angle=0]{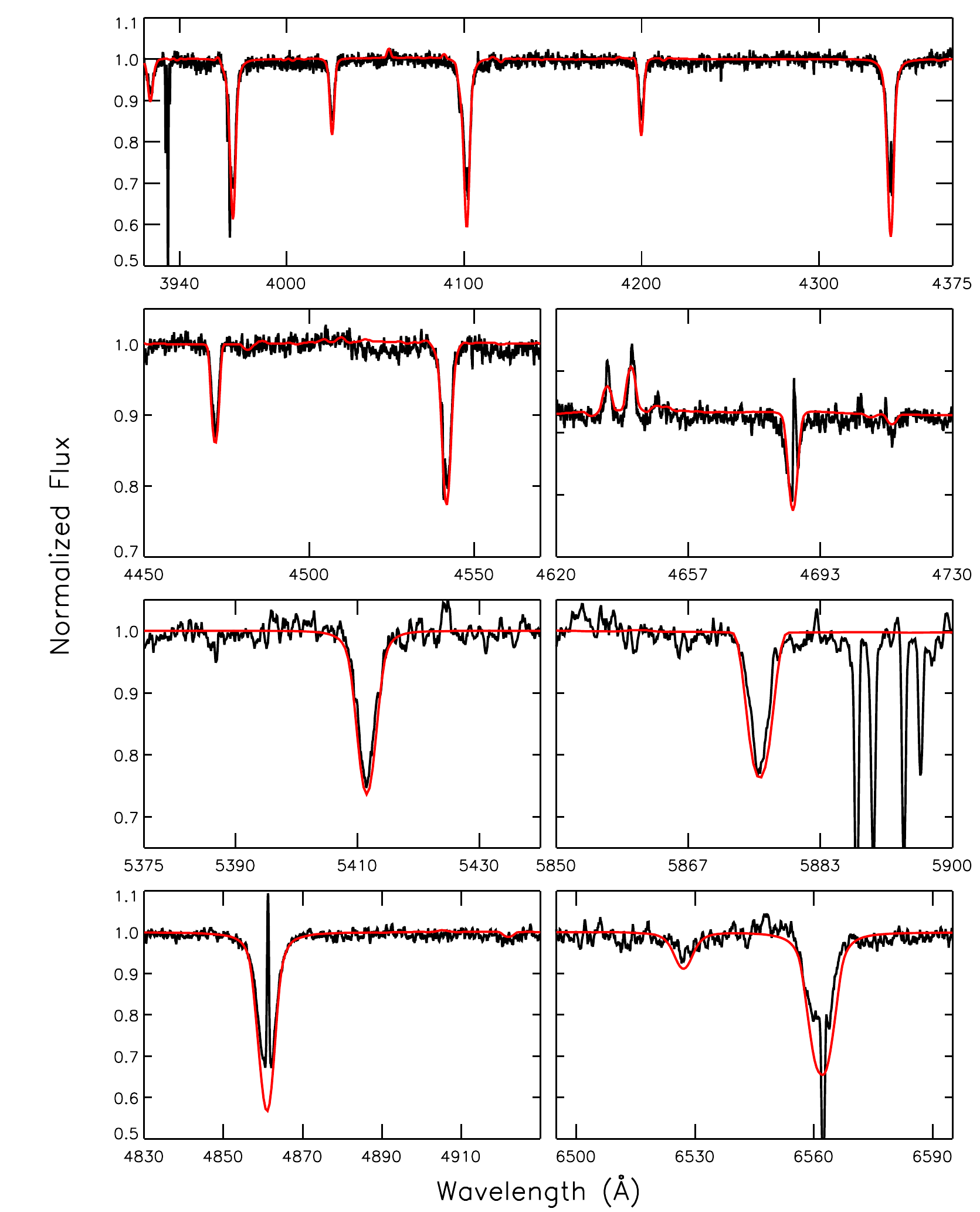}
      \caption[12cm]{Left: Best-fit model for \object{AV 75} (red line) compared to the \stis\ spectrum (black line). Right: Best-fit model for \object{AV 75} (red line) compared to the
      ESO/CASPEC spectrum (black line).}
         \label{Fig_AV75}
   \end{figure*}
   
\begin{figure*}[tbp]
\includegraphics[scale=0.51, angle=0]{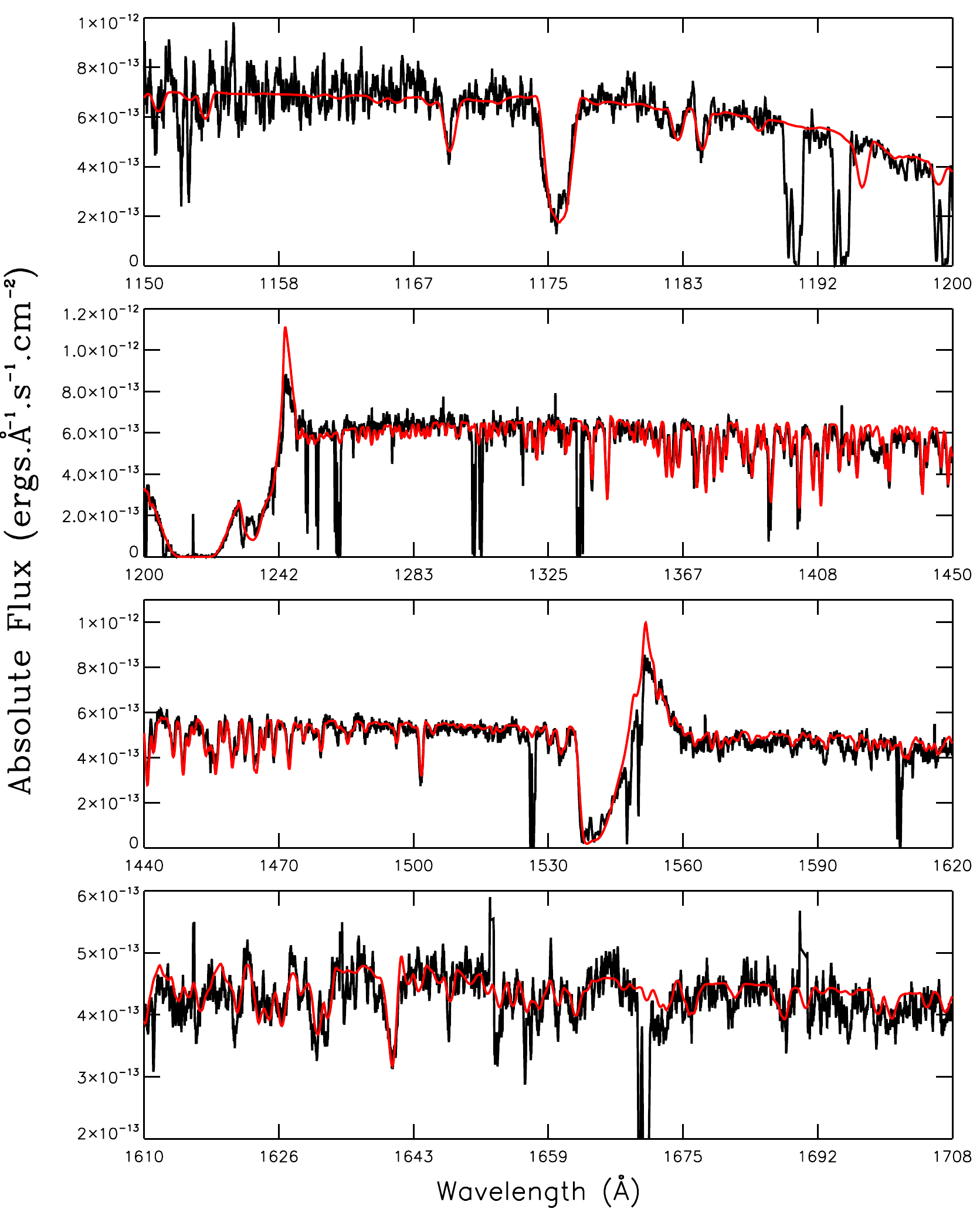}
\includegraphics[scale=0.51, angle=0]{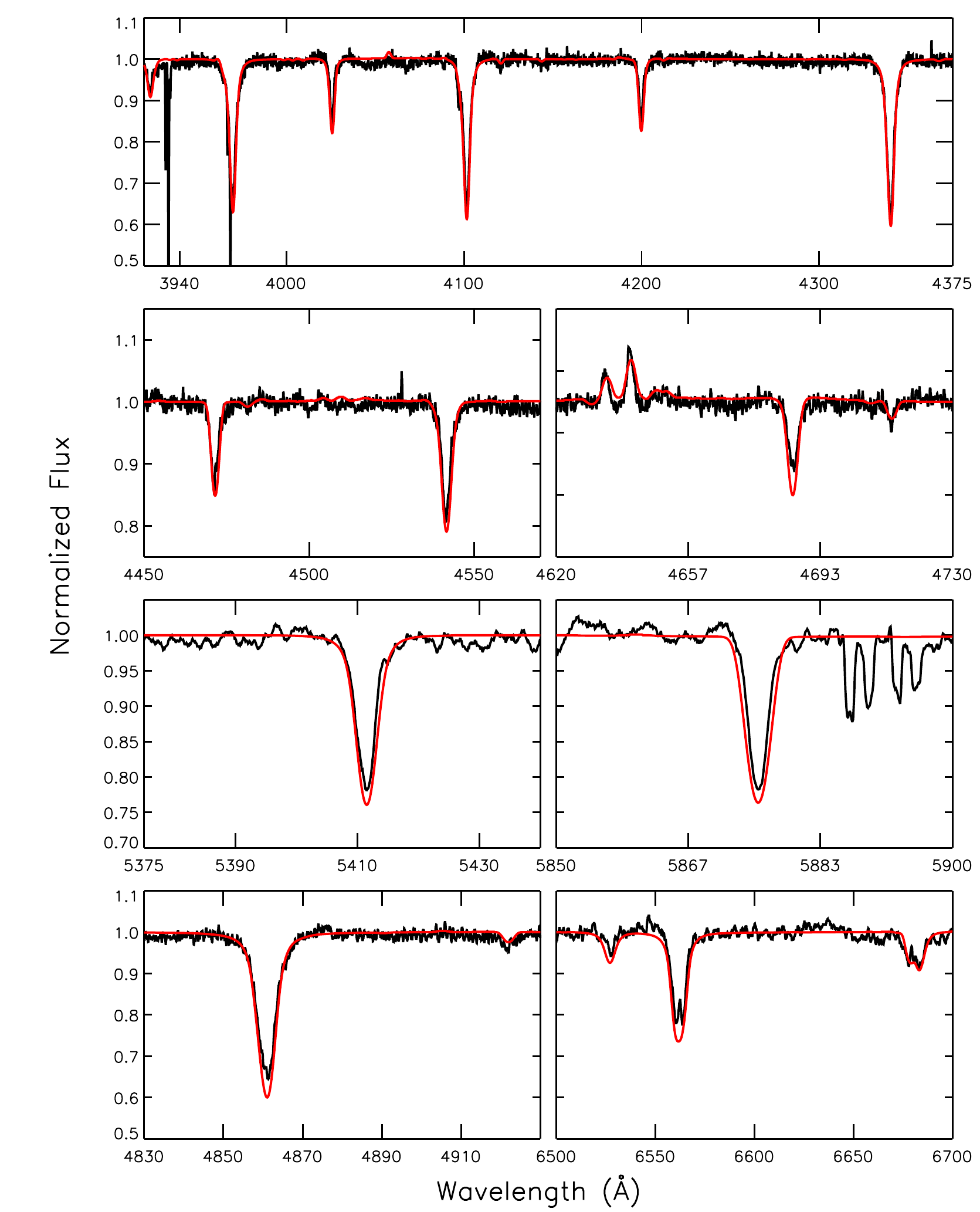}
      \caption[12cm]{Left: Best-fit model for \object{AV 15} (red line) compared to the \stis\ spectrum (black line). Right: Best-fit model for \object{AV 15} (red line) compared to the
      ESO/CASPEC spectrum (black line).}
         \label{Fig_AV15}
   \end{figure*}
   
\begin{figure*}[tbp]
\includegraphics[scale=0.51, angle=0]{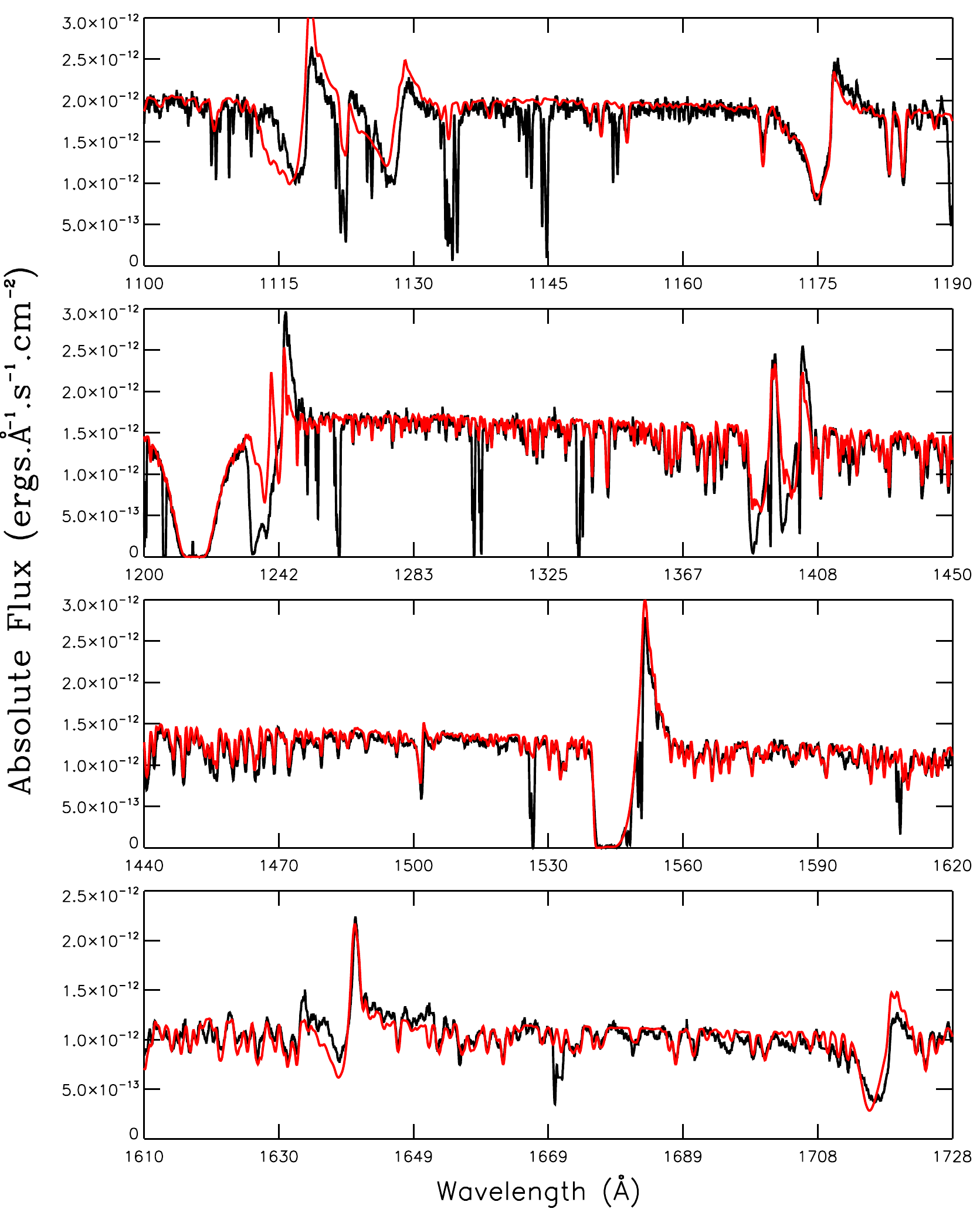}
\includegraphics[scale=0.51, angle=0]{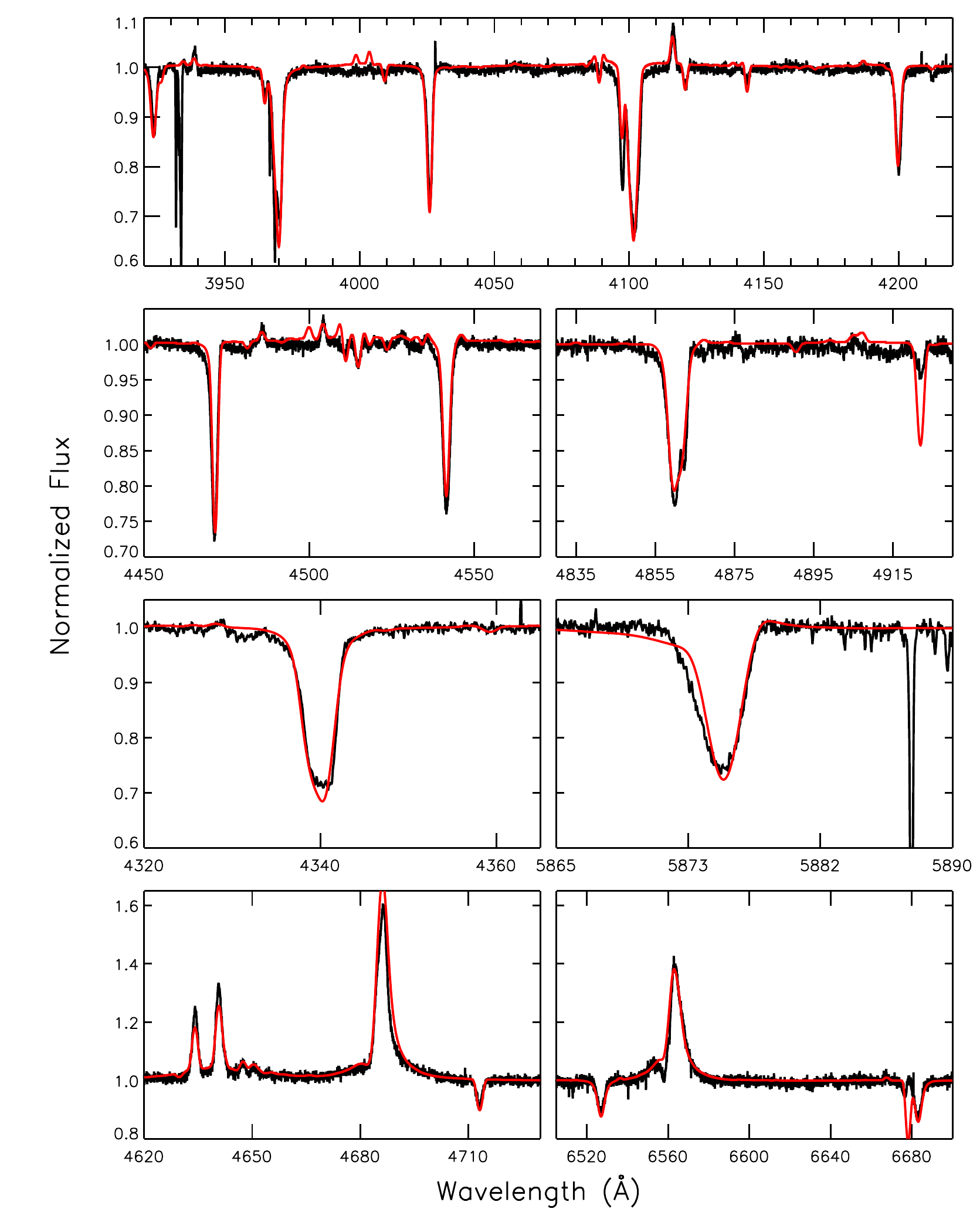}
      \caption[12cm]{Left: Best-fit model for \object{AV 232} (red line) compared to \stis\ spectrum (black line). Right: Best-fit model for \object{AV 232} (red line) compared to the 
      ESO/UVES spectrum (black line).}
         \label{Fig_AV232}
   \end{figure*}

 \begin{figure*}[tbp]
\includegraphics[scale=0.51, angle=0]{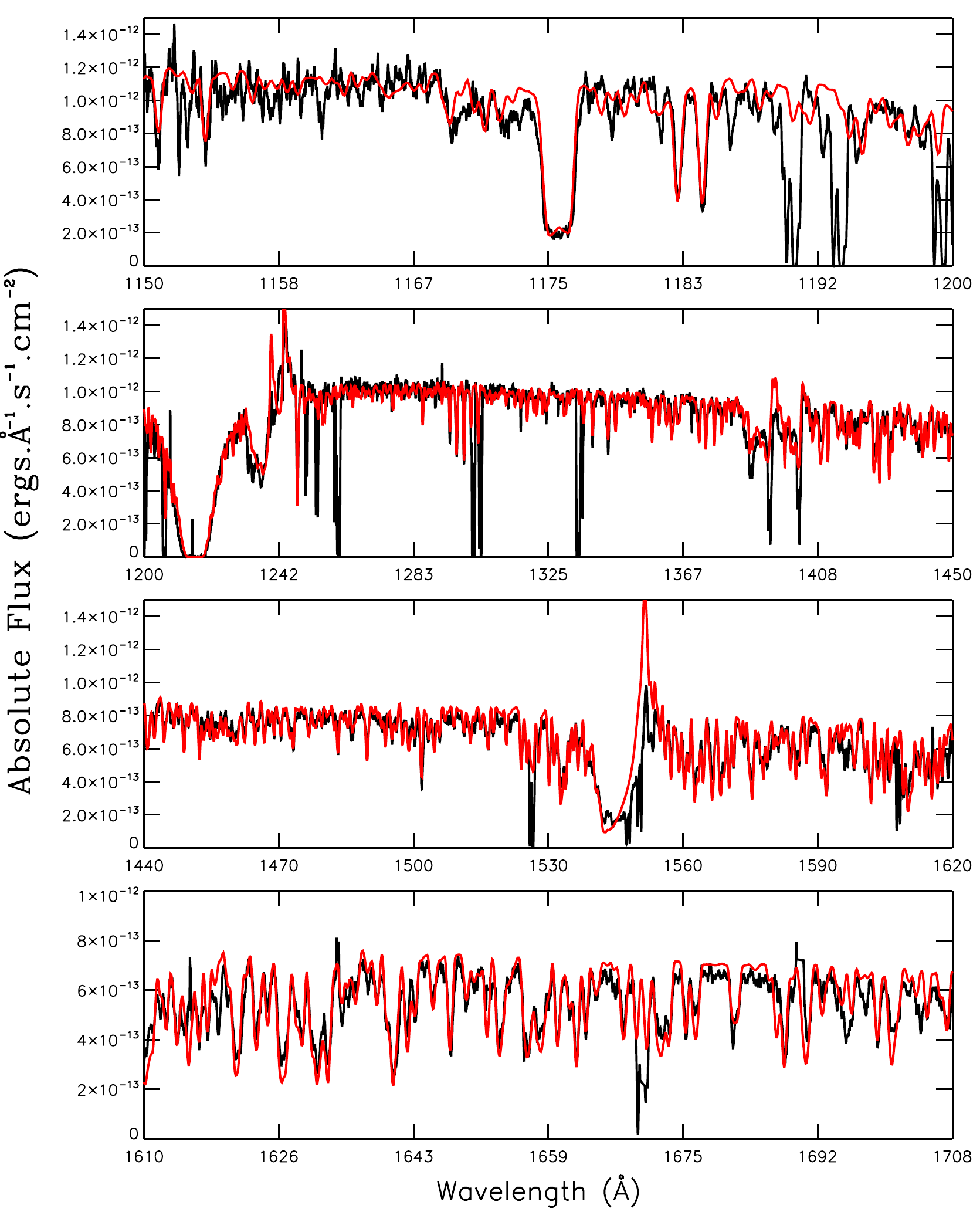}
\includegraphics[scale=0.51, angle=0]{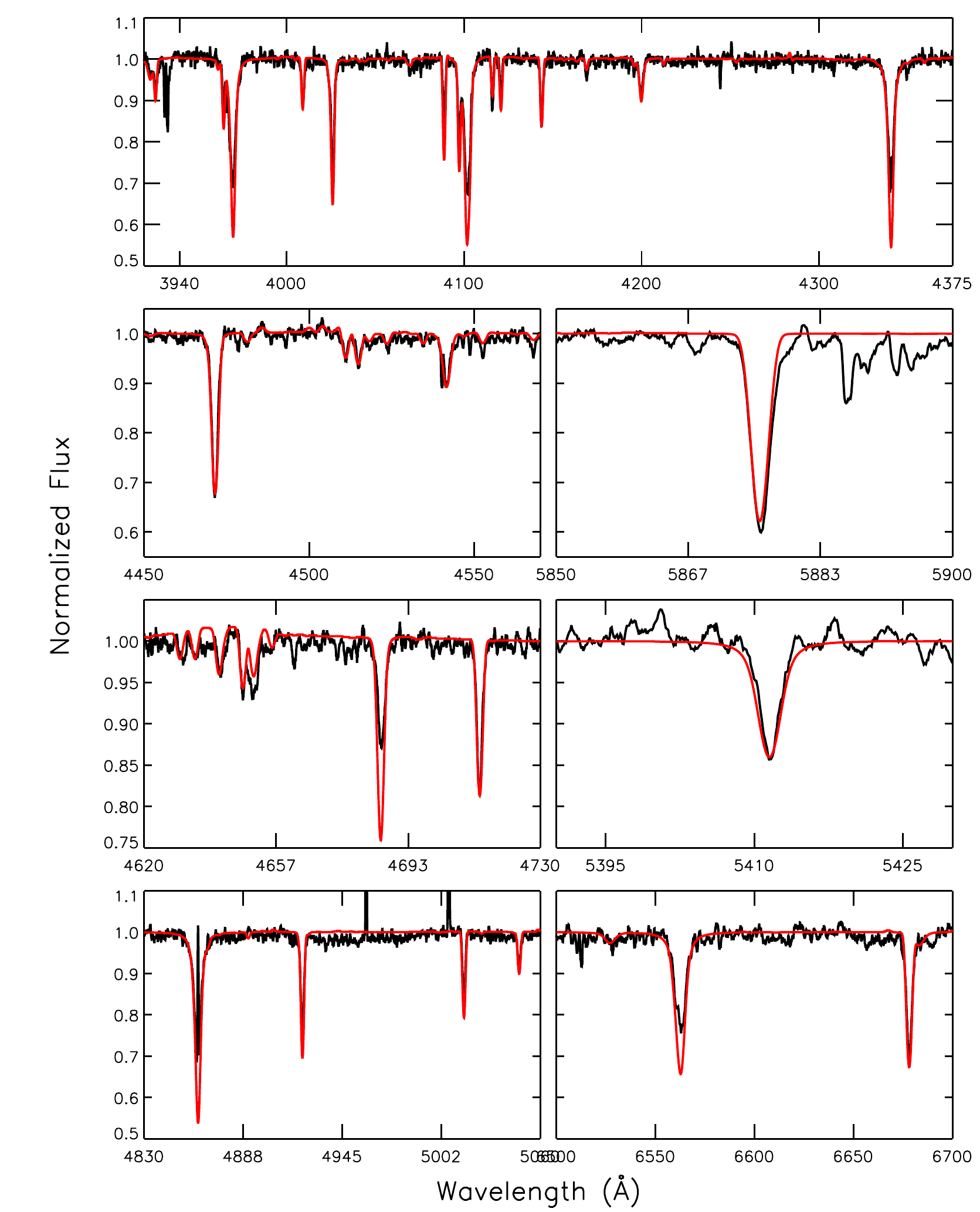}
      \caption[12cm]{Left: Best-fit model for \object{AV 327} (red line) compared to \stis\ spectrum (black line). Right: Best-fit model for \object{AV 327} (red line) compared to the 
      ESO/CASPEC spectrum (black line).}
         \label{Fig_AV327}
   \end{figure*}  
   
\begin{figure*}[tbp]
\includegraphics[scale=0.51, angle=0]{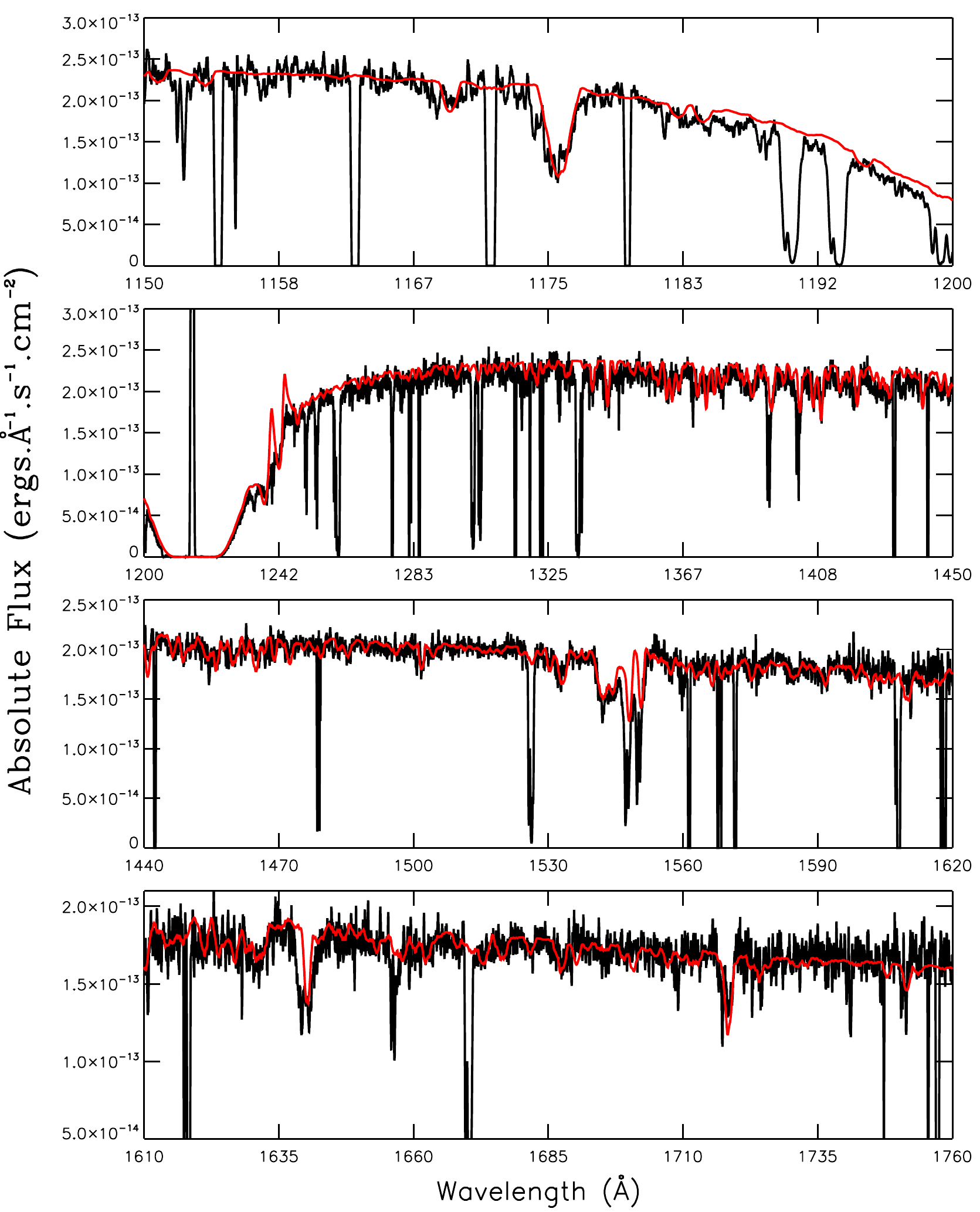}
      \caption[12cm]{Best-fit model for \object{AV 77} (red line) compared to the \cosp\ spectrum (black line).}
         \label{Fig_AV77}
   \end{figure*}

\begin{figure*}[tbp]
\includegraphics[scale=0.51, angle=0]{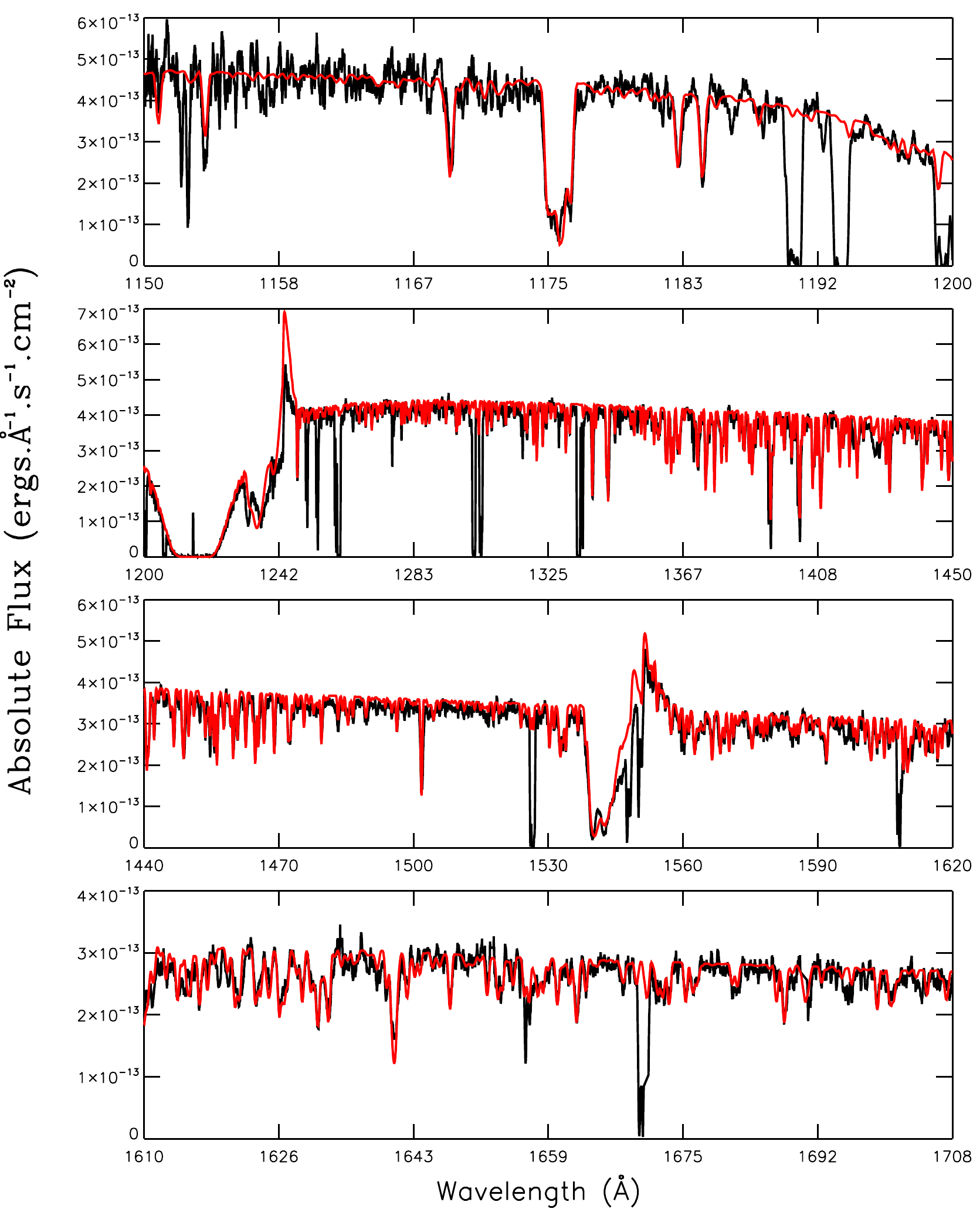}
\includegraphics[scale=0.51, angle=0]{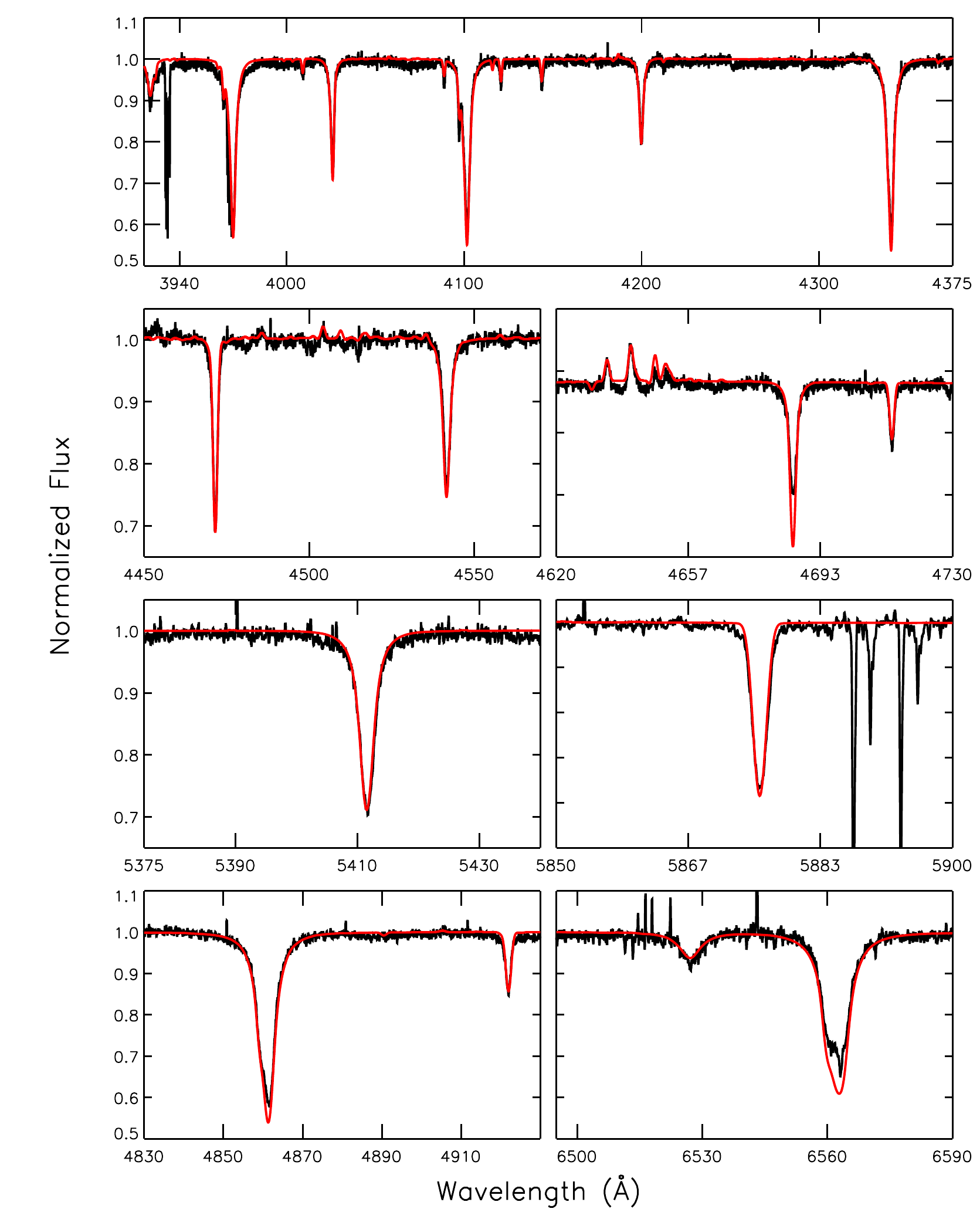}
      \caption[12cm]{Left: Best-fit model for \object{AV 95} (red line) compared to \stis\ spectrum (black line). Right: Best-fit model for \object{AV 95} (red line) compared to  the
      ESO/CASPEC spectrum (black line).}
         \label{Fig_AV95}
   \end{figure*}   

\begin{figure*}[tbp]
\includegraphics[scale=0.51, angle=0]{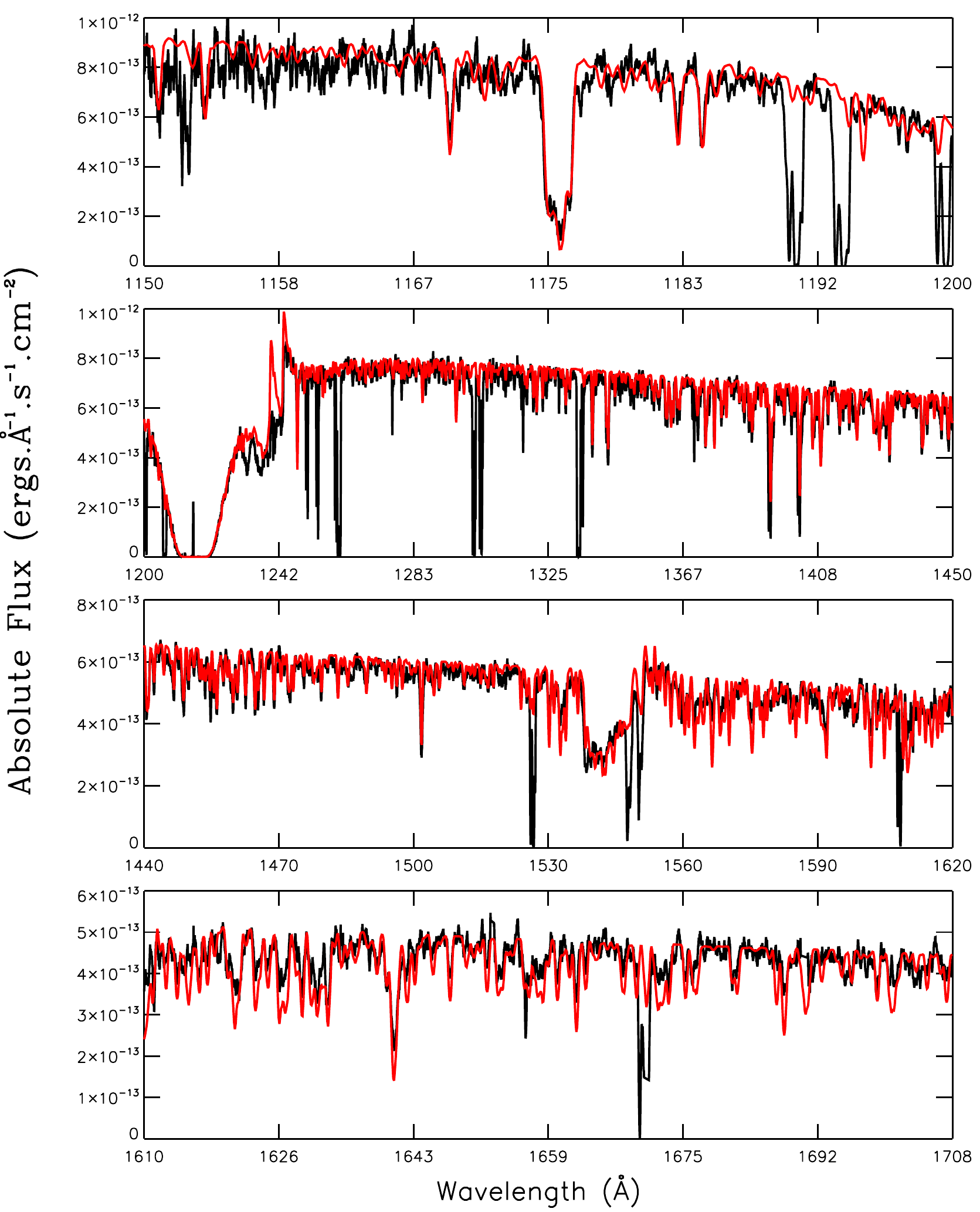}
\includegraphics[scale=0.51, angle=0]{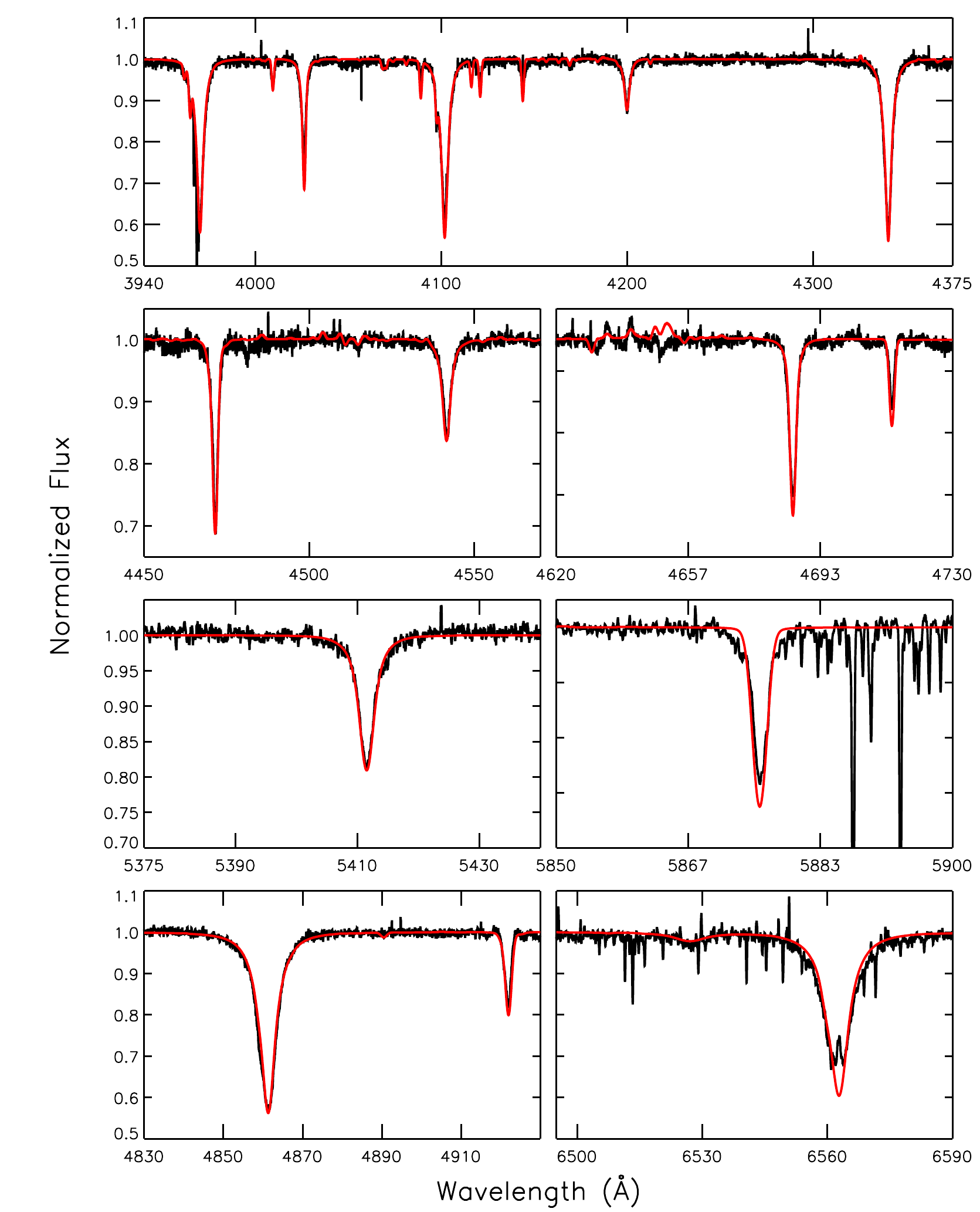}
      \caption[12cm]{Left: Best-fit model for \object{AV 47} (red line) compared to \stis\ spectrum (black line). Right: Best-fit model for \object{AV 47} (red line) compared to the 
      ESO/UVES spectrum (black line).}
         \label{Fig_AV47}
   \end{figure*}
 
 \begin{figure*}[tbp]
\includegraphics[scale=0.51, angle=0]{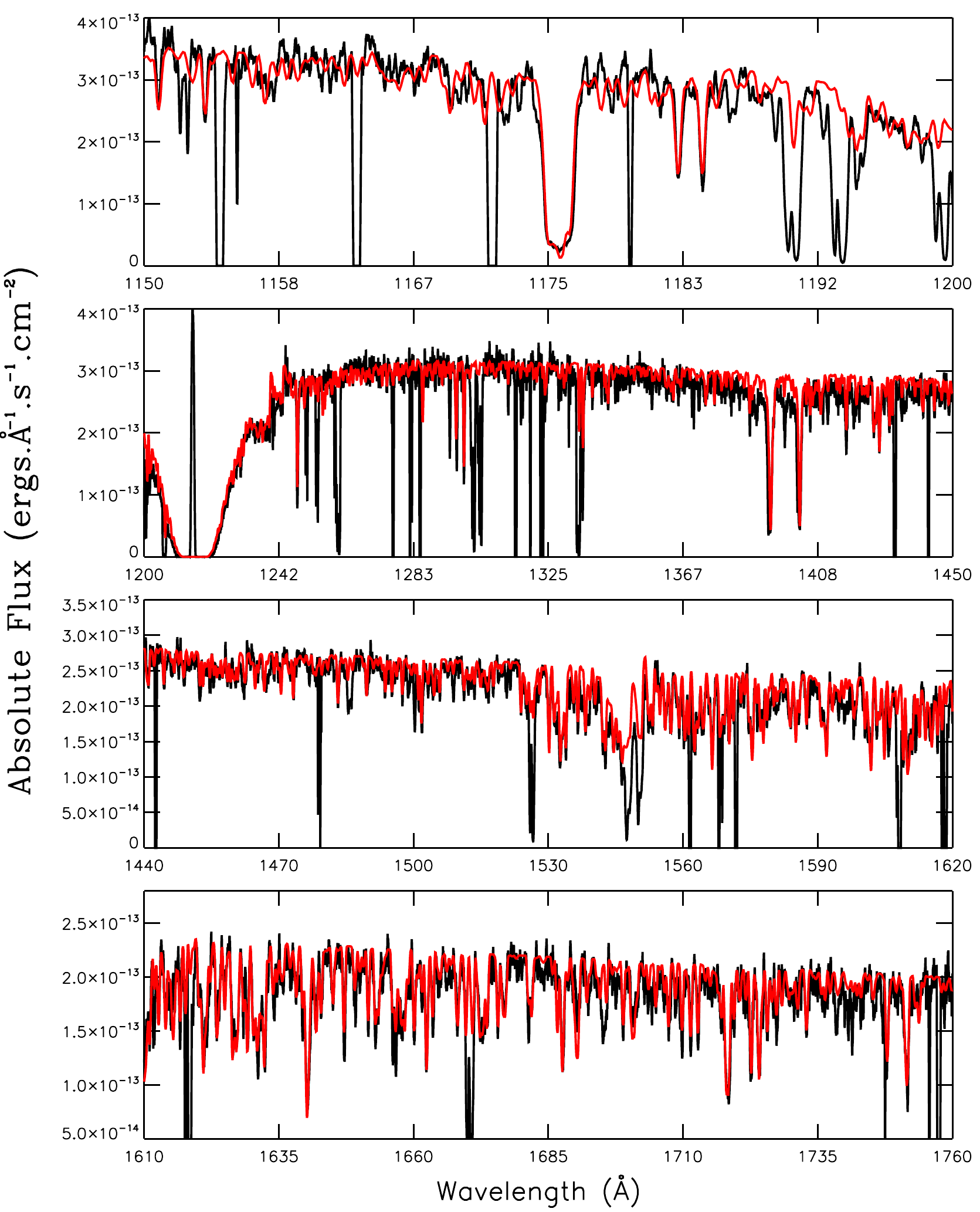}
      \caption[12cm]{Best-fit model for \object{AV 307} (red line) compared to the \cosp\ spectrum (black line).}
         \label{Fig_AV307}
   \end{figure*}   
  
  \begin{figure*}[tbp]
\includegraphics[scale=0.51, angle=0]{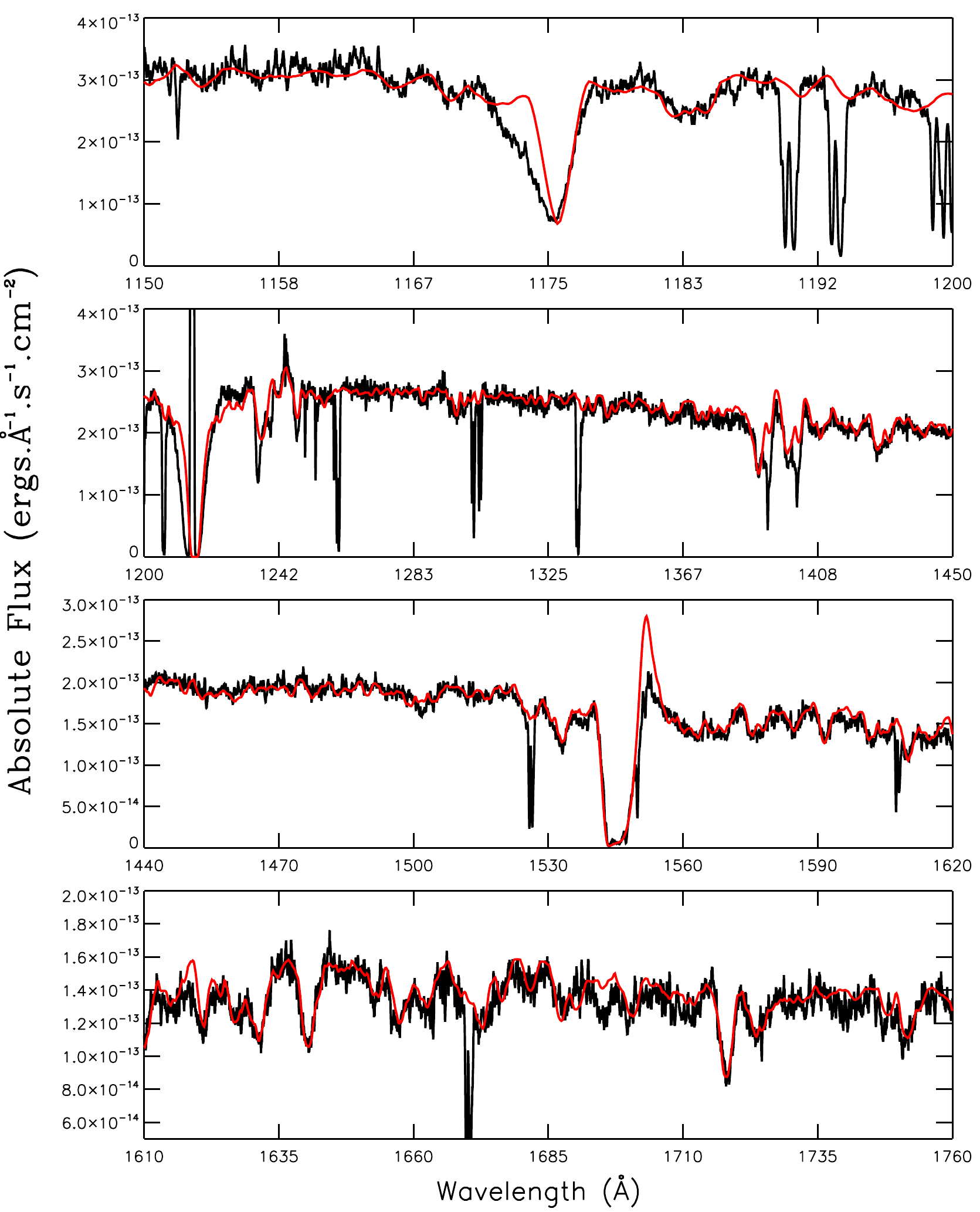}
      \caption[12cm]{Best-fit model for \object{AV 439} (red line) compared to the \cosp\ spectrum (black line).}
         \label{Fig_AV439}
   \end{figure*}

\begin{figure*}[tbp]
\includegraphics[scale=0.51, angle=0]{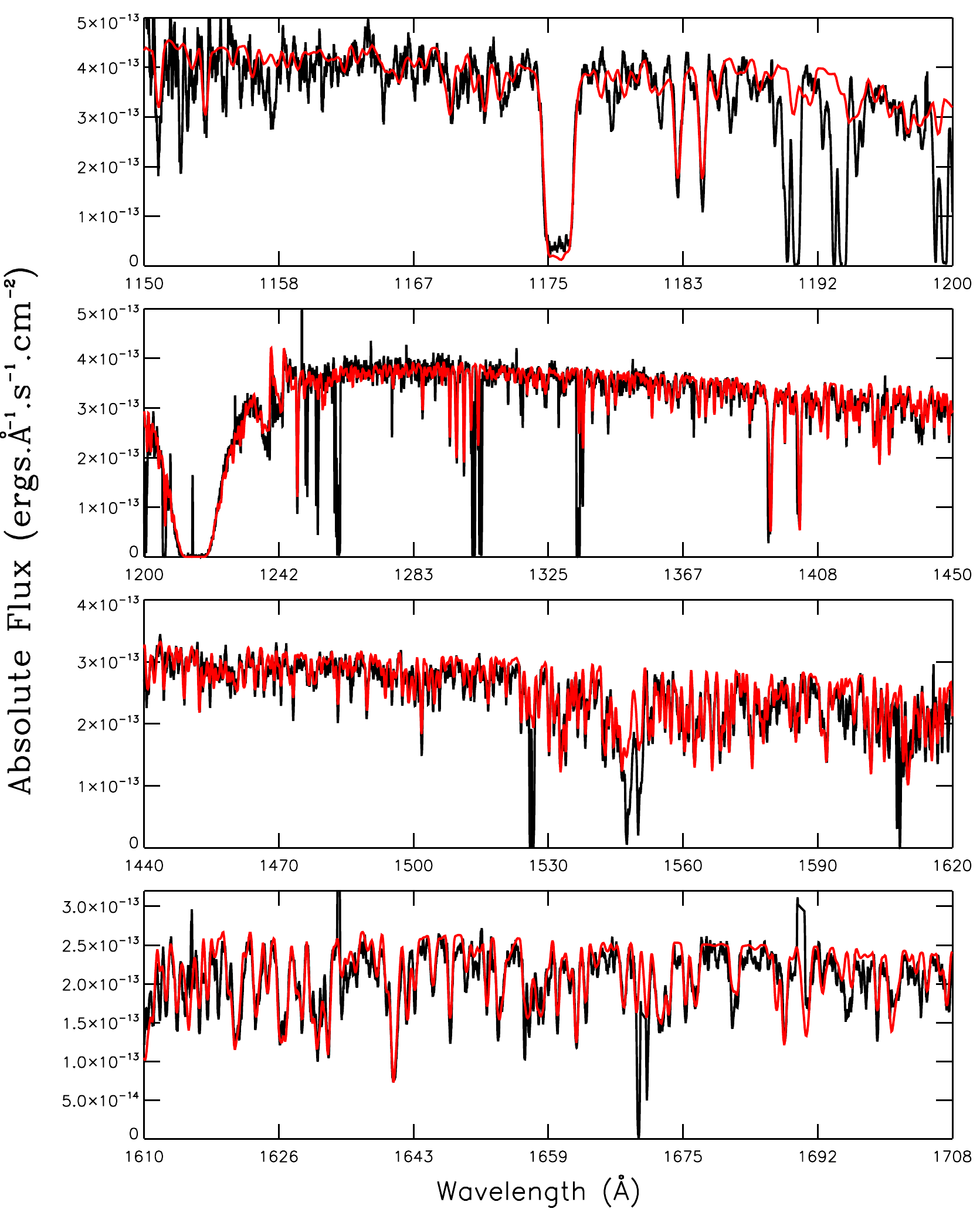}
\includegraphics[scale=0.51, angle=0]{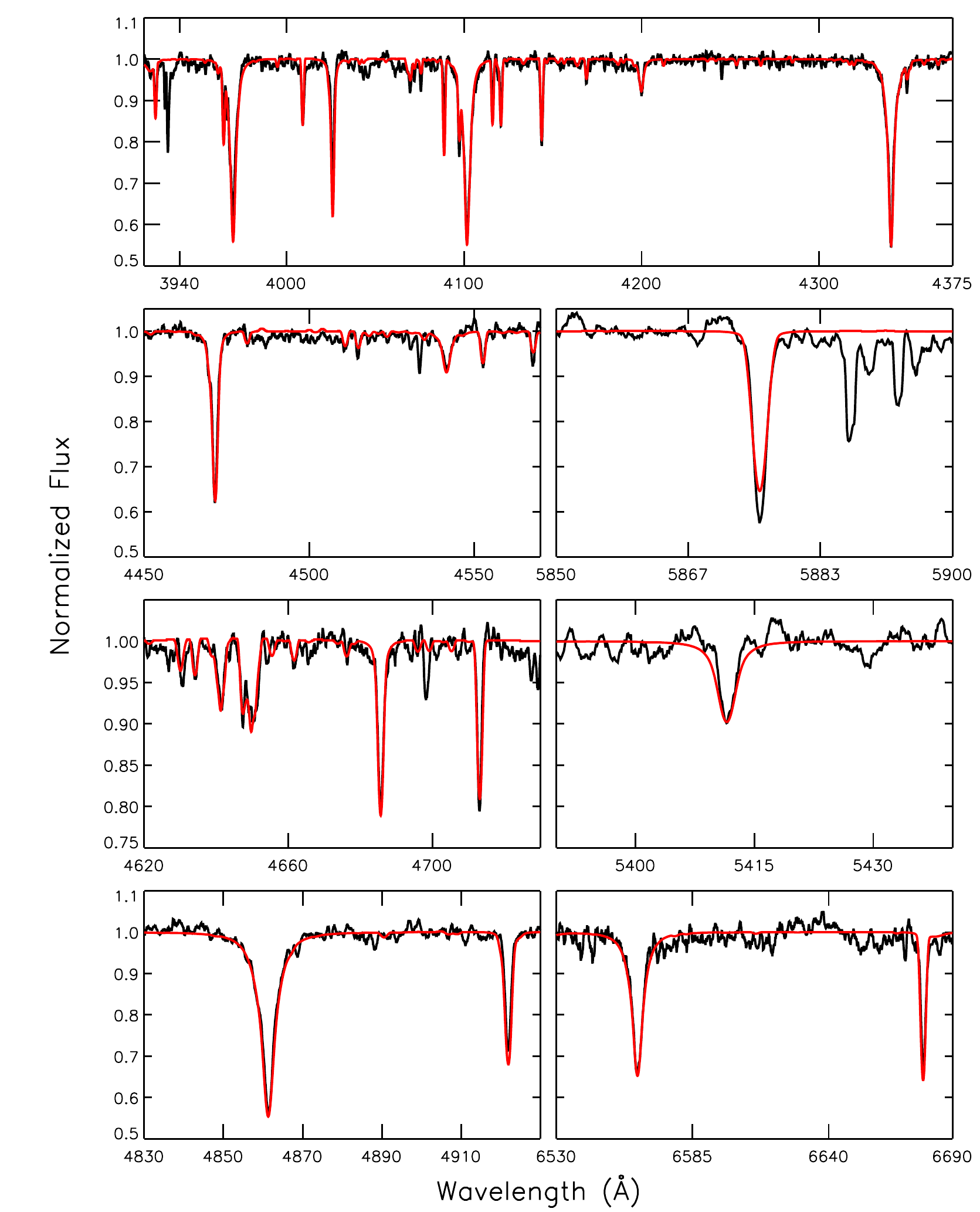}
     \caption[12cm]{Left: Best-fit model for \object{AV 170} (red line) compared to \stis\ spectrum (black line). Right: Best-fit model for \object{AV 170} (red line) compared to the 
      AAT/UCLES spectrum (black line).}
             \label{Fig_AV170}
   \end{figure*}
 
  \begin{figure*}[tbp]
\includegraphics[scale=0.51, angle=0]{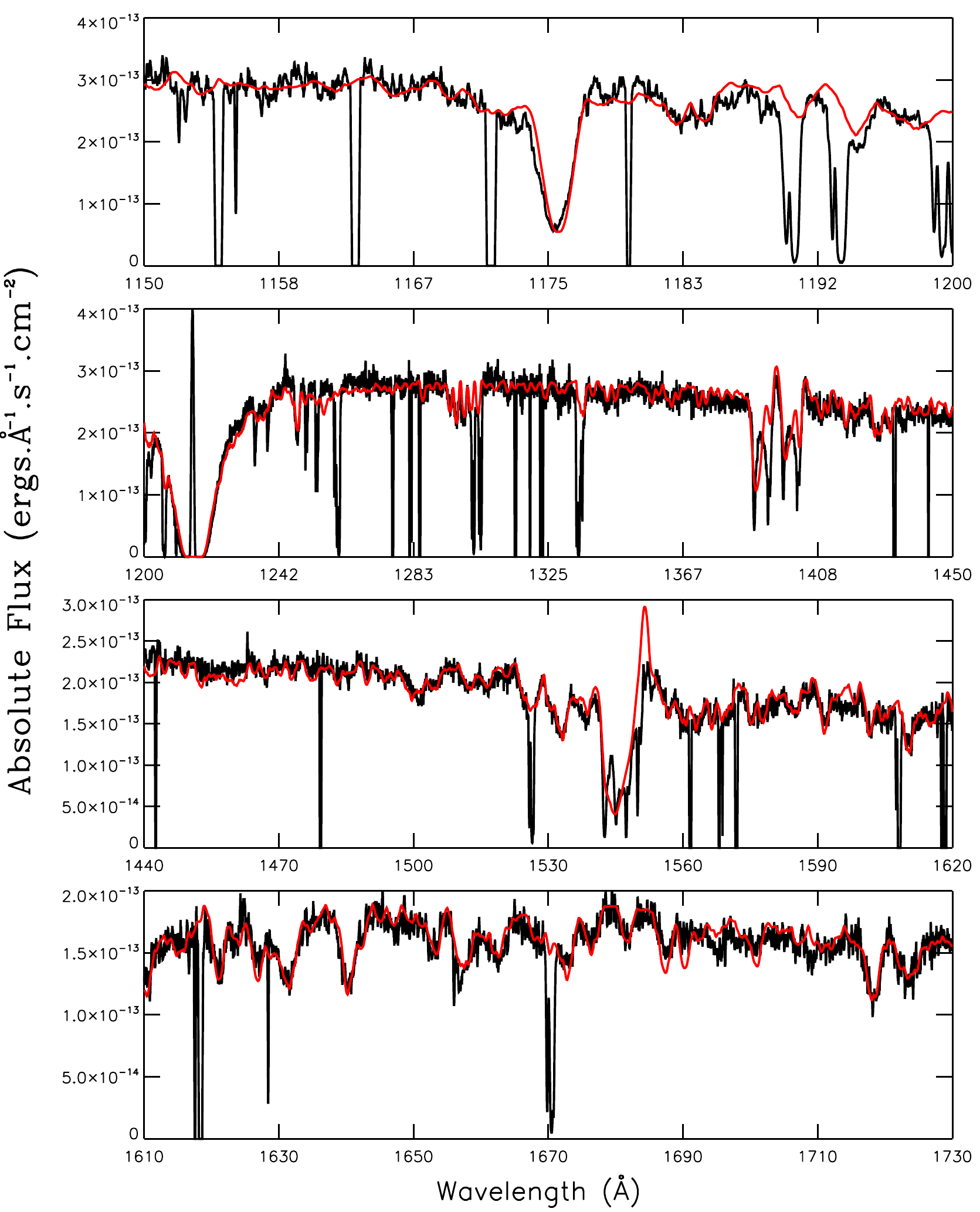}
\includegraphics[scale=0.51, angle=0]{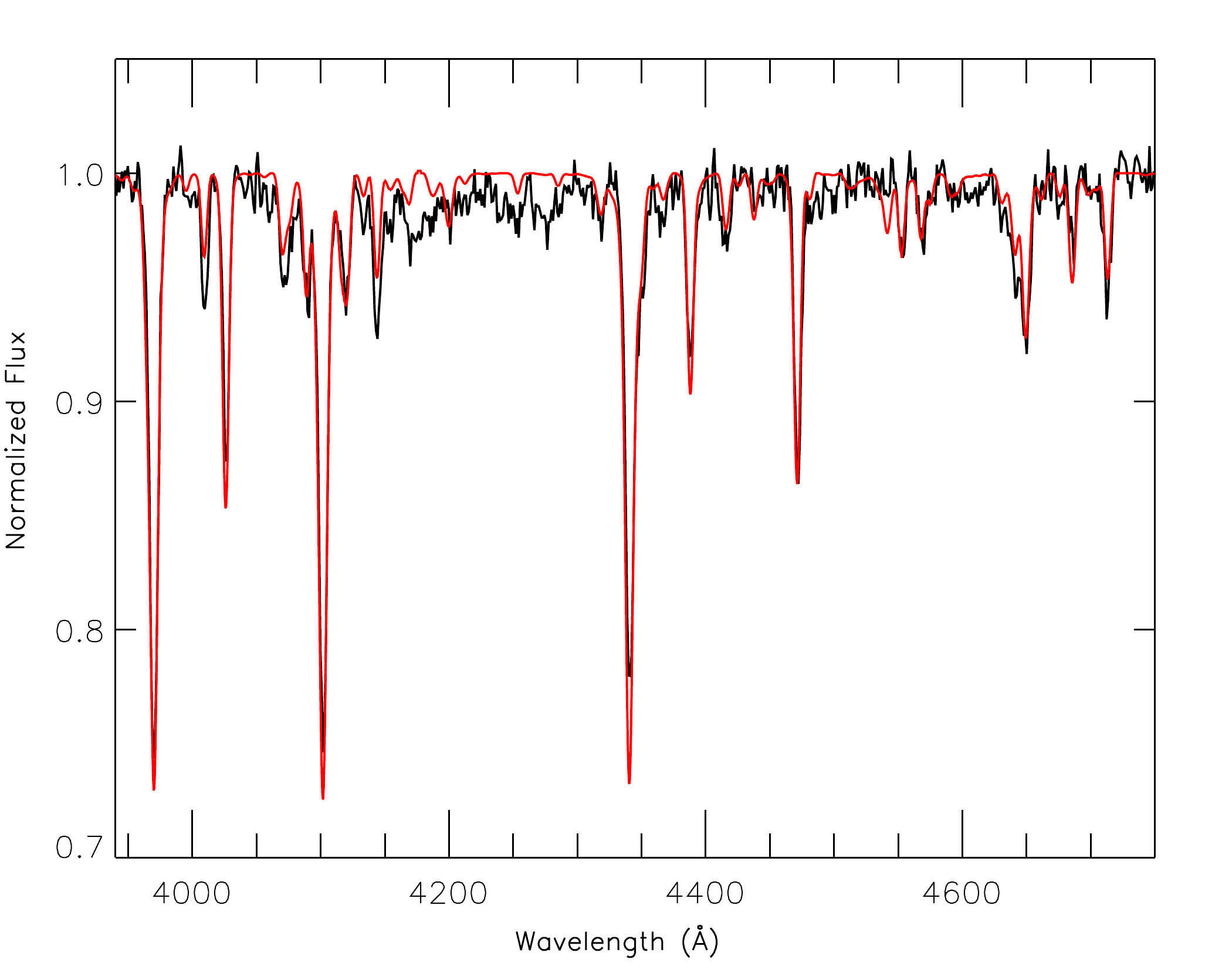}
      \caption[12cm]{Left: Best-fit model for \object{AV 43} (red line) compared to the \cosp\ spectrum (black line). Right: Best-fit model for \object{AV 43} (red line) compared to the
      2dF spectrum (black line).}
         \label{Fig_AV43}
   \end{figure*}

\end{appendix}

%                                     Two column figure (place early!)
%______________________________________________
%______________________________________________

\end{document}